\documentclass{article}
\pdfoutput=1
\usepackage[T1]{fontenc}
\usepackage[utf8x]{inputenc}
\usepackage[text={6in,8in},centering,letterpaper]{geometry}
\usepackage{amsmath,amssymb,amsfonts,amsthm,mathtools,array,booktabs,caption,microtype,graphicx,mathrsfs}
\usepackage{cite}
\usepackage[vcentermath]{youngtab}
\usepackage{tikz}
\usetikzlibrary{calc,trees,arrows,shapes.misc}
\tikzset{level 1/.style={level distance=1.5cm, sibling distance=3.5cm}}
\tikzset{level 2/.style={level distance=1.5cm, sibling distance=2cm}}
\usetikzlibrary{decorations.markings}
\usetikzlibrary{decorations.pathreplacing}
\usepackage[pdfusetitle,hidelinks]{hyperref}
\usepackage{bookmark}
\hypersetup{linktoc = all}
\hypersetup{bookmarksnumbered}

\newtheorem{proposition}{Proposition}[section]
\newcommand\bee{\begin{equation}}
\newcommand\eeq{\end{equation}}

\newcommand{\ZZ}{{\mathbb{Z}}}
\newcommand{\RR}{{\mathbb{R}}}
\newcommand{\CC}{{\mathbb{C}}}
\newcommand{\HH}{{\mathbb{H}}}
\newcommand{\onebf}{{\mathbf{1}}}
\newcommand{\Sbb}{{\mathbb{S}}}

\newcommand{\hfrak}{\mathfrak{h}}
\newcommand{\Nsusy}{\mathcal{N}}

\newcommand{\myoverline}[1]{\mathrlap{\overline{\phantom{#1}}}#1}
\newcommand{\mubar}{\myoverline{\mu}}
\newcommand{\nubar}{\myoverline{\nu}}
\newcommand{\Rbar}{\myoverline{R}}
\newcommand{\Vbar}{\myoverline{V}}
\newcommand{\Vhyp}{V_{\textnormal{hyp}}}

\newcommand{\Nbar}{\myoverline{N}}
\newcommand{\nbar}{\myoverline{n}}
\newcommand{\Mbar}{\myoverline{M}}
\newcommand{\mbf}{\mathbf{m}}

\newcommand{\pbf}{\mathbf{p}}
\newcommand{\zbf}{\mathbf{z}}

\newcommand{\Nst}{N_{\textnormal{s.t.}}}
\newcommand{\crel}{c_{\textnormal{rel}}}
\DeclarePairedDelimiter{\abs}{\lvert}{\rvert}
\DeclareMathOperator{\Tr}{Tr}
\DeclareMathOperator{\Span}{Span}
\DeclareMathOperator{\rank}{rank}
\DeclareMathOperator{\adj}{adj}
\DeclareMathOperator{\roots}{roots}

\DeclareMathOperator{\singlet}{singlet}
\DeclareMathOperator{\PE}{PE}
\numberwithin{equation}{section}

\newcommand{\Ad}{\operatorname{Ad}}

\makeatletter
\newcommand{\yngantisym}{{\let\@nomath\@gobble\let\typeout\@gobble\tiny\yng(1,1)}}
\makeatother

\newcommand{\lie}{\mathfrak}

\newcommand{\Hcal}{\mathcal{H}}
\newcommand{\Mcal}{\mathcal{M}}
\newcommand{\Ocal}{\mathcal{O}}

\newcommand{\half}{\frac{1}{2}}

\DeclareMathOperator{\Hom}{Hom}
\newcommand{\Lambdamon}{\Lambda_{\textnormal{mon}}}
\newcommand{\ZVand}{Z_{\textnormal{Vand}}}
\newcommand{\jH}{j^{\textnormal{H}}}
\newcommand{\jC}{j^{\textnormal{C}}}

\newcommand{\Ithree}{\mathbb{I}_{\textnormal{3d}}}

\newcommand{\Ivecthree}{I^{\textnormal{vec}}_{\textnormal{3d}}}
\newcommand{\Ihypthree}{I^{\textnormal{hyp}}_{\textnormal{3d}}}
\newcommand{\cvhh}{c_{\textnormal{vhh}}}
\newcommand{\nab}{n_{\textnormal{abel}}}
\newcommand{\wab}{w^{\textnormal{abel}}}
\newcommand{\phiab}{\phi_{\textnormal{abel}}}
\newcommand{\nna}{n_{\textnormal{na}}}
\newcommand{\phina}{\phi_{\textnormal{na}}}
\newcommand{\nsec}{n_{\textnormal{sectors}}}
\newcommand{\nCC}{n_{\CC}}
\newcommand{\nRR}{n_{\RR}}
\newcommand{\nHH}{n_{\HH}}
\newcommand{\RCC}{R^{\CC}}
\newcommand{\RRR}{R^{\RR}}
\newcommand{\RHH}{R^{\HH}}
\newcommand{\NCC}{N^{\CC}}
\newcommand{\NRR}{N^{\RR}}
\newcommand{\NHH}{N^{\HH}}
\newcommand{\RCCbar}{\Rbar^{\CC}}
\newcommand{\NCCbar}{\Nbar^{\CC}}
\newcommand{\Nadj}{N^{\adj}}

\newcommand{\pCC}{p^{\CC}}
\newcommand{\pRR}{p^{\RR}}
\newcommand{\pHH}{p^{\HH}}
\newcommand{\tq}{\tilde{q}}
\newcommand{\qCC}{q^{\CC}}
\newcommand{\tqCC}{\tilde{q}^{\CC}}
\newcommand{\Gammared}{\Gamma_{\textnormal{red}}}
\newcommand{\geqrep}{\succcurlyeq}

\newcommand{\Gelec}{G_{\textnormal{elec}}}
\newcommand{\Gmag}{G_{\textnormal{mag}}}

\newcommand{\ttau}{\widetilde{\tau}}

\title{Mixed moduli in 3d \texorpdfstring{${\cal N}=4$}{N=4} higher-genus quivers}
\author{}
\date{November 2022}
\hypersetup{pdfauthor = {Ioannis Lavdas and Bruno Le Floch}}

\begin{document}

\begin{flushright}LMU-ASC 18/22\end{flushright}

{\let\newpage\relax\maketitle}

\begin{center}
Ioannis Lavdas$^a$ and  Bruno Le Floch$^b$\\
\medskip\medskip\medskip

$^a$%
\textit{Arnold-Sommerfeld-Center for Theoretical Physics, Ludwig-Maximilians-Universit{\"a}t, 80333 M{\"u}nchen, Germany}

\medskip

$^b$%
\textit{CNRS, Laboratoire de Physique Théorique et Hautes Energies, Sorbonne Universit\'e, Paris, France}

\vskip 2cm

\begin{abstract}
We analyze exactly marginal deformations of 3d ${\cal N}=4$ Lagrangian gauge theories, especially mixed-branch operators with both electric and magnetic charges.
These mixed-branch moduli can either belong to products of electric and magnetic current supermultiplets, or be single-trace (non-factorizable).
Apart from some exceptional quivers that have additional moduli, 3d ${\cal N}=4$ theories described by genus~$g$ quivers with nonabelian unitary gauge groups have exactly $g$~single-trace mixed moduli, which preserve the global flavour symmetries.
This partly explains why only linear and circular quivers have known AdS$_4$ supergravity duals.
Indeed, for $g>1$, AdS$_4$ gauged supergravities cannot capture the entire $g$-dimensional moduli space even if one takes into account the quantization moduli of boundary conditions.
Likewise, in a general Lagrangian theory, we establish (using the superconformal index) that the number of single-trace mixed moduli is bounded below by the genus of a graph encoding how nonabelian gauge groups act on hypermultiplets.
\end{abstract}

\end{center}

\newpage

\begingroup
\let\oldpdfendlink\pdfendlink
\renewcommand\pdfendlink{\oldpdfendlink\vphantom{Iy}}
\tableofcontents
\endgroup

\newpage

% ===============================================
%					INTRODUCTION
% ===============================================

\section{Introduction}

One of the virtues of holography~\cite{Aharony:1999ti} is that deciphering properties of the theory on the gravitational side of the correspondence becomes feasible if the dual field theory is under strong technical control. 
The present work fits in the framework of the holographic correspondence between 3d $\Nsusy=4$ superconformal field theories (SCFTs) and IIB string theory on a warped AdS$_4$ background. Departing from the standard forms of the holographic dictionary~\cite{Hanany:1996ie,Assel:2011xz,Assel:2012cj}  (which relates linear or circular quiver gauge theories and the well defined corresponding AdS$_4$ backgrounds), we study general Lagrangian 3d $\Nsusy=4$ theories and their exactly marginal deformations in an effort to extract features about their (yet unknown) dual string solutions.

\paragraph{Exactly marginal deformations.}

We explore the moduli space of 3d $\Nsusy=4$ superconformal theories in three dimensions,\footnote{There has been an extensive study of deformations of 4d SCFTs that preserve superconformality~\cite{Leigh:1995ep,Kol:2002zt,Aharony:2002hx,Kol:2010ub,Cordova:2016xhm} and conformal manifolds of these SCFTs~\cite{Benvenuti:2005wi,Green:2010da} have been explored as well.} much beyond the class of linear quivers~$T_\rho^{\check{\rho}}[SU(N)]$ introduced in~\cite{Gaiotto:2008ak}, which we had treated in our previous work~\cite{Bachas:2019jaa}.
Presently, we consider all 3d $\Nsusy=4$ Lagrangian gauge theories that are ``good'' in the sense that their infrared fixed point includes neither free fields nor orbifolds thereof, in other words it is an interacting SCFT or a decoupled product of such SCFTs.\footnote{This notion of goodness is slightly stronger than the standard requirement that monopole operators have dimensions $\Delta\geq 1$.  The difference is exemplified by the abelian theories in \autoref{subsec:abelian} whose infrared limit involves an orbifold of free monopoles.}
Such theories are isolated as 3d $\Nsusy=4$ SCFTs, but they admit $\mathcal{N}=2$ preserving exactly marginal deformations, which are known to belong to three distinct multiplets of the 3d $\mathcal{N}=4$ algebra~\cite{Bachas:2017wva}.
These multiplets, denoted as $B_1[0]^{(2,0)}$, $B_1[0]^{(0,2)}$, $B_1[0]^{(1,1)}$ following~\cite{Cordova:2016emh}, feature (as their bottom components) Higgs, Coulomb and mixed branch operators.
Their number is captured by the superconformal index, or equivalently the partition function of the theory on $S^{2}\times S^{1}$, the computation of which is performed via supersymmetric localization. 
We provide fully explicit formulas for the number of all such operators, culminating in the dimension~\eqref{dimCCMcal} of the superconformal manifold.

Our main focus, specifically, is on analyzing the \emph{mixed} marginal moduli, which lie in $B_1[0]^{(1,1)}$~multiplets.

In the class of linear quiver gauge theories~\cite{Bachas:2019jaa} we had found that linear quivers have no \emph{single-trace mixed marginal moduli}: indeed, all mixed marginal moduli are expressed as a factorized product of electric and magnetic currents.
This was obtained first by an index calculation which shows that $B_1[0]^{(1,1)}$~multiplets transform in the $\adj(\Gelec)\otimes\adj(\Gmag)$ representation of the $\Gelec\times\Gmag$ global symmetry group (consisting of manifest electric flavour symmetries~$\Gelec$ and topological magnetic symmetries~$\Gmag$), modulo some relations between products of electric and magnetic current multiplets.
Secondly, the same result was obtained by expliciting how all $B_1[0]^{(1,1)}$~multiplets constructed from fields in the Lagrangian (and magnetic monopoles) factorize.

\paragraph{Main results on single-trace mixed moduli.}

By following a procedure analogous to the one for the case of linear quiver theories, we determine the low-lying BPS spectrum of general good Lagrangian 3d $\mathcal{N}=4$ theories via an index computation.
Our keystone result, whose precise assumptions are given in \autoref{prop:mixedmoduli}, is:
\begin{equation}\label{Nstgeqg}
  \Nst \geq g \ ,
\end{equation}
which provides a lower bound for the number~$\Nst$ of \emph{\underline{s}ingle-\underline{t}race mixed moduli}\footnote{We derive a stronger bound: there are at least $g$ such moduli with weight~$0$ under the Cartan torus of $\Gelec\times\Gmag$.} in terms of a notion of genus~$g$.
For quiver gauge theories with unitary gauge groups, $g$~simply denotes the genus of the quiver obtained by deleting $U(1)$ gauge group factors, namely it is the number of loops of \emph{nonabelian} gauge-nodes in the quiver.
On the other hand, abelian theories have $g=0$, and we prove that they do not possess single-trace mixed moduli, so that $\Nst=g=0$ saturates the bound.

Apart from the abelian case, and example theories discussed in \autoref{sec:examples}, our analysis only provides a lower bound for the number~$\Nst$ of such moduli, as the counting of their exact number turns out to be rather elaborate due to F-term relations.
As we determine in \autoref{sec:singletrace}, the moduli of interest involve either a vector multiplet or a monopole operator, dressed by a pair of hypermultiplet components to obtain a gauge-invariant combination.

In quiver gauge theories of unitary gauge groups $U(n_i)$, we identify explicitly the single-trace mixed moduli to be of the form
\begin{equation}\label{Be-intro}
  B_e = \Tr_i(\tq_e\phi_j q_e)
\end{equation}
for each oriented edge $e$ of the quiver joining nodes $i$ and~$j$, where $q_e, \overline{\tq_e}$ stand for chiral and anti-chiral components of a hypermultiplet in the bifundamental representation of $U(n_i)\times U(n_j)$, and $\phi_j$ is the $U(n_j)$ vector multiplet scalar.
As described near~\eqref{Be-def}, F-term relations imply that $B_e=-B_{\text{reversed}(e)}$ and that the sum of all $B_e$ starting from a given node~$i$ vanishes.  For instance in a circular quiver all $B_e$ are identified to each other up to signs, leading to one single-trace operator, only.
More generally, the number of linearly independent~$B_e$ is the genus of the quiver.\footnote{For hypermultiplets in other representations there may be several ways to contract indices, leading to multiple single-trace operators counted by the genus.  For instance, each adjoint hypermultiplet of $SU(n)$ increases the genus by one for $n\geq 3$ and zero for $n=2$.  The difference can be tracked down to the identity $\Tr(abc)=-\Tr(acb)$ in $\lie{su}(2)$.}
Whenever a node is abelian, namely $n_j=1$, the manifest factorization $B_e=\phi_j\Tr_i(\tq_e q_e)$ into the product of (gauge-invariant) vector multiplet and hypermultiplet components implies that $B_e$ is no longer single-trace: this explains why $U(1)$ factors must be removed when defining the genus~$g$.

The bound~\eqref{Nstgeqg} is not always an equality: if the quiver has multiple edges $e,e'$ joining the same nodes, operators of the form $\Tr_i(\tq_e\phi_j q_{e'})$ with chiral components of different hypermultiplets provide additional mixed moduli that are not accounted for by the genus.
In \autoref{subsec:quivers}, we analyze the case of \emph{(good) quivers with $U(n_i)$ gauge groups} with fundamental and bifundamental matter, and find that such moduli are necessarily charged under flavour symmetries: to be precise, the number of single-trace mixed moduli that are neutral under the Cartan torus of $\Gelec\times\Gmag$ is exactly~$g$.
We conjecture that this remains valid in the presence of adjoint matter.\footnote{While the adjoint representation of $U(n)$ contains a singlet, we omit the resulting free hypermultiplet from the definition of the quiver theory.}
We also investigate in \autoref{subsec:circular} some circular quivers, for which we determine precisely the representations of $\Gelec\times\Gmag$ in which the single-trace $B_1[0]^{(1,1)}$ multiplets transform.

\paragraph{Holographic interpretation.}

As mentioned above, an important trait of the linear and circular quiver theories is that they admit a dual holographic description. The AdS$_{4}$ is fibered over a 6d base manifold comprised of a product of two-spheres wrapped over a Riemann surface: $\mathcal{M}_{6}=(S^{2}\times\hat{S}^{2})\ltimes \Sigma$.  The local form of the solutions was found in~\cite{DHoker:2007zhm} while the global solutions and the exact holographic dictionary were developed in~\cite{Hanany:1996ie,Assel:2011xz} for duals of linear quivers, and in~\cite{Assel:2012cj} for the duals of circular quivers, see also~\cite{Cottrell:2016nsu,Lozano:2016wrs}.
The solution is labelled by a set of discrete parameters corresponding to the positions and the charges of D3, D5 and NS5 branes supported in the solution. The electric ($B_{1}^{(2,0)}$) and magnetic ($B_{1}^{(0,2)}$) moduli correspond then to open strings on the D5 and NS5 branes, respectively, while the mixed ones ($B_{1}^{(1,1)}$) are either closed string states (massive gravitini superpartners, of which there is at most one) corresponding to the above $B_e$ operators, or double-string states factorizing into electric and magnetic building blocks.

Our previous analysis of linear quiver gauge theories shows that the corresponding four dimensional gauged $\mathcal{N}=4$ supergravity~\cite{Louis:2014gxa} captures the Higgs and Coulomb branch moduli automatically, and captures the factorized mixed moduli through appropriate modification of the AdS boundary conditions~\cite{Witten:2001ua,Berkooz:2002ug}, and not directly because the spectrum does not include massive spin-$\frac{3}{2}$ multiplets~\cite{Bachas:2017wva}.
For the general classes of theories studied in the present work there is a minimal number of single-trace $B_{1}[0]^{(1,1)}$ multiplets given by the lower bound we determine, therefore a putative supergravity dual would not be able in principle to accommodate the full moduli space of gauge theories whenever $g>1$, even after taking into account the quantization moduli of the AdS boundary conditions.

Since the genus is defined while omitting $U(1)$ factors, our statements hold equally for unitary and special unitary gauge groups.  This is consistent with the fact that the holographic dual of \emph{linear} quivers with special unitary gauge groups is known~\cite{Collinucci:2020kdm} in a way that naturally extends to circular quivers, but not to higher genus $g>1$.
Two very interesting classes of quivers with loops were introduced in the study of Argyres--Douglas theories~\cite{Dey:2020hfe}, and of magnetic quivers for $SU(n)$ gauge theories, such as those displayed in Figure~4 of~\cite{Bourget:2021jwo}.  One expects that their string theory origin should lead to an AdS$_4$ holographic dual description, which requires these theories to have no single-trace $B_1[0]^{(1,1)}$ multiplet.  This expectation is borne out: indeed, loops in these quivers always include at least one $U(1)$ gauge node, so that the genus is $g=0$.

In contrast, star-shaped quivers, namely mirrors of the circle reduction of 4d $\Nsusy=2$ class~S theories associated to genus~$g$ punctured Riemann surfaces~\cite{Benini:2010uu}, have $g$~single-trace mixed moduli.\footnote{We thank the JHEP referee for suggesting a discussion of this class of theories.}
Indeed, these quivers consist of unitary gauge groups connected by bifundamental hypermultiplets, together with a central $SU(N)$ node with $g$ adjoint hypermultiplets, and these adjoints lead to mixed moduli of the form~\eqref{Be-intro}.
For $g>1$ this is in tension with our results since these theories definitely have a string theory construction.
The tension is likely resolved by noting that the known M-theory holographic dual of the class~S theory cannot be T-dualized into a IIB holographic dual of the 3d theory without becoming singular.
It would be worthwhile to determine whether a IIB dual exists for $g\leq 1$, as this is not ruled out by our work.

\paragraph{Organization of the paper.}

This work is structured as follows.  We start in \autoref{sec:index} with a review of the superconformal index of a good 3d $\mathcal{N}=4$ theory. We present its definition and the steps comprising its computation via supersymmetric localization. In \autoref{sec:marginal} we perform the computation of the index of a general 3d $\mathcal{N}=4$ theory with a careful treatment of the contribution of each monopole sector. Finally, \autoref{sec:singletrace} is devoted to extracting lower bounds on the number of single-trace $B_{1}^{(1,1)}$ multiplets in general theories, and for special classes of theories: purely abelian, and quivers of unitary gauge groups.  We also present examples in \autoref{sec:examples}.  To support some of the calculations, we complete the note with \autoref{app:representations} in which we outline some representation-theoretic results that are extensively used throughout the main text. While the present analysis is purely field theoretic, it suggests directions that can be explored in the holographically dual solutions.\footnote{C.~Bachas, A.~Bourget, I.~Lavdas and B.~Le Floch, work in progress}

% ===============================================
% 					SECTION 2
% ===============================================

\section{Superconformal index}\label{sec:index}

In this section we review the definition and calculation of  the superconformal index of 3d $\Nsusy=4$ theories. 
We present the explicit localization formula (which could be equivalently obtained by letter counting) and describe how to expand it.

\subsection{The 3d index and its parameters}

\paragraph{Definition of the index.}
The 3d $\Nsusy=2$ superconformal index, defined in terms of the cohomology of the supercharge $Q=Q_-^{++}$, has been extensively studied (see for instance~\cite{Bhattacharya:2008zy,Kim:2009wb,Imamura:2011su,Krattenthaler:2011da,Kapustin:2011jm,Benini:2013yva} and the review~\cite{Willett:2016adv}).
It counts all quantum states of the SCFT on the two-sphere (or equivalently local operators) with suitable powers of fugacities $q, t, e^{-\beta}, x_F$ associated to components of the Cartan subalgebra of the global symmetry algebra:
\begin{equation}\label{sci}
\Ithree = \Tr_{\Hcal_{S^{2}}}\biggl[
(-1)^{F}e^{-\beta(\Delta-J_3 -J_3^H -J_3^C)} q^{\frac{1}{2}(\Delta + J_3)}t^{J_3^H-J_3^C}\prod_F x_F^{K^F} \biggr] \ ,
\end{equation}
where $F$ is the state's fermion number, $J_3$~the third component of its spin, 
$\Delta$ its energy, and $K^F$ are flavour charges, while the $SU(2)_H$ and $SU(2)_C$ R-symmetry spins appear as a $U(1)$ R-charge $J_3^H+J_3^C$ and a $U(1)$ flavour charge $J_3^H-J_3^C$ from the 3d $\Nsusy=2$ point of view.
The index can be further refined by turning on background fluxes on~$S^2$ for the flavour groups~\cite{Kapustin:2011jm} but we will not include this possibility.

The index is independent of the fugacity $\beta$ by the standard argument: states with $\Delta \neq J_3 + J_3^H+ J_3^C$ are paired by~$Q_-^{++}$ and their contribution cancels due to~$(-1)^F$.
This enables us to replace $J_3$ by $\Delta - J_3^H - J_3^C$ in the exponent of~$q$ and write the index as
\begin{equation}\label{sci-2}
\Ithree = \Tr_{\Hcal_{S^{2}}}\biggl[
(-1)^{F} x_+^{2(\Delta-J_3^C)} x_-^{2(\Delta-J_3^H)} \prod_F x_F^{K^F} \biggr] \ ,
\qquad
x_\pm =  q^{\frac{1}{4}}t^{\pm{\frac{1}{2}}} \ .
\end{equation}

The index is thus a series in non-negative integer powers of fugacities $x_\pm$, with coefficients that are Laurent polynomials in the remaining fugacities.
These, in turn, are associated to a maximal commuting set of flavour symmetries that are manifest in the Lagrangian description (thus excluding symmetries that are only present in the infrared).
\begin{itemize}
\item Electric fugacities~$\mu$, in the Cartan torus of the flavour group~$\Gelec$.
\item Magnetic fugacities $w$ for topological symmetries, one per abelian factor $U(1)_i$ of~$G$.
\end{itemize}
Topological symmetries often belong to a larger non-abelian group~$\Gmag$ of magnetic flavour symmetries of the infrared theory~\cite{Borokhov:2002ib, Borokhov:2003yu}.
For certain theories (such as orthosymplectic quivers), the rank of the emergent symmetry~$\Gmag$ is larger than the number of $U(1)$ topological symmetries, and it is not known how to include in the index some fugacities for the additional Cartan generators.

An interesting property of the index is that it reduces~\cite{Razamat:2014pta} to the Coulomb branch Hilbert series~\cite{Hanany:2011db,Cremonesi:2013lqa,Cremonesi:2014kwa,Cremonesi:2014uva,Cremonesi:2016nbo,Cremonesi:2017jrk} (which counts monopole operators parametrizing the Coulomb branch) upon setting $x_+=0$ and to the Higgs branch Hilbert series (which counts gauge-invariant combinations of hypermultiplets parametrizing the Higgs branch) upon setting $x_-=0$. The two branches are exchanged in the case of a pair of theories related by 3d mirror symmetry and this is also reflected on the index~\cite{Okazaki:2019ony}.
% Mixed moduli appear at order $x_+x_-$ and higher,
Finally, the index additionally counts mixed operators~\cite{Carta:2016fjb}.

\paragraph{R-symmetry mixing.}
The index is invariant under continuous deformations such as the RG flow (barring wall-crossing phenomena due to non-compact moduli spaces~\cite{Cecotti:1992rm,Gaiotto:2010okc}).
At first sight this means the index of the IR SCFT is the same as that of the UV SCFT\@.
However, the definition~\eqref{sci} involves the R-symmetry that belongs to the superconformal symmetry algebra.
In general 3d $\Nsusy=2$ theories, the appropriate $U(1)$ R-symmetry in the IR differs from the UV one by a mixing with other (abelian) flavour symmetries, in a way characterized by $F$-extremization~\cite{1012.3210}.

The nonabelian R-symmetry of 3d $\Nsusy\geq 3$ theories makes such a mixing much rarer.
It still occurs in bad theories~\cite{1707.03403,1802.04285} where the R-symmetry mixes with low-energy non-abelian flavour symmetries of hypermultiplets that are free in the infrared, in such a way that unitarity bounds are restored.
Such mixing is expected to be absent in good theories.
As a result, for good 3d $\Nsusy=4$ theories the index $\Ithree$ can be computed in the UV limit of the theory, which is a free gauge theory.

\subsection{Formulas for the 3d index}

In the following we describe the formulas obtained by realizing the superconformal index~\eqref{sci} as a partition function of the theory on $S^2\times S^1$, with suitable background gauge fields, which is evaluated by supersymmetric localization, see the review~\cite{Willett:2016adv}.

\paragraph{Localization formula.}

We consider here a 3d $\Nsusy=4$ gauge theory with gauge group~$G$ of rank $r=\rank G$, and half-hypermultiplet matter in representation~$\Vhyp$ of~$G$.
The twisted partition function on $S^2\times S^1$ can be computed using supersymmetric localization.
This yields a multiple sum and integral over supersymmetric vacua characterized by the $S^1$ holonomy~$\zbf$ of the gauge field and its 2-sphere flux~$\mbf$ (monopole charges of the corresponding local operator in~$\RR^3$) that commute hence can be gauge transformed to the same Cartan torus $T\subset G$.
In this way, $\zbf$~and $\mbf$ are identified with its diagonal components $z_i\in U(1)$ and $m_i\in\ZZ$, where $i=1,\dots,r$ labels the Cartan generators of the gauge group~$G$.
The partition function (or index) is then an integral over $\zbf\in T\simeq U(1)^r$ and a sum over $\mbf\in\Lambdamon\simeq\ZZ^r$ of a product of vector multiplet and hypermultiplet one-loop determinants.

With notation explained below, the index then reads
\begin{equation}\label{fullindex}
  \Ithree
  = \frac{1}{|W(G)|} \sum_{\mbf\in\Lambdamon}
  \!\! x_-^{2\Delta(\mbf)}
  w^{\mbf}
  \int_{\zbf\in T} \!d^r\!\biggl(\frac{\log \zbf}{2\pi i}\biggr)
  \ZVand
  \PE \bigl( \Ivecthree+\Ihypthree \bigr) \ ,
\end{equation}
expressed in terms of
\begin{equation}
  \begin{aligned}
    \ZVand & = \prod_{\alpha\in\roots(G)} \bigl(1-(x_-x_+)^{\abs{\alpha\cdot\mbf}}\zbf^\alpha\bigr) \ ,
    \\
    \Ivecthree & = \frac{x_-^2 - x_+^2}{1-x_-^2 x_+^2} X^{\text{vec}} \ ,
    & X^{\text{vec}} & = \sum_{\omega\preceq \adj(G)} (x_-x_+)^{\abs{\omega\cdot\mbf}}\zbf^\omega \ ,
    \\
    \Ihypthree & = \frac{x_+ (1 - x_-^2)}{1-x_-^2 x_+^2} X^{\text{hyp}} \ ,
    & X^{\text{hyp}} & = \sum_{(\omega,\omega_e)\preceq \Vhyp} (x_-x_+)^{\abs{\omega\cdot\mbf}} \zbf^\omega\mu^{\omega_e} \ .
  \end{aligned}
\end{equation}
The vector multiplet one-loop determinant (together with some integration factors) consists of an analogue $\ZVand$ of Vandermonde determinant and $\PE(\Ivecthree)$, while the matter multiplet one-loop determinant is $\PE(\Ihypthree)$, with $\PE$ defined next.
The sums in $X^{\text{vec}}$ and~$X^{\text{hyp}}$ range over weights of the adjoint representation of~$G$ and of the matter representation of $G\times \Gelec$, respectively.
The exponential notation $\zbf^\omega$ denotes
\begin{equation}
  \zbf^\omega = e^{\omega\cdot\log\zbf} \in U(1)\subset \CC \ ,
\end{equation}
where $\log\zbf$ is any element of the Cartan \emph{algebra} that maps to~$\zbf$ under the exponential map.  Such an element is defined modulo $2\pi i$ times our favorite lattice~$\Lambdamon$, and since $\omega$ takes integer values on this lattice the exponential is well-defined.

This product is integrated with respect to the invariant integration measure on the Cartan torus~$T$ with volume normalized to~$1$, and further quotiented by the cardinal $|W(G)|$ of the Weyl group of~$G$, due to discrete gauge transformations that leave the Cartan torus invariant.
Finally, the integral is summed over monopole charges~$\mbf$, taking into account the monopole dimension\footnote{The formula is usually written for full hypermultiplets $\Vhyp=R\oplus\Rbar$, with a sum over weights of~$R$ only, hence a factor of $1/2$ instead of~$1/4$.  For half-hypermultiplets the dimension remains half-integral because weights of (pseudo)real representations are invariant under $w\mapsto-w$.}
\begin{equation}\label{Deltambf}
  \Delta(\mbf) = \frac{1}{4} \sum_{\omega\preceq\Vhyp} |\omega\cdot\mbf| - \frac{1}{2} \sum_{\alpha\in\roots(G)} |\alpha\cdot\mbf| \ ,
\end{equation}
and the magnetic fugacities $w_i$, raised to powers given by projecting $\mbf$ onto each abelian factor $U(1)_i$ of~$G$:
\begin{equation}
  w^{\mbf} = \prod_{i\in I} w_i^{\Tr_i(\mbf)} \ .
\end{equation}

\paragraph{Plethystic exponential.}
The one-loop determinants are conveniently expressed in terms of the plethystic exponential~($\PE$) operation~\cite{Feng:2007ur},
defined for a power series $f(v_1, v_2, \dots)$ with no constant term (namely such that $f(0,0,\dots)=0$) by
\bee
\PE(f) = \exp\biggl( \sum_{n=1}^\infty {\frac{1}{n}} f(v_1^n, v_2^n, \cdots) \biggr) \ .
\eeq
Equivalently, it is uniquely determined by imposing that $\PE(-\prod_i v_i^{k_i}) = (1-\prod_i v_i^{k_i})$ for any monomial $\prod_i v_i^{k_i}$ (where the integers $k_i$ are not all zero), and that $\PE(f+g)=\PE(f) \,\PE(g)$ for any power series $f,g$ without constant terms.
The idea of the plethystic exponential is that it translates single-particle indices $\Ivecthree$ and~$\Ihypthree$ to multi-particle indices:
for instance, if the one-particle Hilbert space consists of one bosonic state with charges $k_i$ under the $i$-th symmetry, then the one-particle index is $\prod_i v_i^{k_i}$ while the multi-particle index is
\bee
\PE\Bigl(\prod_i v_i^{k_i}\Bigr) = \frac{1}{1-\prod_i v_i^{k_i}} = \sum_{n\geq 0} \Bigl(\prod_i v_i^{k_i}\Bigr)^n .
\eeq
The plethystic exponential gives a very compact way to encode infinite products, for instance
\bee
\PE\Bigl(\frac{-a}{1-q}\Bigr) = \prod_{n\geq 0} (1 - aq^n) = (a,q)_\infty ,
\eeq
and the factors $\PE(\Ivecthree)$ and $\PE(\Ihypthree)$ in~\eqref{fullindex} are finite products of such $q$-Pochhammer symbols $(a,q)_\infty$.
An alternative expression is
\begin{equation}
\PE(f) = \sum_{n\geq 0} \Sbb^n f
\end{equation}
in terms of the (supersymmetric) symmetrized tensor power~$\Sbb^n$, which obeys for instance $\Sbb^0R=1$, or $\Sbb^2(R_1-R_2)=S^2R_1+\Lambda^2R_2-R_1R_2$ in terms of the usual symmetric and antisymmetric powers.

\paragraph{Projection onto gauge singlets.}

In the simplest case $\mbf=0$, the integrand $\PE(\Ivecthree+\Ihypthree)$ is a character of~$G$ (and of global symmetries), which can be decomposed into characters of irreducible representations.
Integrating over the Cartan torus with the Vandermonde determinant~$\ZVand$ and dividing by~$|W(G)|$ has the effect of projecting the operator down to gauge singlets (the trivial character of~$G$).

For $\mbf\neq 0$, the gauge symmetry is broken down to a smaller group $G_{\mbf}\subset G$ consisting of elements that commute with the magnetic flux~$\mbf$ (in particular $G_{\mbf}\supset T$), so that gauge-invariant operators need only be $G_{\mbf}$-invariant rather than $G$-invariant.
This is reflected in $\ZVand$, which splits into two factors according to which roots $\alpha$ have $\alpha\cdot\mbf=0$ or $\neq 0$.
First, we have the Vandermonde determinant of~$G_{\mbf}$, and the integral over~$T$ with this smaller Vandermonde determinant projects down to $G_{\mbf}$ singlets, as expected physically.
Second, we have remaining $\alpha\cdot\mbf\neq 0$ factors.
It is convenient to write them as a plethystic exponential, so as to include it next to $\PE(\Ivecthree+\Ihypthree)$:
\begin{equation}
  \prod_{\substack{\alpha\in\roots(G)\\\alpha\cdot\mbf\neq 0}} \bigl(1-(x_-x_+)^{\abs{\alpha\cdot\mbf}}\zbf^\alpha\bigr)
  = \PE\biggl( - \sum_{\substack{\alpha\in\roots(G)\\\alpha\cdot\mbf\neq 0}} (x_-x_+)^{\abs{\alpha\cdot\mbf}}\zbf^\alpha\biggr) \ .
\end{equation}

\paragraph{Single-particle index and recombination.}

The Hilbert space decomposes into representations of the superconformal algebra $\lie{osp}(4\vert 4)$~\cite{Dolan:2008vc}.
The contribution of each supermultiplet to the index was for example spelled out in Section~3 of~\cite{Bachas:2019jaa}.\footnote{We follow the Cordova-Dumitrescu-Intriligator notations~\cite{Cordova:2016emh} except that all $\lie{su}(2)$ labels are half-integers in our notations rather than being scaled by a factor of~$2$ to be integers as in their notation.}
\begin{itemize}
\item Long multiplets $L[j]^{(\jH,\jC)}$ have dimension $\Delta>j+\jH+\jC+1$ and do not contribute to the index.
  This should not come as a surprise since any non-zero contribution would depend on~$\Delta$, which is not fixed for these multiplets, while we know on general grounds that the index is insensitive to continuous deformations of parameters.

\item Short multiplets $A_1[j]^{(\jH,\jC)}$ with $j>0$ and dimension $\Delta=j+\jH+\jC+1$ have the following index, in terms of~\eqref{indexB1} below:
  \begin{equation}
    \Ithree\bigl(A_1[j]^{(\jH,\jC)}\bigr) = (-1)^{2j+1} \Ithree\bigl(B_1[0]^{(1+j+\jH,1+j+\jC)}\bigr) \ .
  \end{equation}
\item Short multiplets $A_2[0]^{(\jH,\jC)}$ have dimension $\Delta=\jH+\jC+1$ and index
  \begin{equation}
    \Ithree\bigl(A_2[0]^{(\jH,\jC)}\bigr) = - \Ithree\bigl(B_1[0]^{(1+\jH,1+\jC)}\bigr) \ .
  \end{equation}
\item Short multiplets $B_1[0]^{(\jH,\jC)}$ have dimension $\Delta=\jH+\jC$ and index
  \begin{equation}\label{indexB1}
    \Ithree\bigl(B_1[0]^{(\jH,\jC)}\bigr) = x_+^{2\jH} x_-^{2\jC} \frac{(1 - \delta_{\jC>0} \, x_+^2) (1 - \delta_{\jH>0} \, x_-^2)}{1 - \delta_{\jH+\jC>0} \, x_+^2 x_-^2 } \ .
  \end{equation}
\end{itemize}
Recombination rules describe how long multiplets of various spins split into pairs of short multiplets when $\Delta$ reaches the unitarity bound. Schematically: 
\begin{equation*}
L[0]\to A_2[0]\oplus B_1[0], \quad L[\tfrac{1}{2}]\to A_1[\tfrac{1}{2}]\oplus A_2[0] \quad \text{and} \quad L[j]\to A_1[j]\oplus A_1[j-\tfrac{1}{2}] \quad \text{for}\quad j\geq 1
\end{equation*}
Since the long multiplet's index vanishes, the two short multiplets must have opposite indices.
The index expressions above are compatible with, and fully capture, all these recombination rules.
For instance, the $B_1[0]^{(\jH,0)}$ and $B_1[0]^{(0,\jC)}$ multiplets are absolutely protected in the sense that they cannot recombine with any other multiplet, and correspondingly none of the $A$-type multiplets shares the same index (up to a sign) with these two multiplets. 

The index calculated using the general formula~\eqref{fullindex} is a sum of indices of these short multiplets, hence can be decomposed into a linear combination of $\Ithree\bigl(B_1[0]^{(\jH,\jC)}\bigr)$ for integer $2\jH,2\jC\geq 0$.
This decomposition is unique because for any monomial in $x_\pm$ there is exactly one $\Ithree\bigl(B_1[0]^{(\jH,\jC)}\bigr)$ whose leading term is that monomial.

As an illustration, consider the index of $T\big[\text{SU(2)}\big]$, up to order $x_+^4$ and $x_-^4$, with all electric and magnetic flavour fugacities set to~$1$:
\begin{equation}\label{TSU2-index-example}
  \begin{aligned}
    & \Ithree(T\big[\text{SU(2)}\big])
    \\
    & \quad = 1 + 3 x_+^2 + 5x_+^4 + O(x_+^5) + 3x_-^2 - 7 x_+^2x_-^2 - 4x_+^4 x_-^2 + 5x_-^4 - 4 x_+^2x_-^4 + 7x_+^4x_-^4 + O(x_-^5)
    \\
    & \quad = B_1[0]^{(0,0)} + 3 B_1[0]^{(1,0)} + 5 B_1[0]^{(2,0)} + O(x_+^5) + \\
    & \qquad + 3 B_1[0]^{(0,1)} - B_1[0]^{(1,1)} - 3 B_1[0]^{(2,1)}
    + 5 B_1[0]^{(0,2)} - 3 B_1[0]^{(1,2)} + 9 B_1[0]^{(2,2)} + O(x_-^5) \ ,
  \end{aligned}
\end{equation}
where we have denoted the index of a supermultiplet in the same way as the multiplet itself to save space.
(To avoid confusion on the notation, let us point out that $T\big[\text{SU(2)}\big]$ does not feature any half-integer spins nor R charges.)
While the $B_1[0]^{(\jH,0)}$ and $B_1[0]^{(0,\jC)}$ terms guarantee the presence of the corresponding multiplets in the spectrum, the terms $B_1[0]^{(\jH,\jC)}$ with both $\jH,\jC\geq 1$ may arise either from $B_1[0]^{(\jH,\jC)}$ itself, or from $A$-type multiplets.
In fact, whenever the coefficient is negative there \emph{must} be an $A$-type multiplet.

From \eqref{TSU2-index-example} we thus learn that the spectrum must contain exactly
$1$~identity operator,
$3$~electric currents,
$5$~electric marginal deformations,
etc.\ (at higher orders in~$x_+$),
$3$~magnetic currents,
$5$~magnetic marginal deformations,
etc.\ (at higher orders in~$x_-$),
and that it must contain at least
$1$~stress-tensor $A_2[0]^{(0,0)}$,
$3$~multiplets $A_2[0]^{(1,0)}$,
$3$~multiplets $A_2[0]^{(0,1)}$,
and finally $9$~multiplets that are either $B_1[0]^{(2,2)}$ or $A_1[\half]^{(\half,\half)}$.
For the last collection of multiplets, the most natural guess would be $B_1[0]^{(2,2)}$ since all other spins in the theory are integer.
However, supersymmetry does not prevent a long multiplet $L[\half]^{(\half,\half)}$ from going down to the unitarity threshold, splitting into $A_1[\half]^{(\half,\half)}$ and $A_2[0]^{(1,1)}$, and the resulting $A_2$ multiplet to recombine with $B_1[0]^{(2,2)}$ into a long multiplet $L[0]^{(1,1)}$.
We see that the index provides some information, but not all of it, on the $1/2$-BPS multiplet content of the theory.

\subsection{Reduction to order \texorpdfstring{$q$}{q}}

\paragraph{Procedure to expand and decompose the index.}
To determine the protected spectrum up to certain values of R-charges $\jH$ and~$\jC$, we must perform the following steps:
\begin{itemize}
\item list low-lying monopole sectors of the 3d theory, with $2\Delta(\mbf)\leq{}$highest power of $x_-$ of interest;
\item for each sector~$\mbf$, expand $\Ivecthree$ and $\Ihypthree$ to the appropriate order in $x_-$ and~$x_+$ and split them as a sum of characters of the unbroken gauge group $G_{\mbf}\subset G$;

\item perform the plethystic exponential operation up to the appropriate $x_\pm$ order, using suitable (anti)symmetrizations, and include the $\alpha\cdot\mbf\neq 0$ (non-Vandermonde) factors of $\ZVand$;

\item project down to $G_{\mbf}$ gauge singlets and account for the difference between dividing by $|W(G_{\mbf})|$ and by $|W(G)|$;

\item after summing over monopole sectors, reorganize the $x_\pm$ expansion into characters of protected superconformal multiplets.
\end{itemize}

\paragraph{Index at order~$q$.}
In the next section we apply the above procedure to order~$q^1$, namely $x_-^a x_+^{4-a}$ for each $0\leq a\leq 4$.
An important stepping stone is to expand $\Ivecthree$ and~$\Ihypthree$ to this order:
\begin{equation}
  \begin{aligned}
    \Ivecthree & = (x_-^2 - x_+^2) \sum_{\substack{\omega\preceq \adj(G)\\\omega\cdot\mbf=0}} \zbf^\omega
    + (x_-^2 - x_+^2) x_-x_+ \sum_{\substack{\omega\preceq \adj(G)\\|\omega\cdot\mbf|=1}} \zbf^\omega + o(x_{\pm}^4) \ ,
    \\
    \Ihypthree & = x_+ (1 - x_-^2) X^{\text{hyp}} \sum_{\substack{(\omega,\omega_e)\preceq \Vhyp\\\omega\cdot\mbf=0}} \zbf^\omega\mu^{\omega_e}
    + x_- x_+^2 X^{\text{hyp}} \sum_{\substack{(\omega,\omega_e)\preceq \Vhyp\\|\omega\cdot\mbf|=1}} \zbf^\omega\mu^{\omega_e} + o(x_{\pm}^4) \ .
  \end{aligned}
\end{equation}
The various sums actually assemble into characters of representations of $G_{\mbf}\times\Gelec$, distinguished by their charge under the $U(1)_{\mbf}\subset G_{\mbf}$ gauge group generated by~$\mbf$.

% ===============================================
% 					SECTION 3
% ===============================================

\section{Counting marginal operators in Lagrangian 3d theories}
\label{sec:marginal}

\subsection{Notation for index building blocks}

Our goal in this section is to derive general formulas for the low-lying BPS spectrum ($\jH+\jC\leq 2$) of 3d $\Nsusy=4$ Lagrangian gauge theories that are good.
The theories are characterized by a gauge group~$G$ and a quaternionic representation~$\Vhyp$ of~$G$ in which the hypermultiplet scalars transform.
We obtain general group-theoretic formulas in the zero-monopole sector first (\autoref{subsec:Z3d-0}), and then in any given monopole sector (\autoref{subsec:Z3d-m}).
In each sector we discuss some aspects of single-trace $B_1[0]^{(1,1)}$~multiplets.
Assembling these results into the complete index requires a classification of monopole sectors with low dimension, which is not available in general.
We thus specialize the discussion (in \autoref{sec:singletrace}) to various classes of 3d $\Nsusy=4$ quiver gauge theories and determine which ones have single-trace $B_1[0]^{(1,1)}$~multiplets.

To shorten the expressions, we will restrict our attention to theories that are good in the sense that their infrared fixed point has no free fields.  In particular they have \textbf{no free hypermultiplet,} and \textbf{no free monopole operator.}\footnote{The index of theories with free hypermultiplets is easily derived from this one by multiplying by the index of these free hypermultiplets.  In contrast the condition that there are no free monopoles is more restrictive, as one cannot in general explicitly decouple such monopoles in the gauge theory description without introducing twisted hypermultiplets.}
In terms of gauge theory data, we require $\Vhyp$ to feature no gauge singlet, and we require $\Delta(\mbf)\geq 1$ whenever $\mbf\neq 0$.
This will mean that the spectrum has neither $B_1[0]^{(1/2,0)}$ nor $B_1[0]^{(0,1/2)}$ multiplets.

\paragraph{Notation for the index of short multiplets.}

To condense our expressions and make it easier to extract the spectrum, we use a short-hand notation for the index of $B_1[0]$ multiplets given in~\eqref{indexB1}, whose expression we recall:
\begin{equation}\label{BjHjC-notation}
  B^{(\jH,\jC)} = \Ithree\bigl(B_1[0]^{\jH,\jC}\bigr)
  = \begin{cases}
    1 & \jH=0 \ , \ \jC=0 \ , \\
    x_+^{2\jH} (1 - x_-^2)/(1 - x_+^2 x_-^2) & \jH>0 \ , \ \jC=0 \ , \\
    x_-^{2\jC} (1 - x_+^2)/(1 - x_+^2 x_-^2) & \jH=0 \ , \ \jC>0 \ , \\
    x_+^{2\jH} x_-^{2\jC} (1 - x_+^2) (1 - x_-^2) / (1 - x_+^2 x_-^2) & \jH>0 \ , \ \jC>0 \ .
  \end{cases}
\end{equation}
Since the leading term in $B^{(\jH,\jC)}$ is always $x_+^{2\jH}x_-^{2\jC}$, one can iteratively decompose
the whole index, and more generally any series in nonnegative powers of~$x_{\pm}$, into the form
\begin{equation}\label{Ithree-series}
  \Ithree = \sum_{\jH,\jC\in\half\ZZ_{\geq 0}} c_{\jH,\jC} B^{(\jH,\jC)}
\end{equation}
where the coefficients $c_{\jH,\jC}$ are characters of the (electric and magnetic) flavour symmetry groups.
We aim to determine these coefficients for $\jH+\jC\leq 2$.

It is worth pointing out that at the order $x_{\pm}^4$ that we consider, the denominator $1-x_+^2x_-^2$ can be entirely ignored and one has
\begin{equation}
  \begin{aligned}
    & B^{(0,0)} = 1 \ , \qquad
    B^{(1/2,0)} = x_+ (1-x_-^2) + O(x_{\pm}^5) \ , \qquad
    B^{(0,1/2)} = x_-(1-x_+^2) + O(x_{\pm}^5) \ ,
    \\
    & B^{(1,0)} = x_+^2(1-x_-^2) + O(x_{\pm}^5) \ , \qquad
    B^{(1/2,1/2)} = x_+x_- (1 - x_+^2 - x_-^2) + O(x_{\pm}^5) \ , \qquad
    \\
    & B^{(0,1)} = x_-^2(1-x_+^2) + O(x_{\pm}^5) \ , \qquad
    B^{(\jH,\jC)} = x_+^{2\jH} x_-^{2\jC} + O(x_{\pm}^5) \text{ for } \jH+\jC\geq 3/2 \ .
  \end{aligned}
\end{equation}

\paragraph{Notation for an order on representation.}

While at intermediate steps the index is expressed as a sum of monomials in the topological fugacities, these eventually recombine into characters of~$\Gmag$, multiplied by characters of~$\Gelec$ and by the single-particle indices~\eqref{BjHjC-notation}.
In this way, the coefficients $c_{\jH,\jC}$ are characters of the full flavour symmetry $\Gelec\times\Gmag$, which indicate in which representations of $\Gelec\times\Gmag$ the various protected operators transform.
Throughout the paper, \emph{we do not distinguish representations and their characters}.  In particular, we denote by $R_1+R_2$ and $R_1R_2$ the direct sum and the tensor product of two representations.

To be precise, because $A$-type multiplets contribute negatively to the index, $c_{\jH,\jC}$ ($\jH,\jC\geq 1$) can be characters of virtual representations $R - R'$, namely the character of some representation~$R$ minus that of another representation~$R'$.
We introduce the notation
\begin{equation}\label{geqrep}
  R_1\geqrep R_2
\end{equation}
to denote that the right-hand side is a sub-representation of the left-hand side, namely that $R_1 - R_2$ is a non-virtual representation of the symmetry group of interest.  This latter condition generalizes readily to virtual representations: $R_1 - R'_1\geqrep R_2 - R'_2$ if $R_1 + R'_2 - R'_1 - R_2$ is a representation, namely if $R'_1 + R_2$ is a sub-representation of $R_1 + R'_2$.

Because $B_1[0]^{(\jH,\jC)}$ are absolutely protected for $\jH<1$ or $\jC<1$, and recombine monogamously with $A_2[0]^{(\jH-1,\jC-1)}$ when either $\jH=1$ or $\jC=1$, we obtain lower bounds on their numbers.
For a similar reason, since multiplets that recombine with $A_2[0]^{(\jH-1,\jC-1)}$ for $\min(\jH,\jC)\leq\frac{3}{2}$, namely $B_1[0]^{(\jH,\jC)}$ and $A_1[\frac{1}{2}]^{(\jH-\frac{3}{2},\jC-\frac{3}{2})}$, cannot recombine with any other multiplet, we have
\begin{equation}
  \begin{aligned}
    & (\text{representation of $B_1[0]^{(\jH,\jC)}$}) = c_{\jH,\jC} \geqrep 0 , \qquad && \text{if } \jH<1 \text{ or } \jC<1 \ , \\
    & (\text{representation of $B_1[0]^{(\jH,\jC)}$}) \geqrep c_{\jH,\jC} , \qquad && \text{if } \min(\jH,\jC) = 1 \ , \\
    & (\text{representation of $A_2[0]^{(\jH-1,\jC-1)}$}) \geqrep - c_{\jH,\jC} , \qquad && \text{if } \min(\jH,\jC) = 1 \text{ or } \frac{3}{2} \ .
  \end{aligned}
\end{equation}
In particular, a coefficient $\dim c_{1,1}\leq -2$ indicates the presence of more than one stress tensor multiplet, hence of decoupled subsectors: such an approach was used successfully for 4d $\Nsusy=2$ theories in~\cite{Distler:2017xba}.\footnote{Note that if $c_{\jH,\jC}$ with $\min(\jH,\jC)=1$ is a virtual representation with both a positive and a negative terms, it could in principle lead to lower bounds on both $B_1[0]$ and $A_2[0]$ multiplets.  We have not sought an example of theory where this occurs.}

\paragraph{Notation for gauge group and matter content.}

We decompose the gauge group as
\begin{equation}\label{gauge-group-split}
  G = U(1)^{\nab} \times \prod_{i=1}^{\nna} G_i \ ,
\end{equation}
where $\nab$ is the number of abelian factors and $G_i$ are simple non-abelian factors.

We decompose the hypermultiplet representation $\Vhyp$ into irreducible representations~$R_I$.  For such an irreducible representation, the tensor square $R_I^2$ has at most one singlet because such a singlet defines an isomorphism $R_I\simeq\Rbar_I$, which is necessarily unique up to scaling by Schur's lemma.  This motivates us to distinguish in our notation between
\begin{itemize}
\item complex representations~$\RCC_I$, such that neither $S^2\RCC_I$ nor $\Lambda^2\RCC_I$ has any singlet (for instance any representation charged under a $U(1)$ factor),
\item real representation~$\RRR_I$, such that $S^2\RRR_I$~has a singlet,
\item quaternionic representation~$\RHH_I$ (also called pseudoreal), such that $\Lambda^2\RHH_I$~has a singlet.
\end{itemize}
We use the following notation for the decomposition into irreducible representations, and specific names for different types of irreducible representations of~$G$:
\begin{equation}\label{RoplusRbar-decomp}
  \begin{aligned}
    \Vhyp
    & = \sum_{I=1}^n \bigl(\Nbar_I R_I\bigr)
    = \sum_{I=1}^{\nCC} \Bigl( \bigl(\NCCbar_I\RCC_I\bigr)+ \bigl(\NCC_I\RCCbar_I\bigr) \Bigr)
    + \sum_{I=1}^{\nRR} \Bigl( \NHH_I\RRR_I \Bigr)
    + \sum_{I=1}^{\nHH} \Bigl( \NRR_I\RHH_I \Bigr) \ .
  \end{aligned}
\end{equation}
Here, $R_I,\RCC_I,\RCCbar_I,\RRR_I,\RHH_I$ are irreducible representations\footnote{In each conjugate pair of complex representations we select arbitrarily which one to call $\RCC_I$ as opposed to~$\RCCbar_I$.} of the gauge group~$G$ while the multiplicities $\Nbar_I,\NCCbar_I,\NCC_I,\NHH_I,\NRR_I$ are (possibly reducible) representations of the electric flavour symmetry group~$\Gelec$.
Because $\Vhyp$ is a quaternionic representation, the flavour representations $\NHH_I$ and~$\NRR_I$ are direct sums of quaternionic and real representations of the electric flavour symmetry group, respectively (in the opposite order as for the gauge representations).

The theory of interest is assumed to have no free hypermultiplet, so that $\Vhyp$ does not include any copy of the trivial representation of~$G$.
It is useful to introduce a notation~$\Nadj_i$ for the multiplicity of each adjoint representation $\adj(G_i)$ of a simple non-abelian gauge group factor, so
\begin{equation}\label{Nadj-def}
  \Vhyp = \dots + \sum_{i=1}^{\nna} \Bigl( \Nadj_i\adj(G_i) \Bigr) .
\end{equation}
The representation $\adj(G_i)$ is real hence $\Nadj_i$ is one of the quaternionic representations~$\NHH_I$.
Note that since we are decomposing the half-hypermultiplet representation~$\Vhyp$, the dimension of~$\Nadj_i$ is~$2$ for a single adjoint hypermultiplet consisting of a pair of adjoint chiral multiplets.

\subsection{Zero-monopole sector of the index}
\label{subsec:Z3d-0}

\paragraph{Some simplifications.}
We consider first the $\mbf=0$ sector, for which the prefactor $x_-^{2\Delta(\mbf)}$ disappears and the effect of integrating with the integration factor, Weyl group factor, and Vandermonde determinant in~\eqref{fullindex} is simply to project onto gauge-invariant operators, an operation we denote~$\singlet_G$.
The exponents of~$x_-$ are all even (in this zero-monopole sector), and in the notation of~\eqref{Ithree-series}, we therefore have
\begin{equation}\label{Ithree-series-m0}
  \Ithree^{\mbf=0} = \sum_{\jH,\jC\geq 0} c^{\mbf=0}_{\jH,\jC} B^{(\jH,\jC)} , \qquad
  \text{with } \ c^{\mbf=0}_{\jH,\jC} = 0 \quad \text{for } 2\jC \text{ odd.}
\end{equation}
The explicit expressions of coefficients are given in \eqref{cm000-cm001}--\eqref{cm020-cm011} below.

% \eqref{cm000-cm001}, \eqref{cm002}, \eqref{cm0120}, \eqref{cm010}, \eqref{cm0121}, \eqref{cm0320}, and~\eqref{cm020-cm011} below.

In the zero-monopole sector, the vector multiplet and hypermultiplet contributions are simply characters of representations of~$G$ or $G\times \Gelec$, which we denote as a short-hand notation by the name of the representation itself:
\begin{equation}\label{Xvechyp-char}
  X^{\text{vec}} = \sum_{\omega\preceq \adj(G)} \zbf^\omega = \adj(G) , \qquad
  X^{\text{hyp}} = \sum_{(\omega,\omega_e)\preceq \Vhyp} \zbf^\omega\mu^{\omega_e} = \Vhyp \ .
\end{equation}
(We use $+$ for the addition of characters, which amounts to direct sum $\oplus$ of representations.)
We are only interested in powers up to~$x_+^k x_-^{4-k}$, so we expand to this order:
\begin{equation}
  \Ithree^{\mbf=0} = \singlet_G \Bigl( \PE \Bigl( (x_-^2 - x_+^2) \adj(G) + x_+ (1 - x_-^2) \Vhyp + O(x_{\pm}^5) \Bigr) \Bigr) \ .
\end{equation}
To evaluate this we remember that the plethystic exponential is the sum of (super\mbox{-})symmetrized tensor powers $\Sbb^n$, $n\geq 0$.
For each $n\in\{0,1,2,3,4\}$ we rewrite these powers in terms of~$B^{(\jH,\jC)}$.

\paragraph{Symmetric powers.}
The $n=0$ contribution is simply the identity operator~$B^{(0,0)}$.
The $n=1$ term is the one-particle contribution
\begin{subequations}\label{ithreem0-pieces}
\begin{equation}\label{ithreem0-linear}
  \begin{aligned}
    & (x_-^2 - x_+^2) \adj(G) + x_+ (1 - x_-^2) \Vhyp \\
    & \quad = B^{(1/2,0)} \Vhyp + (B^{(0,1)} - B^{(1,0)}) \adj(G) + O(x_{\pm}^5) \ ,
  \end{aligned}
\end{equation}
where there is a cancellation between $x_+^2x_-^2$ terms appearing in $B^{(0,1)}$ and~$B^{(1,0)}$.
Next, the symmetrized square is computed using $\Sbb^2(R_1-R_2)=S^2R_1+\Lambda^2R_2-R_1R_2$:
\begin{equation}
  \begin{aligned}
    & \Sbb^2 \bigl( (x_-^2 - x_+^2) \adj(G) + x_+ (1 - x_-^2) \Vhyp \bigr) \\
    & \quad = x_+^4 \Lambda^2(\adj(G)) - x_+^2 x_-^2 \adj(G)^2 + x_-^4 S^2(\adj(G)) \\
    & \qquad + x_+ (x_-^2 - x_+^2) \adj(G) \Vhyp + x_+^2 S^2\Vhyp - x_+ x_-^2 \Vhyp^2 + O(x_{\pm}^5)
    \\[1ex]
    & \quad = B^{(1,0)} S^2\Vhyp + (B^{(1/2,1)} - B^{(3/2,0)}) \Vhyp \adj(G)
    \\
    & \qquad + B^{(2,0)} \Lambda^2(\adj(G)) - B^{(1,1)} \bigl( \adj(G)^2 + \Lambda^2\Vhyp \bigr) + B^{(0,2)} S^2(\adj(G)) + O(x_{\pm}^5) \ ,
  \end{aligned}
\end{equation}
The symmetrized cube and fourth power are simpler because most terms are absorbed in $O(x_{\pm}^5)$ and one can directly replace powers of $x_{\pm}$ by suitable~$B^{(\jH,\jC)}$ at this order:
\begin{equation}\label{ithreem0-S3S4}
  \begin{aligned}
    & \Sbb^3 \bigl( (x_-^2 - x_+^2) \adj(G) + x_+ (1 - x_-^2) \Vhyp \bigr) \\
    & \quad = B^{(3/2,0)} S^3\Vhyp + (B^{(1,1)} - B^{(2,0)}) \adj(G)\, S^2\Vhyp + O(x_{\pm}^5) \ ,
    \\[1ex]
    & \Sbb^4 \bigl( (x_-^2 - x_+^2) \adj(G) + x_+ (1 - x_-^2) \Vhyp \bigr)
    \\
    & \quad = B^{(2,0)} S^4\Vhyp + O(x_{\pm}^5) \ .
  \end{aligned}
\end{equation}
\end{subequations}
Summing the above symmetric products and projecting to gauge singlets gives a formula for the zero-monopole sector contribution to the index of a theory, and we now begin extracting coefficients~$c^{\mbf=0}$ of the various indices~$B^{(\jH,\jC)}$, as defined in~\eqref{Ithree-series-m0}.

\paragraph{Magnetic part: $c^{\mbf=0}_{0,0},c^{\mbf=0}_{0,1},c^{\mbf=0}_{0,2}$.}

We begin with the purely magnetic contributions, with $\jH=0$.
The adjoint representation of each~$U(1)$ gauge group factor is a singlet, while each $\adj(G_i)$ is an irreducible representation hence has no gauge singlet.
We thus obtain our first characters, and their interpretation:
\begin{equation}\label{cm000-cm001}
  \begin{aligned}
    c^{\mbf=0}_{0,0} & = 1 \ , & & \text{(a single identity operator),} \\
    c^{\mbf=0}_{0,1} & = \singlet_G(\adj(G)) = \nab & & \text{(a topological symmetry for each $U(1)$).}
  \end{aligned}
\end{equation}
We emphasize that many more $B^{(0,1)}$ multiplets can arise in monopole sectors.

The remaining magnetic term that we can access counts $B^{(0,2)}$ multiplets, whose coefficient in~\eqref{ithreem0-pieces} is
(later we are also interested in $\Lambda^2\adj(G)$ obtained by replacing all $S^2$ by~$\Lambda^2$)
\begin{equation}
  S^2\adj(G)
  = S^2\adj\bigl(U(1)^{\nab}\bigr)
  + \sum_{i=1}^{\nna} \Bigl( \adj\bigl(U(1)^{\nab}\bigr) \, \adj(G_i) + S^2 \adj(G_i) \Bigr)
  + \sum_{i<j}^{\nna} \adj(G_i) \adj(G_j) \ .
\end{equation}
To project this to gauge singlets, we use the fact that $\adj(G_i)$ is a real representation:
indeed, (the inverse of) the Killing form of~$G_i$ is a $G$-invariant element of $(\adj(G_i))^2$ that is symmetric.
Thus, $S^2\adj(G_i)$ has a singlet while $\Lambda^2\adj(G_i)$ does not, and of course $\adj(G_i)\adj(G_j)$ is irreducible and not a singlet.
Altogether,
\begin{equation}\label{cm002}
  \begin{aligned}
    \singlet_G\bigl(\Lambda^2\adj(G)\bigr) & = \tfrac{1}{2} \nab(\nab-1) \ , \\
    c^{\mbf=0}_{0,2} = \singlet_G\bigl(S^2\adj(G)\bigr) & = \tfrac{1}{2}\nab(\nab+1) + \nna \ .
  \end{aligned}
\end{equation}
Recall now that this character counts some exactly marginal operators.
We recognize here the number $\tfrac{1}{2}\nab(\nab+1)$ of products of magnetic topological currents, while the $\nna$ remaining $B^{(2,0)}$ multiplets are not factorizable in this way.

\paragraph{Hypermultiplet decomposition: $c^{\mbf=0}_{1/2,0}$.}

Our assumption that the theory has no free hypermultiplet means $\Vhyp$~has no singlet, and we immediately find that there are no $B^{(1/2,0)}$ (free hypermultiplets),
\begin{equation}\label{cm0120}
  c^{\mbf=0}_{1/2,0} = \singlet_G(\Vhyp) = 0 \ .
\end{equation}

At higher orders we will need to determine singlets in $S^2\Vhyp$ and $\Lambda^2\Vhyp$.
To work them out, note that for two irreducible representations $R_1,R_2$, the representation $R_1 R_2$ contains exactly one singlet if $R_1$ and $R_2$ are conjugate from each other.
Thus, the gauge singlets in the (anti)symmetric square of the decomposition~\eqref{RoplusRbar-decomp} only come from terms where a representation $\RRR_I$ or $\RHH_I$ is tensored with itself, or $\RCC_I$ is tensored with~$\RCCbar_I$:
\begin{subequations}
\begin{align}
  \label{singletGS2RRbar}
  \singlet_G\bigl(S^2\Vhyp\bigr)
  & = \sum_{I=1}^{\nCC} \Bigl(\NCCbar_I\NCC_I\Bigr) + \sum_{I=1}^{\nRR} \Bigl(S^2 \NHH_I\Bigr) + \sum_{I=1}^{\nHH} \Bigl(\Lambda^2\NRR_I\Bigr) \ ,
  \\
  \singlet_G\bigl(\Lambda^2\Vhyp\bigr)
  & = \sum_{I=1}^{\nCC} \Bigl(\NCCbar_I\NCC_I\Bigr) + \sum_{I=1}^{\nRR} \Bigl(\Lambda^2 \NHH_I\Bigr) + \sum_{I=1}^{\nHH} \Bigl(S^2\NRR_I\Bigr) \ .
\end{align}
\end{subequations}
Here, we used the facts that $S^2(\Nbar_I R_I)=(S^2\Nbar_I)(S^2R_I)+(\Lambda^2\Nbar_I)(\Lambda^2R_I)$ and that $S^2(\Nbar_IR_I)=(S^2\Nbar_I)(\Lambda^2R_I)+(\Lambda^2\Nbar_I)(S^2R_I)$, and our knowledge that gauge singlets are in $S^2\RRR_I$ and $\Lambda^2\RHH_I$.

\paragraph{Electric currents: $c^{\mbf=0}_{1,0}$.}

At order~$x_+^2$, the index features electric current multiplets~$B^{(1,0)}$ that are gauge-invariant products of two hypermultiplets, namely are gauge singlets in $S^2\Vhyp$.
As counted by the $-B^{(1,0)}\singlet_G(\adj(G))$ term in~\eqref{ithreem0-linear}, one must subtract one F-term relation per abelian factor of~$G$: all gauge group factors of~$G$ leads to one F-term relation, but the relations coming from nonabelian groups are not gauge-invariant hence only have an effect once multiplied by further factors.
We deduce the representation in which electric conserved currents transform:
\begin{equation}\label{cm010}
  \begin{aligned}
    c^{\mbf=0}_{1,0} & = \singlet_G\bigl(S^2\Vhyp\bigr) - \nab \\
    & = \sum_{I=1}^{\nCC} \bigl(\NCCbar_I\NCC_I\bigr) + \sum_{I=1}^{\nRR} \bigl(S^2 \NHH_I\bigr) + \sum_{I=1}^{\nHH} \bigl(\Lambda^2\NRR_I\bigr) - \nab \ .
  \end{aligned}
\end{equation}

As an aside,
let us verify that (as expected) this character $c^{\mbf=0}_{1,0}\geqrep 0$ does not have a negative part.
By assumption the theory is not bad, so $\Delta(\mbf)>0$ for any non-zero~$\mbf$, and in particular there cannot be a monopole charge vector orthogonal to all weights of~$R$.
Restricting to the abelian part, this means that there must be hypermultiplets with at least $\nab$~different (linearly independent) charge vectors under $U(1)^{\nab}$.
Any representation charged under the abelian group factors must be complex since conjugation flips the sign of $U(1)$ charges.
We thus have $\nCC\geq\nab$, so that the $\Gelec$-singlets in each $\NCCbar_I\NCC_I$ are enough to compensate for the $-\nab$ term in~\eqref{cm010}.

At higher orders in the index, we encounter the gauge singlets of $S^3\Vhyp$ and $S^4\Vhyp$, and no simple formula is available for these.
However, it should also be possible to prove the overall positivity of coefficients in similar ways.

\paragraph{Third order: $c^{\mbf=0}_{3/2,0},c^{\mbf=0}_{1/2,1}$.}

We turn to operators with $\jH+\jC=3/2$.  Recall that $\jC$ must be an integer in the zero-monopole sector, so there are two terms, $(\jH,\jC)=(1/2,1)$ and $(3/2,0)$.
The first one is easily found by splitting $\adj(G)$ into abelian and non-abelian parts and using the notation in~\eqref{Nadj-def}:
\begin{equation}\label{cm0121}
  c^{\mbf=0}_{1/2,1} = \singlet_G\bigl(\adj(G)\,\Vhyp\bigr)
  = \sum_{i=1}^n \Nadj_i \ .
\end{equation}
The other one is more elaborate because it involves cubic combinations $S^3\Vhyp$ of hypermultiplets, and gauge singlets therein,
\begin{equation}\label{cm0320}
  \begin{aligned}
    c^{\mbf=0}_{3/2,0}
    & = \singlet_G\bigl(S^3\Vhyp - \adj(G)\,\Vhyp \bigr)
    \\
    & = \singlet_G\bigl(S^3\Vhyp\bigr) - \sum_{i=1}^n \Nadj_i \ .
  \end{aligned}
\end{equation}

\paragraph{Fourth order: $c^{\mbf=0}_{2,0},c^{\mbf=0}_{1,1}$.}

We are left with studying the $(\jH,\jC)=(2,0)$ and $(1,1)$ multiplets:
\begin{equation}\label{cm020-cm011}
  \begin{aligned}
    c^{\mbf=0}_{2,0}
    & = \singlet_G\Bigl( \Lambda^2(\adj(G)) + S^4\Vhyp - \adj(G)\, S^2\Vhyp \Bigr) \\
    & = \singlet_G\bigl(S^4\Vhyp\bigr) + \tfrac{1}{2} \nab(\nab-1) - \cvhh \ ,
    \\
    c^{\mbf=0}_{1,1} & = \singlet_G\Bigl(\adj(G)S^2\Vhyp-\adj(G)^2-\Lambda^2\Vhyp\Bigr) \\
    & = \cvhh
    - \nab^2 - \nna
    - \singlet_G\bigl(\Lambda^2\Vhyp\bigr)
    \ ,
  \end{aligned}
\end{equation}
where we have introduced the notation
\begin{equation}\label{cvhh}
  \cvhh = \singlet_G\bigl(\adj(G)\, S^2\Vhyp\bigr) ,
\end{equation}
in which ``vhh'' refers to the fact that it counts gauge-invariant products of one vector multiplet scalar and two hypermultiplet scalars.
This concludes our calculation of coefficients in~\eqref{Ithree-series-m0} for the zero-monopole index.
We return later to expliciting the character~$\cvhh$ of vector-hypermultiplet combinations, and bounding it from below.

\subsection{Monopole sectors of the index}
\label{subsec:Z3d-m}

We now consider a fixed nonzero monopole sector $\mbf\in\Lambdamon\setminus\{0\}$.
In a good theory, one has $\Delta(\mbf)\geq 1$ for $\mbf\neq 0$, as otherwise the monopole operator would give a free multiplet in the infrared.
The corresponding term in the full index~\eqref{fullindex} is $x_-^{2\Delta(\mbf)}$ times a series in~$x_{\pm}$, so that at the $x_{\pm}^4$ order we care about, we can focus on sectors with $\Delta(\mbf)\in\{1,3/2,2\}$.
It is not feasible to classify very explicitly the low-dimension sectors~$\mbf$ for general Lagrangian 3d $\Nsusy=4$ theories.
We return to this classification question in concrete theories in \autoref{sec:singletrace}.

For now, fixing a sector~$\mbf$, we expand the index~$\Ithree^{\mbf}$ in this sector, up to order~$x_{\pm}^4$.
Given the prefactor $x_-^{2\Delta(\mbf)}$, numerous terms in the index~\eqref{fullindex} simplify.

\paragraph{Commutant.}

To best express the index, we consider the commutant of~$\mbf$ in~$G$ and in the Lie algebra~$\lie{g}$:
\begin{equation}
  \begin{aligned}
    G_{\mbf} & = \{g\in G, \ \Ad_g(\mbf)=0 \} \subseteq G , \\
    \adj(G_{\mbf}) = \lie{g}_{\mbf} & = \{x\in\lie{g}, \ [\mbf,x]=0 \} \subseteq \lie{g} .
  \end{aligned}
\end{equation}
The group~$G_{\mbf}$ shares a Cartan torus with~$G$ (and likewise the Lie algebras share their Cartan subalgebra), hence weights of a representation~$R$ of~$G$ or of its induced representation of~$G_{\mbf}$ coincide.
It is useful to decompose~$R$ into eigenspaces~$R_{\mbf}^{(p)}$ of~$\mbf$ with eigenvalue~$p$: since the action of $G_{\mbf}$ on~$R$ commutes with that of~$\mbf$, these eigenspaces\footnote{To obtain~\eqref{Rmbfp-def} we identified the weights of~$R_{\mbf}^{(p)}$ by working in a basis of~$R$ in which the Cartant algebra~$\lie{h}$ of $\lie{g}$ acts diagonally (with eigenvalues giving the weights), and noting that $\mbf\in\lie{h}$.} are automatically representations of~$G_{\mbf}$,
\begin{equation}\label{Rmbfp-def}
  \begin{gathered}
    R_{\mbf}^{(p)} = \bigl\{x\in R, \ \mbf\cdot x = p x\bigr\} \subseteq R \ , \\
    \omega\preceq R_{\mbf}^{(p)} \iff \bigl(\omega\preceq R \text{ and } \omega\cdot\mbf = p \bigr) \ .
  \end{gathered}
\end{equation}
The zero eigenspace $R_{\mbf}=R_{\mbf}^{(0)}$ is particularly interesting to us.
Applying this operation to the representation $R=\adj(G)=\lie{g}$ reproduces~$\lie{g}_{\mbf}$, whose roots are thus roots of~$\lie{g}$ orthogonal to~$\mbf$.

\paragraph{Low-order index.}

At low orders, the ingredients in~\eqref{fullindex} are conveniently written as
\begin{equation}
  \begin{aligned}
    \ZVand & = \biggl( 1 - x_- x_+ \sum_{\omega\preceq\lie{g},\ \omega\cdot\mbf=1} \zbf^\omega \biggr) \prod_{\alpha\in\roots(G_{\mbf})} \bigl(1-\zbf^\alpha\bigr) + o(x_{\pm}^2) \ ,
    \\[1ex]
    \Ivecthree & = (x_-^2 - x_+^2) \adj(G_{\mbf}) + o(x_{\pm}^2) \ ,
    \qquad\qquad
    \Ihypthree = x_+ \Vhyp^{\mbf} + o(x_{\pm}^2) \ ,
  \end{aligned}
\end{equation}
in which $\adj(G_{\mbf})$ and $\Vhyp^{\mbf}$ denote characters of the commutant gauge group~$G_{\mbf}$ times the electric flavour symmetry group~$\Gelec$.
The $\mbf$ contribution to the index involves an integral over the Cartan torus, weighted by a product over roots of~$G_{\mbf}$ (Vandermonde determinant), thus it projects to singlets of~$G_{\mbf}$ up to a factor of $|W(G_{\mbf})|$.
For $\Delta(\mbf)\geq 1$,
\begin{equation}
  \begin{aligned}
    \frac{|W(G)|}{|W(G_{\mbf})|} \Ithree^{\mbf}
    & =
    x_-^{2\Delta(\mbf)} w^{\mbf}
    \singlet_{G_{\mbf}}
    \biggl(
    1  + x_+ \Vhyp^{\mbf} + x_+^2 S^2(\Vhyp^{\mbf})
    \\[-2ex]
    & \qquad\qquad\qquad\qquad\qquad
    + (x_-^2 - x_+^2) \adj(G_{\mbf})
    - x_- x_+ \sum_{\omega\preceq\lie{g},\omega\cdot\mbf=1} \zbf^\omega
    \biggr)
    + o(x_{\pm}^4)
    \ .
  \end{aligned}
\end{equation}
The weights $\omega\preceq\lie{g}$ such that $\omega\cdot\mbf=1$ are weights of the representation
\begin{equation}
  \lie{g}_{\mbf}^{(1)} = \bigl\{x\in\lie{g}, \ \mbf\cdot x = x \bigr\} .
\end{equation}
In particular, $\mbf$ acts non-trivially on every vector in this representation (except $x=0$), so that $\lie{g}_{\mbf}^{(1)}$ has no $G_{\mbf}$~singlet.
The explicit sum of $\zbf^\omega$ in the above expression thus drops out.

Another observation explains the $|W(G)|/|W(G_{\mbf})|$ factor.
Every monopole sector~$\mbf'$ conjugate to $\mbf\in\lie{h}$ under the Weyl group (gauge transformations) contributes equally to the index.
Since $W(G_{\mbf})$ is the centralizer of~$\mbf$, the orbit of~$\mbf$ under the Weyl group has $|W(G)|/|W(G_{\mbf})|$ elements.

\paragraph{Index of monopole sectors up to order~$4$.}

Altogether, for $\Delta(\mbf)\geq 1$ we find
\begin{equation}\label{sumithree}
  \sum_{\mbf'\sim\mbf} \Ithree^{\mbf'}
  = \prod_{i\in I} w_i^{\Tr_i(\mbf)}
  \begin{cases}
    \begin{aligned}[b]
      & \textstyle
      B^{(0,1)}
      + B^{(1/2,1)} \singlet_{G_{\mbf}}(\Vhyp^{\mbf})
      + B^{(0,2)} \nab^{\mbf}
      \\
      & \textstyle \quad
      + B^{(1,1)} \bigl( \singlet_{G_{\mbf}}(S^2\Vhyp^{\mbf}) - \nab^{\mbf} + 1\bigr)
      + o(x_{\pm}^4)
      \ ,
    \end{aligned}
    & \Delta(\mbf)=1 \ ,
    \\
    B^{(0,3/2)} + B^{(1/2,3/2)} \singlet_{G_{\mbf}}(\Vhyp^{\mbf})
    + o(x_{\pm}^4)
    \ ,
    & \Delta(\mbf)=3/2 \ ,
    \\
    B^{(0,2)} + o(x_{\pm}^4) \ ,
    & \Delta(\mbf)=2 \ ,
    \\
    o(x_{\pm}^4) \ ,
    & \Delta(\mbf)>2 \ ,
  \end{cases}
\end{equation}
where $\nab^{\mbf}$ is the number of $\lie{u}(1)$ factors in~$\lie{g}_{\mbf}$
and we have recast powers of $x_{\pm}$ as characters~$B^{(\jH,\jC)}$, for instance $x_-^2=B^{(0,1)}+B^{(1,1)}+o(x_{\pm}^4)$.

\subsection{Low-lying BPS spectrum from the 3d index}

We are ready to combine the zero-monopole results of \autoref{subsec:Z3d-0} and the monopole sectors in \autoref{subsec:Z3d-m}.
The full index is
\begin{equation}
  \Ithree = \Ithree^{\mbf=0} + \sum_{\mbf\in\Lambdamon/W,1\leq\Delta(\mbf)\leq 2} \biggl(\sum_{\mbf'\sim\mbf} \Ithree^{\mbf'} \biggr) + o(x_{\pm}^4) \ ,
\end{equation}
where the sum over monopoles has one term for each conjugacy class of $\mbf\in\Lambdamon$, and is restricted to dimensions $\Delta(\mbf)\in\{1,3/2,2\}$ due to our goodness assumption and our truncation to~$o(x_{\pm}^4)$.
To shorten later expressions we introduce a notation $\sum_{\Delta(\mbf)=s}$ for the sum over monopoles with a given dimension, counting each conjugacy class only once, so that
\begin{equation}
  \Ithree = \Ithree^{\mbf=0} + \sum_{s=1,3/2,2} \sum_{\Delta(\mbf)=s} \eqref{sumithree} + o(x_{\pm}^4) \ .
\end{equation}

\paragraph{Explicit expressions for the low-lying spectrum.}

We deduce the total number of various low-lying $B^{(\jH,\jC)}$ multiplets for $\jH+\jC\leq 2$.
First the identity operator and electric and magnetic currents,
\begin{equation}\label{c10c01c11}
  \begin{aligned}
    c_{0,0} & = 1 \ , \qquad c_{1/2,0} = 0 \ , \qquad c_{0,1/2} = 0 \ ,
    \\
    c_{1,0}
    & = \singlet_G\bigl(S^2\Vhyp\bigr) - \nab \ ,
  \end{aligned}
  \qquad
  c_{0,1}
  = \nab + \sum_{\Delta(\mbf)=1} w^{\mbf} \ .
\end{equation}
Next the operators with $\jH+\jC=3/2$, where the sums over~$i$ run over nonabelian gauge group factors~$G_i$ and we recall that $\Nadj_i$~counts the number of adjoint representations $\adj(G_i)$ in~$\Vhyp$,
\begin{equation}\label{c320-c032}
  \begin{aligned}
    c_{3/2,0}
    & = \singlet_G\bigl(S^3\Vhyp\bigr) - \sum_{1\leq i\leq n} \Nadj_i \ ,
    \\
    c_{1,1/2} & = 0 \ ,
    \\
    c_{1/2,1}
    & = \sum_{1\leq i\leq n} N_i^{\adj} + \sum_{\Delta(\mbf)=1} w^{\mbf} \singlet_{G_{\mbf}}(\Vhyp^{\mbf}) \ ,
    \\
    c_{0,3/2}
    & = \sum_{\Delta(\mbf)=3/2} w^{\mbf} \ .
  \end{aligned}
\end{equation}
Interestingly, since $c_{1,1/2}$~vanishes while $c_{1/2,1}$ does not, the class of good \emph{Lagrangian} gauge theories is not stable under mirror symmetry.  An important caveat is that these calculations do not prevent a good gauge theory from having as its mirror the interacting sector of an ugly theory.

Finally the operators with $\jH+\jC=2$, where we refer to~\eqref{cvhh} for the expression of~$\cvhh$:
\begin{equation}\label{c20-c02}
  \begin{aligned}
    c_{2,0}
    & = \singlet_G\bigl(S^4\Vhyp\bigr) + \tfrac{1}{2} \nab(\nab-1) - \cvhh \ ,
    \\
    c_{3/2,1/2}
    & = 0 \ ,
    \\
    c_{1,1}
    & = \cvhh
    - \nab^2 - \nna
    - \singlet_G\bigl(\Lambda^2\Vhyp\bigr) \\
    & \quad + \sum_{\Delta(\mbf)=1} \bigl( \singlet_{G_{\mbf}}(S^2\Vhyp^{\mbf}) - \nab^{\mbf} + 1 \bigr) w^{\mbf} \ ,
    \\
    c_{1/2,3/2} & = \sum_{\Delta(\mbf)=3/2} w^{\mbf} \singlet_{G_{\mbf}}(\Vhyp^{\mbf}) \ ,
    \\
    c_{0,2} & = \tfrac{1}{2}\nab(\nab+1) + \nna + \sum_{\Delta(\mbf)=1} w^{\mbf} \nab^{\mbf} + \sum_{\Delta(\mbf)=2} w^{\mbf} \ .
  \end{aligned}
\end{equation}
This completes our derivation of the low-lying protected BPS spectrum, up to the order that includes exactly marginal deformations.

For $\jH\leq 1/2$ or $\jC\leq 1/2$ (or both) the coefficient $c_{\jH,\jC}$ counts the absolutely protected multiplets $B_1[0]^{(\jH,\jC)}$, which ensures that the count is unambiguous.
On the other hand, one should interpret $c_{1,1}$ (and higher coefficients) with care, because $B_1[0]^{(1,1)}$ participates in a recombination rule with the stress-tensor.
Denoting by $n_T$ the number of stress-tensors ($n_T=1$ unless the theory has decoupled subsectors), the number of $B_1[0]^{(1,1)}$ multiplets is $c_{1,1}+n_T$.

\paragraph{Conformal manifold.}

The dimension of the conformal manifold is in fact simpler than the coefficients~$c_{\jH,\jC}$ may suggest, thanks to some cancellations:
\begin{equation}\label{dimCCMcal}
  \begin{aligned}
    \dim_{\CC}\Mcal_{\textnormal{SC}}
    & = n_T - c_{1,0} - c_{0,1} - 1 + \sum_{\jH+\jC=2} c_{\jH,\jC}
    \\
    & = (n_T - 1) + \singlet_G\bigl(S^4\Vhyp - \Lambda^2\Vhyp - S^2\Vhyp\bigr)
    + \sum_{\Delta(\mbf)=1} \singlet_{G_{\mbf}}(S^2\Vhyp^{\mbf}) \\
    & \quad + \sum_{\Delta(\mbf)=3/2} \singlet_{G_{\mbf}}(\Vhyp^{\mbf})
    + \sum_{\Delta(\mbf)=2} 1
    \ ,
  \end{aligned}
\end{equation}
in which we implicitly set all flavour fugacities to~$1$, for instance $\singlet_G(S^4\Vhyp)$ is not a representation (or character) of~$\Gelec$ but rather just its dimension in this formula.

% ===============================================
% 					SECTION 4
% ===============================================

\section{Single-trace mixed moduli}
\label{sec:singletrace}

\subsection{Main results}

\paragraph{Results on non-factorized $B_1[0]^{(1,1)}$ multiplets.}

We have now reached the main focus of our work, mixed moduli of 3d $\Nsusy=4$ SCFTs, which sit in $B_1[0]^{(1,1)}$ multiplets.
We consider the corresponding chiral ring elements, namely the $SU(2)_H\times SU(2)_C$ lowest-weight state in the bottom component of the multiplet, and whether it is the product of chiral ring elements corresponding to electric and magnetic current multiplets.
We introduce the notation 
\begin{equation}
  \Nst = \text{single-trace $B_1[0]^{(1,1)}$ multiplets}
  \geqrep 0
\end{equation}
for the representation of the electric and magnetic flavour symmetries under which the single-trace mixed moduli multiplets transform.
Note that $\Nst\geqrep 0$ because it is counting some actual operators in the theory.
As usual we also denote by~$\Nst$ the character, which reduces to the dimension upon setting all fugacities to one.\footnote{In many theories, the mixed moduli are flavour singlets, so that the character is equal to the dimension $\dim\Nst$.}

The index calculations allow us to place a lower bound on~$\Nst$.  Then, in several classes of concrete theories, we use F-term relations to show explicitly how almost all $B_1[0]^{(1,1)}$ multiplets factorize: the number of remaining multiplets is by definition~$\Nst$, which in many examples matches with the lower bound, so that the inequality is tight.

\begin{figure}\centering
  \begin{tikzpicture}
    \node(A)[rectangle,draw] at (1,.8) {\!$4$\!};
    \node(B)[rounded rectangle,inner sep=2pt,draw] at (1,0) {\!$U(1)$\!};
    \node(C)[rounded rectangle,inner sep=2pt,draw] at (2.3,0) {\!$U(2)$\!};
    \node(D)[rounded rectangle,inner sep=2pt,draw] at (3.6,0) {\!$U(3)$\!};
    \node(E)[rounded rectangle,inner sep=2pt,draw] at (4.9,0) {\!$U(1)$\!};
    \node(F)[rounded rectangle,inner sep=2pt,draw] at (6.2,0) {\!$U(3)$\!};
    \node(G)[rounded rectangle,inner sep=2pt,draw] at (7,1) {\!$U(3)$\!};
    \node(H)[rounded rectangle,inner sep=2pt,draw] at (7.8,0) {\!$U(3)$\!};
    \node(I)[rectangle,draw] at (7.8,.8) {\!$3$\!};
    \draw(A)--(B);
    \draw(B)--(C);
    \draw(C) to[in=60,out=120,loop] (C);
    \draw(C) to[bend right=30](D);
    \draw(C) to[bend left=30](D);
    \draw(D) to[in=60,out=120,loop] (D);
    \draw(D)--(E);
    \draw(E)--(F);
    \draw(F)--(H);\draw(7,0)--(G);
    \draw(G)--(H);
    \draw(H)--(I);
  \end{tikzpicture}\quad
  \begin{tikzpicture}
    \node(C)[circle,draw] at (2.3,0) {};
    \node(D)[circle,draw] at (3.6,0) {};
    \node(E)[circle,draw] at (5.1,0) {};
    \node(F)[circle,draw] at (5.7,1) {};
    \node(G)[circle,draw] at (6.5,0) {};
    \foreach \x/\y in {1.35/.6, 1.75/.6, 1.35/1, 1.75/1}
    { \draw(\x,\y) [fill] circle (.1); }
    \draw(1.65,0) [fill] circle (.1)--(C);
    \draw(C)--(2.3,.6) [fill] circle (.1);
    \draw(C)--(2.95,.3) [fill] circle (.1)--(D);
    \draw(C)--(2.95,-.3) [fill] circle (.1)--(D);
    \draw(D) to [bend right=30] (3.6,.6); \draw(D) to [bend left=30] (3.6,.6); \draw (3.6,.6) [fill] circle (.1);
    \draw(D)--(4.2,0) [fill] circle (.1);
    \draw(E)--(4.5,0) [fill] circle (.1);
    \draw(F)--(5.7,0) [fill] circle (.1);
    \draw(E)--(G);
    \draw(F)--(6.1,.5) [fill] circle (.1)--(G);
    \foreach \x/\y in {6.3/.8, 6.7/.8, 7.1/.8}
    { \draw (G)--(\x,\y) [fill] circle (.1); }
  \end{tikzpicture}
  \caption{\label{fig:gauge-graphs}Left: quiver depiction of an example $U(1)\times U(3)^5$ gauge theory where edges are hypermultiplets: fundamental, bifundamental, adjoint, and trifundamental (denoted by multi-edge).  Right: bipartite graph~$\Gamma$ (here, disconnected and of genus $g=2$) depicting which bifundamentals are charged under which simple gauge group factors; this drops any $U(1)$ factor.  The graph~$\Gammared$ omits the four isolated black nodes.}
\end{figure}
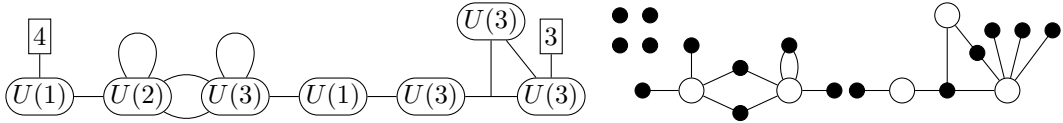

\paragraph{A bipartite graph.}
The lower bound we derive is the genus~$g$ of a bipartite graph~$\Gamma$ that encapsulates some aspects of the quiver.\footnote{By genus, we mean the number of loops in the graph, namely the dimension of its first homology when seen as a topological space.  To accomodate half-hypermultiplets the definition must be altered as described below~\eqref{K00-geq}.}  The graph, depicted in \autoref{fig:gauge-graphs} in an example, has
\begin{itemize}
\item one vertex for each nonabelian simple gauge group~$G_i$;
\item one vertex for each \emph{full} hypermultiplet, namely for each summand~$R_I$ in the decomposition of~$\Vhyp$ into irreducible representations of $\prod_i G_i$ (equivalently of the whole gauge group~$G$), where non-trivial flavour symmetries lead to multiple vertices;
\item $p_{Ii}$ edges joining each \emph{full} hypermultiplet in representation~$R_I$ to each~$G_i$, where the \emph{complexity} $p_{Ii}$ is the number of copies of $\adj(G_i)$ in $R_I\otimes\Rbar_I$, which is at least $1$ whenever $R_I$ is charged under~$G_i$, as discussed further in \autoref{subsec:sec4-zero}.
\end{itemize}
As explained in \autoref{app:tensor-adjoint}, denoting by~$\mu$ the highest weight of~$R_I$ (as a representation of~$G$, or simply of~$G_i$) the number of edges joining that hypermultiplet to~$G_i$ is the number of simple roots~$\alpha_j$ of~$G_i$ such that $\langle\mu,\alpha_j\rangle>0$.
In particular, hypermultiplets are connected precisely to the groups~$G_i$ that they are charged under.
Hypermultiplets that are only charged under abelian gauge groups thus become isolated vertices in~$\Gamma$, and $\Gamma$~may well be disconnected even in a fully interacting theory.

For instance, if $G_i=SU(N)$, the number of edges joining a hypermultiplet to~$G_i$ is the number of distinct lengths of rows of the Young diagram: a single edge for fundamental or (anti)symmetric representations, a pair of edges for an adjoint representation (provided $N\geq 3$), etc., as listed in \autoref{tab:pI} on \autopageref{tab:pI}.
Consider the special case of a quiver gauge theory with unitary or special unitary gauge groups and hypermultiplets in fundamental, (bi/tri/\dots)fundamental, and adjoint representations, such as depicted in \autoref{fig:gauge-graphs}.
Each adjoint hypermultiplet of some $SU(N)$, drawn as a loop in the usual quiver description, gives rise to a pair of edges\footnote{For $N=2$ an adjoint of $SU(2)$ only gives a single edge rather than a pair, hence does not increase the genus.} joining $SU(N)$ to some hypermultiplet node, which contributes in the same way to the genus.
The edges depicting (bi/tri/\dots)fundamental hypermultiplets in the original quiver are simply subdivided with the addition of a vertex in~$\Gamma$ representing the hypermultiplet, again leaving the genus unchanged.
Finally, collections of fundamental hypermultiplets do not introduce further loops in~$\Gamma$.
We conclude that the genus of~$\Gamma$ typically agrees with the genus of the standard quiver depiction of the theory, decreased by the following effects:
\begin{itemize}
\item $U(1)$ gauge groups are omitted in the graph~$\Gamma$, which may reduce the genus;
\item adjoint matter of a $U(2)$ or $SU(2)$ gauge group does not contribute to the genus of~$\Gamma$.
\end{itemize}

\paragraph{Main statement.}
We are now ready to state our lower bound on~$\Nst$, and related results.
The assumption (in bold face) on monopole sectors is explained further below, near~\eqref{tech-assumption}.

\begin{proposition}[Counting single-trace mixed moduli]\label{prop:mixedmoduli}
  Consider a 3d $\Nsusy=4$ Lagrangian gauge theory that is good, has no accidental decoupled subsectors, and \textbf{does not have free $U(1)$ vector multiplets in any of its $\Delta(\mbf)=1$ monopole sectors.}\footnote{We prove in \autoref{subsec:quivers-proof} (resp.\ \autoref{subsec:abelian}) that this technical assumption is satisfied by quivers with unitary gauge group factors (resp.\ \emph{good} abelian theories).}
  Let $g$ be the genus of the bipartite graph~$\Gamma$ whose vertices are the nonabelian simple gauge groups~$G_i$, and the hypermultiplets (one per irreducible summand $R_I\subset\Vhyp$, with repetition), with one edge connecting $R_I$ to~$G_i$ for each copy of $\adj(G_i)$ in $R_I\otimes\Rbar_I$.
\begin{enumerate}
\item The multiplicity of the weight zero (of $\Gelec\times\Gmag$) in $\Nst$ is at least~$g$.
  In particular there are $\dim\Nst\geq g$ single-trace mixed moduli.
\item\label{item:ab-fac} In an abelian theory, all mixed moduli are factorizable, namely $\Nst=0$.
\item\label{item:quiv} In a quiver gauge theory with fundamental, bifundamental or adjoint matter\footnote{By adjoint matter we mean a hypermultiplet in the $\adj(U(n))-1$ representation of $U(n)$, omitting the free hypermultiplet that naturally appears for the complete adjoint representation of $U(n)$.} of \emph{nonabelian unitary} gauge groups $U(n_i\geq 2)$, such that none of the $U(2)$ factors have adjoint matter, the genus~$g$ is equal to the genus of the quiver.
\end{enumerate}
\end{proposition}

An aspect to be stressed again at this point is that the genus~$g$ is insensitive to abelian nodes.
If some hypermultiplets are only charged under abelian factors, they lead to disconnected nodes in the graph~$\Gamma$.  It is sometimes convenient to introduce the graph~$\Gammared$ in which only the hypermultiplets charged under at least one~$G_i$ are kept.  Its genus is the same as~$\Gamma$.  An example of graph associated to a gauge theory is given in \autoref{fig:gauge-graphs}.

For quivers with $U(n_i)$ gauge groups and (bi)fundamental hypermultiplets without adjoint matter\footnote{We conjecture that the equality remains valid with adjoint matter.}, we determine in \autoref{subsec:quivers} that the multiplicity of the weight zero (under electric and magnetic flavour symmetries) in~$\Nst$ is exactly the genus~$g$.
This generalizes the case of linear quivers that we have analyzed in~\cite{Bachas:2019jaa}, where we found that every $B_1[0]^{(1,1)}$ multiplet is factorizable, namely $\Nst=0$ (\autoref{prop:mixedmoduli} has $g=0$ in that case).
\label{forcircularquivers}Another standard case is that of circular quivers, which have
\begin{itemize}
\item enhanced $SU(2)_{\textnormal{elec}}$ symmetry acting on the pair of bifundamental hypermultiplets when there are exactly two nodes, and
\item enhanced $SU(2)_{\textnormal{mag}}$ symmetry due to $\Delta(\mbf=(1,1,\dots,1))=1$ when the total number of fundamental hypermultiplets is two (when it is lower, the theory is ugly or bad).
\end{itemize}
Our calculations for two-node circular quivers in \autoref{subsec:circular} (we also outline their generalization to any circular quiver) show that $\Nst$~reflects these enhanced symmetries,
\begin{equation}\label{Nst-circular}
  \Nst = \begin{cases}
    0 & \text{if any } n_i = 1 \ , \\
    r_{\textnormal{elec}}\otimes r_{\textnormal{mag}} & \text{if all } n_i \geq 2 \ ,
  \end{cases}
\end{equation}
where $r_X$ is $\adj(SU(2)_X)$ if that symmetry exists and is otherwise~$1$.
The weight-zero term is exactly equal to the genus ($0$ or~$1$ depending on $\min(n_i)$), consistent with our general analysis.

\paragraph{Counting products of currents.}

The proof of \autoref{prop:mixedmoduli} begins by rewriting~$\Nst$ in terms of known quantities.
A good theory has no $B_1[0]^{(\jH,1/2)}$ multiplets, as they would have to be dressings of a monopole of dimension $\Delta(\mbf)=1/2$, which does not exist in such a theory.
Thus, the mixed moduli multiplets $B_1[0]^{(1,1)}$ can only factor as products of $B_1[0]^{(\jH,0)}$ and~$B_1[0]^{(1-\jH,1)}$ multiplets.  The absence of free hypermultiplets $B_1[0]^{(1/2,0)}$ forbids $\jH=1/2$, and the case $\jH=0$ is trivial, so that the only possible factorization is into an electric current multiplet~$B_1[0]^{(1,0)}$ and a magnetic current multiplet~$B_1[0]^{(0,1)}$.

The number (or rather, the character) of $B_1[0]^{(1,1)}$ multiplets that are such products of electric and magnetic currents is
\begin{equation}
  (\text{factorized } B_1[0]^{(1,1)})
  = c_{1,0}c_{0,1} - \crel \ ,
\end{equation}
where the relations $\crel\geqrep 0$ arise in the gauge theory as F-term relations.
The character $\Nst$ is then
\begin{equation}\label{Nst-value}
  \Nst = n_T + c_{1,1} - c_{1,0} c_{0,1} + \crel \ .
\end{equation}
Here the $n_T$ contribution comes from the recombination with the $n_T\geq 1$ stress-tensor multiplet(s), where $n_T=1$ unless the theory splits into decoupled sectors.
Note that it can only increase~$\Nst$, so for the purpose of establishing lower bounds we do not need to determine $n_T$ beyond the fact that $n_T\geq 1$.

\paragraph{Technical assumption on flavours and monopole sectors.}

We have evaluated~$c_{1,1}$ in \autoref{sec:marginal} and found that its contribution from a monopole sector~$\mbf$ with $\Delta(\mbf)=1$ is
\begin{equation}\label{c11mbf}
  c_{1,1}^{\mbf} = \bigl( \singlet_{G_{\mbf}}(S^2\Vhyp^{\mbf}) - \nab^{\mbf} + 1 \bigr) w^{\mbf} \ .
\end{equation}
This is very similar to the formula $c_{1,0}=\singlet_G(S^2\Vhyp)-\nab$ for the electric flavour symmetries.
On the face of it, $c_{1,1}^{\mbf}$~counts flavour symmetries of an auxiliary theory with gauge group $G_{\mbf}/U(1)_{\mbf}$ and hypermultiplets~$\Vhyp^{\mbf}$, where the quotient (by the gauge group generated by~$\mbf$) is allowed because $\mbf$ acts trivially on~$\Vhyp^{\mbf}$, by construction.

Throughout this section we make the following technical assumption: in all $\Delta(\mbf)=1$ sectors, the auxiliary theory has no decoupled $U(1)$ vector multiplet.  In other words, we assume that the only $U(1)$ factor of~$G_{\mbf}$ that acts trivially on all of~$\Vhyp^{\mbf}$ (and on~$G_{\mbf}$) is the group~$U(1)_{\mbf}$.
Our assumption is thus that the weights of~$\Vhyp^{\mbf}$ and roots of~$G_{\mbf}$ generate the whole space orthogonal to~$\mbf$.
\begin{equation}\label{tech-assumption}
  \begin{aligned}
    & \textbf{Technical assumption:} \ \ \text{for $\Delta(\mbf)=1$,} \\
    & \qquad\qquad \Span \Bigl( \mbf^\perp \cap \bigl(\{w\preceq\Vhyp\}\cup\{\alpha\in\roots(G)\}\bigr)\Bigr) = \mbf^\perp \subset \lie{h}^* \ .
  \end{aligned}
\end{equation}
A violation of this assumption is equivalent to the existence of $x\in\lie{h}$ orthogonal to every weight or root in $\mbf^\perp$, namely such that $w\cdot\mbf=0$ implies $w\cdot x=0$ for all $w\preceq\Vhyp+\adj(G)$.
Based on the requirement that $\Delta(k\mbf+lx)\geq 1$ for any $(k,l)\neq(0,0)$ due to goodness of the theory, we prove the technical assumption in \autoref{subsec:quivers-proof} and \autoref{subsec:abelian} for quivers with unitary gauge group factors and for abelian theories, respectively.

It would be very interesting to find a counterexample to this assumption, namely a good theory with a decoupled $U(1)$ vector multiplet in some $\Delta(\mbf)=1$ monopole sector.
This could lead to a negative contribution $c_{1,1}^{\mbf}$ to the index, which would be either an indication for an accidental $A_2[0]^{(0,0)}$ stress-tensor multiplet (hence a decoupled subsector), or for a cancellation with some other monopole sector with the same $w^{\mbf}$~fugacities.
Without the condition on the monopole dimension, a simple counterexample would be the $\mbf=(1,2)$ sector of $U(2)$ SQCD with $4$ flavours: in that sector $\Vhyp^{\mbf}=0$ so all of $G_{\mbf}=U(1)^2$ acts trivially, and the combination $\singlet_{G_{\mbf}}(S^2\Vhyp^{\mbf}) - \nab^{\mbf} + 1$ in~\eqref{c11mbf} would then be negative.

\subsection{Abelian gauge theories}

\paragraph{Set-up.}
As a practice run, we consider the interesting case of $G=U(1)^{\nab}$ gauge theories.
We show first by a quick argument that every $B_1[0]^{(1,1)}$ multiplet factorizes thanks to F-term relations, then we more slowly verify that the index calculation confirms this.
As always, we require the theory to be good.

All representations of~$G$ are complex, except the trivial representation which we have excluded (we have assumed the theory has no free hypermultiplet).
Thus the quaternionic representation $\Vhyp$ has to be a sum (as usual we denote by $+$ and $\sum$ the direct sum of representations)
\begin{equation}
  \Vhyp = R + \Rbar \ , \qquad
  R = \sum_{I=1}^{\nCC} \bigl(\NCCbar_I\RCC_I\bigr) \ ,
\end{equation}
where the splitting into $R$ and~$\Rbar$ is arbitrary and we arrange for the gauge representations~$\RCC_I$ and $\RCCbar_J$ to be pairwise distinct.
We could easily recast this into a matrix of charges of hypermultiplets under~$G$, but we will rather keep the same notation as the rest of the paper.
Incidentally, one finds
\begin{equation}\label{abel-S2Vhyp}
  \singlet_G(S^2\Vhyp) = \singlet_G(\Lambda^2\Vhyp) = \sum_{I=1}^{\nCC} \NCCbar_I\NCC_I \ .
\end{equation}

\paragraph{Quick argument for factorization.}

Before imposing relations, bottom components of $B_1[0]^{(1,1)}$ multiplets can arise in two ways (see for instance the positive terms in $c_{1,1}$ in~\eqref{c20-c02}):
\begin{itemize}
\item In the zero-monopole sector: $G$-invariant contractions of a vector multiplet scalar and two hypermultiplet scalars.  In any such contraction, the vector multiplet scalars~$\phi$ can be factorized out because they are singlets under all symmetry groups ($G$~is abelian).  Chiral ring operators arising in this way are thus products of a component of~$\phi$ (part of a $U(1)_T$ current multiplet) and a contraction of two hypermultiplet scalars (based on $R$-charges these must be part of an electric current).
\item In a monopole sector with $\Delta(\mbf)=1$: $G_{\mbf}$-invariant dressings of this monopole by a pair of hypermultiplet scalars in $\Vhyp^{\mbf}$, namely that are not lifted by the monopole background.  Since $G$~is abelian, $\mbf$~commutes with all of it, so $G_{\mbf}=G$.  Thus, the $G_{\mbf}$-invariant pairs of hypermultiplets in $\Vhyp^{\mbf}\subseteq\Vhyp$ form a subset of those in~$\Vhyp$.  The dressed monopole is thus a product of the bare monopole (part of a magnetic current) and a pair of hypermultiplets (part of an electric current).
\end{itemize}
This shows that abelian 3d $\Nsusy=4$ Lagrangian gauge theories have no single-trace mixed moduli, establishing point~\ref{item:ab-fac} in \autoref{prop:mixedmoduli}.
We now give a different derivation starting from the superconformal index, which generalizes more robustly to the nonabelian setting.

\paragraph{Index calculation.}

Let us insert explicit expressions of $c_{\jH,\jC}$ from \eqref{c10c01c11}--\eqref{c20-c02} into the expression~\eqref{Nst-value} of~$\Nst$.
This involves the character $\cvhh = \singlet_G(\adj(G)\, S^2\Vhyp)$ counting certain combinations of vector and hypermultiplets.
For an abelian gauge group, we simply have
\begin{equation}
  \cvhh = \nab \singlet_G(S^2\Vhyp) \ .
\end{equation}
Collecting terms according to the monopole sector, and using that $G_{\mbf}=G$ and $\nab^{\mbf}=\nab$ for an abelian theory, one finds
\begin{equation}\label{Nst-abel}
  \begin{aligned}
    \Nst & = n_T + c_{1,1} - c_{1,0} c_{0,1} + \crel
    \\
    & = n_T + \crel - \singlet_G\bigl(\Lambda^2\Vhyp\bigr) + \sum_{\Delta(\mbf)=1} \bigl( \singlet_G(S^2\Vhyp^{\mbf} - S^2\Vhyp) + 1 \bigr) w^{\mbf} \ .
  \end{aligned}
\end{equation}
If the term $\crel$ was absent, this would typically give a negative result, since $\Lambda^2\Vhyp$ contains singlets and $S^2\Vhyp^{\mbf}\subseteq S^2\Vhyp$.
To get a sensible value for $\Nst\geqrep 0$, we must account for relations.

\paragraph{F-term relations and currents.}

Denote by $(q_I,\tq_I)$ the scalars in the chiral multiplets composing the $I$-th hypermultiplet, with $q_I \in \NCCbar_I\RCC_I$ and $\tq_I\in\NCC_I\RCCbar_I$.
An element $x\in\lie{g}$ of the gauge Lie algebra acts on~$q_I$ by multiplication by a scalar $\omega_I\cdot x$, where $\omega_I$~is the (unique) weight of~$G$ on~$\RCC_I$.
When seen as a 3d $\Nsusy=2$ theory, our 3d $\Nsusy=4$ gauge theory has the superpotential
\begin{equation}
  W = \sum_{I=1}^{\nCC} \Tr\bigl(\tq_I (\phi\cdot q_I)\bigr)
  = \sum_{I=1}^{\nCC} (\omega_I\cdot\phi) \Tr(\tq_Iq_I) \ , \qquad
\end{equation}
where $\phi$~is the vector multiplet scalar in the gauge Lie algebra, which acts on~$q_I$, and $\Tr$~is the pairing of dual representations.
In the second expression, $\omega_I\cdot\phi$ and $\tq_Iq_I$ are scalars so that every term is factorized.

In the chiral ring, every derivative of~$W$ vanishes.  This yields:
\begin{equation}\label{abel-Fterm}
  \begin{aligned}
    0 & = \sum_{I=1}^{\nCC} (\omega_I\cdot x) \Tr(\tq_Iq_I) \ , \qquad x\in\lie{g} \ , \\
    0 & = (\omega_I\cdot\phi) q_I \ , \quad 0 = (\omega_I\cdot\phi) \tq_I \ , \qquad 1\leq I\leq\nCC \ .
  \end{aligned}
\end{equation}
These relations allow us to retrieve the counting of electric currents in~\eqref{c10c01c11}, namely
\begin{equation}\label{abel-c10}
  c_{1,0} = \singlet_G(S^2\Vhyp) - \nab
  = \sum_{I=1}^{\nCC} \NCCbar_I\NCC_I - \nab \ .
\end{equation}
Indeed, each $\NCCbar_I\NCC_I$ consists of $\tq_Iq_I$ defined by contracting gauge indices (but not flavour indices), and these operators are subject to the constraints~\eqref{abel-Fterm}, namely $\nab$ linear constraints on the traces~$\Tr(\tq_Iq_I)$.
These F-term relations are linearly independent, otherwise one $U(1)$ gauge group factor would act trivially on all hypermultiplets and the theory would not be good.

\paragraph{F-term relations for zero-monopole terms.}

The character $\crel$ counts chiral ring relations among products of electric and magnetic currents.  Such products are of the form $\tq_Iq_I$ (modulo the $\nab$ aforementioned F-term relations) times components of~$\phi$, or times a monopole operator of dimension $\Delta(\mbf)=1$.  For now, we consider $\crel^{\mbf=0}$ which counts relations involving~$\phi$.
The second F-term constraints in~\eqref{abel-Fterm} imply that
\begin{equation}
  (\omega_I\cdot\phi) \tq_Iq_I = 0 \ , \qquad 1\leq I\leq\nCC \ ,
\end{equation}
which amounts to $\NCCbar_I\NCC_I$ relations for each~$I$, namely to $\singlet_G(\Lambda^2\Vhyp)$ relations in total.
Because the $\tq_Iq_I$ are themselves subject to the (first) F-term relation in~\eqref{abel-Fterm}, the overall trace
\begin{equation}
  \sum_{I=1}^{\nCC} (\omega_I\cdot\phi) \Tr(\tq_Iq_I)
  = \phi \cdot \sum_{I=1}^{\nCC} \Tr(\tq_Iq_I)\,\omega_I = 0
\end{equation}
involves an ``electric current'' $\sum_{I=1}^{\nCC} \Tr(\tq_Iq_I)\,\omega_I$ that was already removed by F-term conditions in~\eqref{abel-c10}.  It does not correspond to a relation on products of actual magnetic and electric current multiplets.
If the theory consists of several explicitly non-interacting subsectors, namely $G=H_1\times \dots\times H_{\nsec}$ with each hypermultiplet being charged only under one~$H_i$, then one relation is redundant in each sector.
Altogether we obtain
\begin{equation}
  \crel^{\mbf=0} = \singlet_G\bigl(\Lambda^2\Vhyp\bigr) - \nsec \ .
\end{equation}
Note that this combines neatly with the zero-monopole contributions in~\eqref{Nst-abel}:
\begin{equation}
  \Nst = n_T + \crel^{\mbf=0} - \singlet_G\bigl(\Lambda^2\Vhyp\bigr) + \text{monopoles}
  = n_T - \nsec + \text{monopoles} \ ,
\end{equation}
in which $n_T \geq \nsec$ because each sector has (at least) one stress-tensor.
In \autoref{prop:mixedmoduli} we assume that there are no accidental decoupled subsectors, so that $n_T=\nsec$.

\paragraph{F-term relations in monopole backgrounds.}

In a monopole background~$\mbf$, any hypermultiplet $(q_I,\tq_I)$ charged under $\mbf\in\lie{g}$ (namely such that $\omega_I\cdot\mbf\neq 0$) is lifted, so that the product of the bare monopole operator with the electric current $\tq_Iq_I$ vanishes.

Consider specifically the case $\Delta(\mbf)=1$.
Given the expression of~$\Delta$ (note that the sum ranges over~$R$ and not $\Vhyp=R+\Rbar$),
\begin{equation}
  1 = \Delta(\mbf) = \frac{1}{2} \sum_{\omega\preceq R} |\omega\cdot\mbf| \ ,
\end{equation}
all $\omega\cdot\mbf$ vanish, except for a single hypermultiplet with $|\omega\cdot\mbf|=2$, a hypermultiplet with $|\omega\cdot\mbf|=1$ and multiplicity~$2$, or two distinct hypermultiplets with $|\omega\cdot\mbf|=1$.

\

\noindent\underline{\textbf{Case~1}}: one has $|\omega_J\cdot\mbf|=2$ with $\dim\NCCbar_J=1$, and all other $\omega_I\cdot\mbf=0$.
The F-term condition~\eqref{abel-Fterm}, evaluated with $x=\mbf$, reduces to a single term,
\begin{equation}
  \pm \Tr(\tq_Jq_J)
  = \sum_{I=1}^{\nCC} (\omega_I\cdot\mbf) \Tr(\tq_Iq_I)
  = 0 \ .
\end{equation}
Thus, $\tq_Jq_J$ actually vanishes and the fact that its product with the monopole operator vanishes does not yield any relation.

\

\noindent\underline{\textbf{Case~2}}: $R$~includes that weight with multiplicity exactly two: $\dim\NCCbar_J=2$ and $|\omega_J\cdot\mbf|=1$ for some~$J$, while all other $\omega_I\cdot\mbf=0$.  This occurs for instance in the $T[SU(2)]$ theory.
The F-term condition~\eqref{abel-Fterm}, evaluated with $x=\mbf$, reduces to a single term,
\begin{equation}
  \pm \Tr(\tq_Jq_J)
  = \sum_{I=1}^{\nCC} (\omega_I\cdot\mbf) \Tr(\tq_Iq_I)
  = 0 \ ,
\end{equation}
hence $\tq_Jq_J$ is an $SU(2)$ electric current rather than~$U(2)$.
Its product with the monopole operator vanishes, which gives three relations.

\

\noindent\underline{\textbf{Case~3}}: two distinct hypermultiplets have $|\omega\cdot\mbf|=1$, namely $\dim\NCCbar_J=\dim\NCCbar_K=1$ and $|\omega_J\cdot\mbf|=|\omega_K\cdot\mbf|=1$ for some $J\neq K$, while all other $\omega_I\cdot\mbf=0$.
The F-term condition~\eqref{abel-Fterm}, evaluated with $x=\mbf$, reduces to two terms hence
\begin{equation}
  \tq_Jq_J = \pm \tq_Kq_K \ ,
\end{equation}
where we have omitted the traces because $\NCCbar_J,\NCCbar_K$ are one-dimensional.
Thus the two currents $\tq_Jq_J\in\NCCbar_J\NCC_J=\onebf$ and $\tq_Kq_K\in\NCCbar_K\NCC_K=\onebf$ coincide (up to sign), and we obtain a single relation when writing that its product with the magnetic monopole vanishes.

We obtain zero, one or three relations for each monopole sector:
\begin{equation}
  \crel^{\Delta(\mbf)=1} = \sum_{\Delta(\mbf)=1} w^{\mbf} \begin{cases}
    0 & \text{if } \dim\NCCbar_J=1 \ , \ |\omega_J\cdot\mbf|=2 \ , \\
    \NCCbar_J\NCC_J - 1 & \text{if } \dim\NCCbar_J=2 \ , \ |\omega_J\cdot\mbf|=1 \ , \\
    1 & \text{otherwise.}
  \end{cases}
\end{equation}

\paragraph{Collecting all the terms.}

The expression~\eqref{Nst-abel} of~$\Nst$ involves $\singlet_G(S^2\Vhyp^{\mbf} - S^2\Vhyp)$, which can be computed in the same three cases considered above.
We recall from~\eqref{abel-S2Vhyp} that $\singlet_G(S^2\Vhyp)=\sum_I\NCCbar_I\NCC_I$.  Likewise $\singlet_G(S^2\Vhyp^{\mbf})$ is obtained by omitting the hypermultiplets that are charged under~$\mbf$.
In the first case, that's a single hypermultiplet and the difference is~$\onebf$.
In the second case (with multiplicity) the difference is $\NCCbar_J\NCC_J$, and in the last case it is $(\NCCbar_J\NCC_J)+(\NCCbar_K\NCC_K)\simeq\onebf+\onebf$.
In all cases this is $\crel^{\Delta(\mbf)=1}+\onebf$ (the shift by~$\onebf$ appears in~\eqref{Nst-abel}).
All terms cancel and we are left with
\begin{equation}
  \Nst = n_T - \nsec \ ,
\end{equation}
which vanishes under our assumption in \autoref{prop:mixedmoduli} that no subsector accidentally decouples.

\subsection{Zero-monopole sector}
\label{subsec:sec4-zero}

We attack the nonabelian setting in several steps.
Based on the explicit expression~\eqref{Nst-value} of~$\Nst$, our tasks are: to evaluate~$\crel$, to expand $c_{\jH,\jC}$ in terms of gauge theory data using \eqref{c10c01c11}--\eqref{c20-c02}, and to further expand a combination $\cvhh$ appearing in these formulas.
We do these steps separately depending on the monopole sector~$\mbf$, namely we split
\begin{equation}\label{Nst-split}
  \Nst = \Nst^{\mbf=0} + \Nst^{\Delta(\mbf)=1} \ , \qquad
  \begin{aligned}
    \Nst^{\mbf=0} & = n_T + \crel^{\mbf=0} + c^{\mbf=0}_{1,1} - c_{1,0} c^{\mbf=0}_{0,1} \ , \\
    \Nst^{\Delta(\mbf)=1} & = \crel^{\Delta(\mbf)=1} + c^{\Delta(\mbf)=1}_{1,1} - c_{1,0} c^{\Delta(\mbf)=1}_{0,1} \ .
  \end{aligned}
\end{equation}
Importantly, the signs of $\Nst^{\mbf=0}$ and~$\Nst^{\Delta(\mbf)=1}$ are not known at this stage.
We evaluate first $\Nst^{\mbf=0}$, postponing the other contribution to \autoref{subsec:b11-monop}.

\paragraph{Combinations of vector and hypermultiplets~$\cvhh$.}

The term $c^{\mbf=0}_{1,1}$ involves the character $\cvhh = \singlet_G(\adj(G)\, S^2\Vhyp)$, which we now bound below.
We decompose $G = U(1)^{\nab} \times \prod_{i=1}^{\nna} G_i$ according to~\eqref{gauge-group-split} into abelian and non-abelian factors, and split the adjoint representation correspondingly into irreducible components.
From the abelian part, one has $\nab$ singlets in~$\adj(G)$, which contribute to $\cvhh$ as
\begin{equation}
  \cvhh
  = \nab \, \singlet_G\bigl( S^2\Vhyp \bigr)
  + \sum_{i=1}^{\nna} \singlet_G\bigl(\adj(G_i)\, S^2\Vhyp \bigr)\ .
\end{equation}
The number of singlets in $S^2\Vhyp$ was expressed in~\eqref{singletGS2RRbar}, whose detailed expression is not needed here because this term will cancel with part of the current-current contribution $c_{1,0}c^{\mbf=0}_{0,1}$.

We turn to a given non-abelian factor~$\adj(G_i)$.
We decompose $\Vhyp$ into irreducible components~\eqref{RoplusRbar-decomp}, and expand out $S^2\Vhyp$ using that $S^2(R+R')=S^2R+(R R')+S^2R'$ and $S^2(NR)=(S^2N)(S^2R)+(\Lambda^2 N)(\Lambda^2 R)$.  Dropping terms that involve distinct representations leads to the lower bound
\begin{equation}
  S^2\Vhyp
  \geqrep \sum_{I=1}^{\nCC} \NCC_I \NCCbar_I \RCC_I \RCCbar_I
  + \sum_{I=1}^{\nRR} \Bigl( S^2\NHH_I S^2\RRR_I + \Lambda^2\NHH_I \Lambda^2\RRR_I \Bigr)
  + \sum_{I=1}^{\nHH} \Bigl( S^2\NRR_I S^2\RHH_I + \Lambda^2\NRR_I \Lambda^2\RHH_I\Bigr) \ .
\end{equation}
We then tensor with $\adj(G_i)$ and seek gauge singlets in $\adj(G_i)\RCC_I\RCCbar_I$, and $\adj(G_i)S^2\RRR_I$, and so on.

\paragraph{Complexity of a representation.}

We analyse the number $p_{Ii}$ of gauge singlets in $\adj(G_i)R_I\Rbar_I$ for arbitrary irreducible representations in \autoref{app:tensor-adjoint}.
The highest weight~$\mu_I$ of~$R_I$ under the gauge group~$G$ necessarily lies in the Weyl chamber defined by $\langle\mu_I,\alpha\rangle\geq 0$ for all simple roots $\alpha$ of~$G$, and we show that $p_{Ii}$ counts how many boundaries $\mu_I$ does not belong to: we dub it the \emph{complexity} of $R_I$ under~$G_i$,
\begin{equation}
  p_{Ii} = \singlet_G\bigl(\adj(G_i)R_I\Rbar_I\bigr)
  = \#\bigl\{\alpha\text{ simple root of }G_i \bigm| \langle\alpha,\mu_I\rangle>0 \bigr\} \ .
\end{equation}
It is positive unless $R_I$ is neutral under~$G_i$, in which case $p_{Ii}=0$.
We also introduce the total complexity $p_I\in\ZZ_{\geq 0}$,
\begin{equation}
  p_I = \sum_{i=1}^{\nna} p_{Ii} \geq (\text{number of $G_i$ acting on $R_I$}) \ ,
\end{equation}
which is positive except for hypermultiplets that only transform under abelian gauge groups.
Since the theory has no free hypermultiplets, any hypermultiplet with $p_I=0$ must transform under some $U(1)$ factors, hence the representation must be complex.
Denoting by superscripts $\CC$, $\RR$, and~$\HH$ the quantities pertaining to complex, real, and quaternionic representations as before, we have learned that
\begin{equation}
  \pRR_I \geq 1 \ , \qquad \pHH_I \geq 1
\end{equation}
(with equality if and only if $\mu_I$ is a multiple of a fundamental weight of~$G$), while $\pCC_I$ may vanish.
For reference we summarize in \autoref{tab:pI} the complexity of various representations.
For commonly used representations they are often~$1$, with the notable exception of $SU(n\geq 3)$ adjoints.

\begin{table}\centering
  \caption{\label{tab:pI}Complexity $p(R)$ of the most commonly used representations of classical Lie groups.  Most of these representations have complexity~$1$, with the notable exceptions of the adjoint representation of $SU(n)$ for $n\geq 3$ (the other case listed in the table is the adjoint representation of $SO(6)=SU(4)/\ZZ_2$).  The complexity is additive under tensoring representations of different groups, insensitive to abelian factors, invariant under conjugation, and vanishes for the trivial representation.}
  \begin{tabular}{lll}
    \toprule
    Group & representation & complexity \\
    \midrule
    $SU(2)$
          & $R\neq\boldsymbol{1}$ & $1$ \\
    \midrule
    $SU(n)$, $n\geq 3$
          & $\boldsymbol{n}$ or $S^k\boldsymbol{n}$ & $1$ \\
          & $\Lambda^k\boldsymbol{n}$, $1\leq k\leq n-1$ & $1$ \\
          & $\textbf{adj}$ & $2$ \\
    \midrule
    $SO(n)$, $n\geq 5$
          & $\boldsymbol{n}$ or $(S^k\boldsymbol{n}-S^{k-2}\boldsymbol{n})$ & $1$ \\
          & $\Lambda^k\boldsymbol{n}$, $1\leq k\leq n/2$, $k\neq n/2-1$, including $\textbf{adj}$ for $n\neq 6$ & $1$ \\
          & $\Lambda^k\boldsymbol{n}$, $k=n/2-1$, including $\textbf{adj}$ for $n=6$ & $2$ \\
          & Chiral $(\Lambda^{n/2}\boldsymbol{n})^{\pm}$ (for even~$n$) & $1$ \\
          & Spinor & $1$ \\
    \midrule
    $USp(2n)$, $n\geq 2$
          & $\boldsymbol{2n}$ or $(\Lambda^k(\boldsymbol{2n})-\Lambda^{k-2}(\boldsymbol{2n}))$ & $1$ \\
          & $S^k(\boldsymbol{2n})$ including $\textbf{adj}$ & $1$ \\
    \bottomrule
  \end{tabular}
\end{table}

When $R_I$ is real or quaternionic, one has $R_I\Rbar_I=R_I^2=S^2R_I+\Lambda^2R_I$, so that $p_{Ii}$ splits into the (anti)symmetric complexities
\begin{equation}
  p_{Ii} = p_{Ii}^S + p_{Ii}^\Lambda \ , \qquad
  p_{Ii}^S = \singlet_G\bigl(\adj(G_i)S^2R_I\bigr) \ , \qquad
  p_{Ii}^\Lambda = \singlet_G\bigl(\adj(G_i)\Lambda^2R_I\bigr) \ .
\end{equation}
The explicit values of $p_{Ii}^S$ and~$p_{Ii}^\Lambda$ are determined in~\cite{LeFloch-Smilga}, as we outline in \autoref{app:tensor-adjoint-explicit}.
For our purposes later on, the key outcome is that for a real representation $R_I=\RRR_I$ or quaternionic representation $R_I=\RHH_I$,
\begin{equation}\label{pHH-S-gt-Lambda}
  p_{Ii}^{\RR,\Lambda} \geq \max(1, p^{\RR,S}_{Ii}) \ , \qquad
  p^{\HH,S}_{Ii} \geq 1 + p^{\HH,\Lambda}_{Ii} \qquad \text{for representations charged under } G_i \ .
\end{equation}
In particular, the complexities $p_I^{\HH,S}$ and $p_I^{\RR,\Lambda}$ (obtained by summing over gauge factors) are bounded below by the number of non-abelian gauge factors under which the given self-dual representation is charged.

Altogether we deduce the lower bound
\begin{equation}\label{cvhh-nna-geq}
  \begin{aligned}
    & \sum_{i=1}^{\nna} \singlet_G\bigl(\adj(G_i)S^2\Vhyp\bigr) \\
    & \quad \geqrep \sum_{I=1}^{\nCC} \pCC_I \NCC_I \NCCbar_I
    + \sum_{I=1}^{\nRR} \Bigl( p_I^{\RR,S} S^2\NHH_I + p_I^{\RR,\Lambda}\Lambda^2\NHH_I \Bigr)
    + \sum_{I=1}^{\nHH} \Bigl( p_I^{\HH,S} S^2\NRR_I + p_I^{\HH,\Lambda} \Lambda^2\NRR_I \Bigr) \ .
  \end{aligned}
\end{equation}
The flavour symmetry representations $\NCC_I \NCCbar_I$, $\Lambda^2\NHH_I$, and $S^2\NRR_I$ are each multiplied by at least the number of gauge groups under which the corresponding representations $\RCC_I$, $\RRR_I$, $\RHH_I$ are charged.
Interestingly, these flavour representations are the same as those that appear in $\singlet_G(\Lambda^2\Vhyp)$, see~\eqref{singletGS2RRbar}.

\paragraph{Evaluating the lower bound.}

Having worked out one of the ingredients in~$c^{\mbf=0}_{1,1}$, we are ready to calculate the number of single-trace mixed moduli one would have if there were no relations between products of currents.
Specifically, we evaluate the term $c^{\mbf=0}_{1,1} - c_{1,0} c^{\mbf=0}_{0,1}$ appearing in~\eqref{Nst-value} in terms of gauge theory data.
Taking into account \eqref{c10c01c11}--\eqref{c20-c02}, and collecting the monopole terms together yields some simplifications,
\begin{equation}\label{Nst-bound-2}
  \begin{aligned}
    c^{\mbf=0}_{1,1} - c_{1,0} c^{\mbf=0}_{0,1}
    & = \cvhh - \nab^2 - \nna - \singlet_G\bigl(\Lambda^2\Vhyp\bigr)
    - \bigl( \singlet_G\bigl(S^2\Vhyp\bigr) - \nab \bigr) \nab
    \\
    & = - \nna - \singlet_G\bigl(\Lambda^2\Vhyp\bigr)
    + \sum_{i=1}^{\nna} \singlet_G\bigl(\adj(G_i)\, S^2\Vhyp\bigr) \ .
  \end{aligned}
\end{equation}
We then use the lower bound \eqref{cvhh-nna-geq} on singlets of $\adj(G_i)\, S^2\Vhyp$, and the counting \eqref{singletGS2RRbar} of singlets in $\Lambda^2\Vhyp$ to obtain
\begin{equation}\label{Nst-bound-3}
  \begin{aligned}
    c^{\mbf=0}_{1,1} - c_{1,0} c^{\mbf=0}_{0,1}
    \geqrep - \nna
    + \sum_{I=1}^{\nCC} (\pCC_I - 1) \NCCbar_I\NCC_I
    & + \sum_{I=1}^{\nRR} \Bigl(p^{\RR,S}_I S^2\NHH_I + (p^{\RR,\Lambda}_I - 1) \Lambda^2\NHH_I \Bigr)
    \\
    & + \sum_{I=1}^{\nHH} \Bigl( (p^{\HH,S}_I - 1) S^2\NRR_I + p^{\HH,\Lambda}_I \Lambda^2\NRR_I \Bigr) \ .
  \end{aligned}
\end{equation}
On its own, this coarse lower bound of $\Nst$ is often useless because it is often negative (for instance for abelian theories).
That said, the contributions of real and quaternionic representations is always non-negative: $p^{\RR,\Lambda}$ and $p^{\HH,S}$ could only vanish if the representations were neutral under all non-abelian gauge groups hence under~$G$ itself (since pseudoreality forbids abelian charges), which would correspond to free hypermultiplets.

\paragraph{F-term relations.}

The superpotential and F-term relations are pretty similar to the abelian case.
Again, one has two types of relations depending on whether one differentiates the superpotential with respect to a vector multiplet or hypermultiplet scalars.
We recall the decomposition \eqref{RoplusRbar-decomp} $\Vhyp = \sum_{I=1}^n(\Nbar_I R_I)$ and we denote by $q_I\in \Nbar_I R_I$ the corresponding hypermultiplet scalars.
We also denote by $\tq_I\in N_I\Rbar_I$ the conjugate scalar, which lies in the same hypermultiplet.

The superpotential reads
\begin{equation}
  W = \sum_{I=1}^n \Tr(\tq_I\phi q_I) \ ,
\end{equation}
where $\phi q_I$ involves the action of $\lie{g}$ on~$\Vhyp$, and the trace denotes the pairing\footnote{Since the sum over~$I$ covers all of~$\Vhyp$, the sum is redundant for full hypermultiplets, as each term appears twice.  The action of $\phi$ on conjugate representations involves opposite signs, which would suggest that terms cancel.  However, the trace notation is hiding an additional sign: the same \emph{antisymmetric} pairing $\Lambda^2\Vhyp\to\onebf$ as in the kinetic term (in which the antisymmetry is compensated by the R-symmetry representations involved, so that the kinetic term remains symmetric).} of conjugate representations $\Nbar_I R_I$ and $N_I\Rbar_I$.
The first F-term relation is
\begin{equation}\label{Fterm-1}
  \sum_{I=1}^n \Tr(\tq_I x q_I) = 0 \ , \qquad x \in \lie{g} \ .
\end{equation}
More abstractly, the first F-term relation states that suitably contracting indices of $q^2\in S^2\Vhyp$ yields $0\in\lie{g}$.
The second relation states that any hypermultiplet scalar $q_I\in \Nbar_I R_I\subseteq\Vhyp$ is annihilated by the action of $\phi\in\lie{g}$ on that representation:
\begin{equation}\label{Fterm-2}
  0 = \phi q_I \ , \qquad 1\leq I\leq n \ .
\end{equation}

\paragraph{Electric currents.}

We first specialize~\eqref{Fterm-1} to $x\in\lie{u}(1)^{\nab}$ in the abelian part of~$\lie{g}$.
In terms of the charge vector~$\wab_I$ (abelian part of the weight) of~$R_I$ under $U(1)^{\nab}$, we have $xq_I=(\wab_I\cdot x)q_I$ and the factor $(\wab_I\cdot x)$ can be pulled out of the trace.
This yields a vector relation
\begin{equation}\label{Fterm-3}
  \sum_{I=1}^{\nCC} \Tr(\tqCC_I \qCC_I) \wab_I = 0 \in \lie{u}(1)^{\nab} \ .
\end{equation}
Only complex representations contribute because only these can be charged under $U(1)$ factors.
We thus have $\nab$~linear relations between the electric currents $\tqCC_I \qCC_I$, which correctly reproduces the counting $c_{1,0}=\singlet_G(S^2\Vhyp)-\nab$ of these currents.

\paragraph{F-term relations for cubic gauge-invariants.}

We are interested in relations on products of electric and magnetic currents, so we are only interested in gauge-invariant combinations of the hypermultiplets.
The F-term relations are thus only relevant through their gauge-invariant contractions with further vector or hypermultiplets:%
\begin{subequations}\label{Fterm-11}%
\begin{align}
  & & 0 & = \sum_{I=1}^n (\wab_I\cdot\phiab^{(i)})\Tr(\tq_Iq_I) \ , & & 1\leq i\leq\nab \ , & \label{F11-first} \\
  & & 0 & = \sum_{I=1}^n \Tr(\tq_I\phina^{(i)}q_I) \ , & & 1\leq i\leq\nna \ , \label{F11-second} \\
  & & 0 & = \tq_I \phi q_I
          = \sum_{i=1}^{\nab} (\wab_I\cdot\phiab^{(i)}) \tq_I q_I
          + \sum_{i=1}^{\nna} \tq_I \phina^{(i)} q_I \ , & & 1\leq I\leq n \ , \label{F11-third}
\end{align}%
\end{subequations}%
in which we decomposed $\phi\in\lie{g}$ into $\phiab^{(i)}\in\lie{u}(1)_i$ for $1\leq i\leq\nab$ and $\phina^{(i)}\in\adj(G_i)$ for each simple gauge factor.
(In the last relation, we have simply explicited this decomposition.)
An important word of warning: the first of these equations is actually stating the absence of certain electric currents, while the others are stating relations between cubic combinations of $\phi,\tq,q$, from which we will extract relations between currents.

\paragraph{Type 1.  Abelian relations.}

The current-current relations are all the linear combinations of~\eqref{Fterm-11} in which all $\phina^{(i)}$ terms cancel out.
One obvious class consists of the relation~\eqref{F11-third} $\tq_I\phi q_I=0$ for each hypermultiplet that only transforms under abelian gauge factors, namely such that $p_I=0$.
Such a hypermultiplet must have nontrivial $U(1)^{\nab}$ charges hence lie in a complex representation, and thus the relation is
\begin{equation}\label{F11-result-1}
  \sum_{i=1}^{\nab} (\wab_I\cdot\phiab) \tq_I q_I = 0 \in \NCC_I\NCCbar_I \ , \qquad \pCC_I = 0 \ , \qquad 1\leq I\leq\nCC \ .
\end{equation}
In view of~\eqref{F11-first}, the sum over~$I$ of traces of these relations is in fact not a relation between products of currents, because the involved ``electric currents'' in fact vanish.
More precisely, we have one such redundancy for each manifestly decoupled sector in the theory.
Overall, this gives the following number of relations:
\begin{equation}
  \sum_{1\leq I\leq\nCC} \delta_{\pCC_I=0} (\NCC_I\NCCbar_I) - \nsec \ .
\end{equation}

\paragraph{Type 2.  Connected components of the reduced graph.}

The second option is to cancel $\phina^{(i)}$ terms between \eqref{F11-second} and~\eqref{F11-third}.
This will be well described in terms of the graph~$\Gammared$ whose vertices are simple gauge nodes~$G_i$ and hypermultiplets charged under at least one~$G_i$, and with some number of edges (specifically~$p_{Ii}$) connecting each hypermultiplet to the~$G_i$ it is charged under.

Starting from the F-term relation~\eqref{F11-second} for a given~$G_i$, we must subtract the trace of~\eqref{F11-third} for every hypermultiplet charged under~$G_i$.  This, in turn, introduces $\phina^{(j)}$ for any $G_j$ under which these hypermultiplets are charged.  Compensating the $\phina^{(j)}$ with~\eqref{F11-second} and continuing back and forth, we find out that we need to include \eqref{F11-second} for all simple gauge nodes~$G_j$ that are in a given connected component of the graph~$\Gammared$, as well as \eqref{F11-third} for all hypermultiplets in the connected component.

We thus consider the connected components of~$\Gammared$.
Recall that hypermultiplets with $p_I=0$ are omitted in this graph.
For each representation~$R_I$ with $p_I\geq 1$, the $\dim\Nbar_I$ vertices corresponding to hypermultiplets in that representation must belong to the same component since they are connected to the same gauge nodes.
We can thus characterize a connected component by a non-empty subset $S\subset\{1,\dots,\nna\}$ and a subset $T\subset\{1,\dots,n\}$, such that
(1)~every hypermultiplet $q_I$, $I\in T$, is charged only under gauge groups $G_i$, $i\in S$;
(2)~every hypermultiplet $q_I$, $I\not\in T$ is neutral under all gauge groups $G_i$, $i\in S$; and
(3)~there is no proper subset of vertices with that property.

For any such connected component, consider the sum of traces of~\eqref{F11-third} for all $I\in T$, minus the sum of~\eqref{F11-second} for all $i\in S$:
\begin{equation}\label{F11-result-2}
  \begin{aligned}
    0 & = \sum_{I\in T} \sum_{i=1}^{\nab} (\wab_I\cdot\phiab^{(i)}) \Tr(\tq_I q_I)
    + \sum_{I\in T} \sum_{i=1}^{\nna} \Tr(\tq_I \phina^{(i)} q_I)
    - \sum_{i\in S} \sum_{I=1}^n \Tr(\tq_I\phina^{(i)}q_I) \\
    & = \sum_{I\in T} \sum_{i=1}^{\nab} (\wab_I\cdot\phiab^{(i)}) \Tr(\tq_I q_I) \ ,
  \end{aligned}
\end{equation}
where we have used that the two $\Tr(\tq_I\phina^{(i)}q_I)$ terms cancel out because they both reduce to a sum over $i\in S$ and $I\in T$ since hypermultiplets $q_I$, $I\in T$, are only charged under $G_i$, $i\in S$, and conversely the only hypermultiplets charged under any $G_i$, $i\in S$, are the $q_I$, $I\in T$.
The calculation~\eqref{F11-result-2} results in a relation between products of $U(1)_T$ magnetic currents and electric currents, for each connected component of~$\Gammared$.

\paragraph{Total number of relations.}

If we had been working with the full graph~$\Gamma$ instead of~$\Gammared$, there would be some obvious overlap between the two sets of relations: the trace of~\eqref{F11-result-1} is equal to the expression~\eqref{F11-result-2} for $S=\emptyset$ and $T=\{I\}$.
It is more convenient to neatly separate these by considering the reduced graph~$\Gammared$.
In conclusion, taking into account both types of relations, we find that
\begin{equation}\label{final-crel-m0}
  \crel^{\mbf=0} = H_0(\Gammared) - \nsec + \sum_{I=1}^{\nCC} \delta_{\pCC_I=0} (\NCC_I\NCCbar_I) \ ,
\end{equation}
where $H_0(\Gammared)$ denotes the number of connected components of the graph~$\Gammared$.
Incidentally, let us show that this number of relations is always non-negative.
Among the $\nsec$ manifestly non-interacting parts of the theory, distinguish those that involve at least one non-abelian gauge factor, from the purely abelian ones.  The former consist of one or more connected components of~$\Gammared$, together with some additional abelian factors and hypermultiplets, so there are at most $H_0(\Gammared)$ of these parts.
A decoupled sector of the latter type consists of abelian gauge groups and hypermultiplets charged only under these (which are thus complex and have $p_I=0$), and there must be at least one hypermultiplet, so that the last term in~\eqref{final-crel-m0} is enough to compensate for the $-1$ contribution of each such abelian sector.

\paragraph{Lower bound on zero-monopole single-trace mixed moduli.}

We now combine~\eqref{final-crel-m0} with~\eqref{Nst-bound-3} to obtain a lower bound on~$\Nst^{\mbf=0}$.
It is worth noting how the $\NCC_I\NCCbar_I$ terms combine: if $\pCC_I=0$ then the positive contribution from~\eqref{final-crel-m0} cancel the negative one from~\eqref{Nst-bound-3}, so that only terms with $\pCC_I\geq 2$ remain in this sum.
(We also recall that $p^{\RR,\Lambda}_I\geq 1$ and $p^{\HH,S}_I\geq 1$.)
Altogether,
\begin{equation}\label{Nst-bound-4}
  \begin{aligned}
    \Nst^{\mbf=0} & = n_T + \crel^{\mbf=0} + c^{\mbf=0}_{1,1} - c_{1,0} c^{\mbf=0}_{0,1} \\
    & \geqrep n_T - \nsec + H_0(\Gammared) \\
    & \quad - \nna
    + \sum_{\substack{1\leq I\leq\nCC\\\pCC_I\geq 2}} (\pCC_I - 1) \NCCbar_I\NCC_I
    \begin{aligned}[t]
      & + \sum_{1\leq I\leq\nRR} \bigl(p^{\RR,S}_I S^2\NHH_I + (p^{\RR,\Lambda}_I - 1) \Lambda^2\NHH_I \bigr)
      \\
      & + \sum_{1\leq I\leq\nHH} \bigl( (p^{\HH,S}_I - 1) S^2\NRR_I + p^{\HH,\Lambda}_I \Lambda^2\NRR_I \bigr) \ .
    \end{aligned}
  \end{aligned}
\end{equation}
Each decoupled sector has (at least) one stress-tensor so $n_T-\nsec\geq 0$.
We will show in \autoref{subsec:b11-genus} that the zero-weight coefficient of the last two lines is bounded below by minus the Euler characteristic of~$\Gammared$.  Combining with $H_0(\Gammared)$ yields the genus $H_1(\Gammared)$ of the reduced graph, hence of the full graph~$\Gamma$.

\subsection{Monopole contributions}
\label{subsec:b11-monop}

\paragraph{Evaluating the lower bound.}

We have evaluated and bounded below the zero-monopole part of~$\Nst$.
It is now time to evaluate $\Nst^{\Delta(\mbf)=1}$ given in~\eqref{Nst-split}, which we repeat here:
\begin{equation}
  \Nst^{\Delta(\mbf)=1} = \crel^{\Delta(\mbf)=1} + c^{\Delta(\mbf)=1}_{1,1} - c_{1,0} c^{\Delta(\mbf)=1}_{0,1} \ .
\end{equation}
We evaluate the term $c^{\Delta(\mbf)=1}_{1,1} - c_{1,0} c^{\Delta(\mbf)=1}_{0,1}$, taking into account \eqref{c10c01c11}--\eqref{c20-c02},
\begin{equation}\label{Nst-bound-Delta1}
  \begin{aligned}
    & c^{\Delta(\mbf)=1}_{1,1} - c_{1,0} c^{\Delta(\mbf)=1}_{0,1} \\
    & \quad = \sum_{\Delta(\mbf)=1} \Bigl( \singlet_{G_{\mbf}}\bigl(S^2\Vhyp^{\mbf}\bigr) - \singlet_G\bigl(S^2\Vhyp\bigr) - \nab^{\mbf} + \nab + 1 \Bigr) w^{\mbf} \ .
  \end{aligned}
\end{equation}
In general this has no sign, but adding the number of relations will make the monopole sector~$\mbf$ contribute non-negatively to~$\Nst$.

\paragraph{F-term relations in monopole backgrounds.}

In a given monopole background, one must keep only the hypermultiplets $\Vhyp^{\mbf}\subseteq\Vhyp$ that are not lifted by~$\mbf$, namely on which $\mbf\in\hfrak$ acts trivially.
Contrarily to the abelian case where hypermultiplets got lifted wholesale, the irreducible representations $\Nbar_I R_I\subseteq\Vhyp$ typically split into representations of~$G_{\mbf}$ as
\begin{equation}
  R_I = \sum_{\ell\in\ZZ} R_I^{\mbf,(\ell)} \ , \qquad
  R_I^{\mbf,(\ell)} = \bigl\{v\in R_I\bigm| \mbf\cdot v=\ell v\bigr\} \ .
\end{equation}
Each $R_I^{\mbf,(\ell)}$ can itself be further reducible as a representation of~$G_{\mbf}$.
Only the $R_I^{\mbf,(0)}$ piece remains in~$\Vhyp^{\mbf}$.
We denote by $q_I^{\mbf}\in R_I^{\mbf,(0)}$ the component of~$q_I$ that is not lifted.

Consider next the electric current $\tq_Iq_I\in N_I \Nbar_I$ (for real/quaternionic~$R_I$, symmetries restrict the current to $\Lambda^2N_I$ or $S^2N_I$).
When multiplied by the monopole operator~$\Ocal_{\mbf}$, the current $\tq_Iq_I$ reduces to
\begin{equation}\label{in the chiral ring.}
  (\tq_Iq_I) \Ocal_{\mbf}
  = \tq_I^{\mbf} q_I^{\mbf} \Ocal_{\mbf} \qquad \text{in the chiral ring.}
\end{equation}
If $q_I$ is entirely lifted (as happened in the abelian case), then $q_I^{\mbf}=0$ and the equation states that $\tq_Iq_I$ multiplied by the monopole~$\Ocal_{\mbf}$ gives zero.

\paragraph{Case 1. Abelian monopole charge.}
We consider first the case where $\mbf$~is entirely contained in the abelian part of~$\lie{g}$, so that the monopole background does not break~$G$, namely $G_{\mbf}=G$.
In that case, the representations $R_I$~are not decomposed: for each one, either the $\mbf$ charge vanishes and the representation is kept in~$\Vhyp^{\mbf}$, or it does not vanish and the hypermultiplet is lifted.
The lifted hypermultiplets must then be in complex representations (because they have a non-trivial charge under the abelian~$\mbf$), and they give the following relations: for $1\leq I\leq\nCC$,
\begin{equation}\label{quasi-abel-qtqO}
  (\tqCC_I\qCC_I) \Ocal_{\mbf} = 0 \in \NCC_I\NCCbar_I \ , \qquad \text{if } \wab_I\cdot\mbf\neq 0 \ .
\end{equation}
We check below which of these relations are actually redundant with the $\nab$ F-term constraints~\eqref{Fterm-3} that restrict the set of electric currents.

Before this, observe that $\Delta(\mbf)=1$ is very constraining.
Because $\mbf\in\lie{u}(1)^{\nab}\subset\lie{g}$, one has $\alpha\cdot\mbf=0$ for all roots~$\alpha$ of~$G$ so the monopole dimension formula reduces to a sum over hypermultiplet weights.
In addition, only complex representations can be charged under~$\mbf$, so the sum restricts further to
\begin{equation}
  \Delta(\mbf) = \frac{1}{2} \sum_{I=1}^{\nCC} \bigl|\wab_I\cdot\mbf\bigr| (\dim\NCC_I) \ .
\end{equation}
This can only be~$1$ if all $\wab_I\cdot\mbf=0$ except one or two hypermultiplets.
There are three subcases, mimicking those for abelian gauge theories:

\

\noindent\underline{\textbf{Subcase~1}}: one hypermultiplet with $|\wab_J\cdot\mbf|=2$ and without multiplicity.

\

\noindent\underline{\textbf{Subcase~2}}: a doublet with $|\wab_J\cdot\mbf|=1$ and $\dim\NCC_J=2$.

\

\noindent\underline{\textbf{Subcase~3}}: two hypermultiplets with $|\wab_J\cdot\mbf|=|\wab_K\cdot\mbf|=1$ and $\dim\NCC_J=\dim\NCC_K=1$.

\

In the three subcases there are $\onebf$, $\NCC_J\NCCbar_J$, and $\onebf+\onebf$ relations~\eqref{quasi-abel-qtqO}, respectively.
In all subcases a linear combination of these relations is
\begin{equation}
  0 = \sum_{I=1}^{\nCC} (\wab_I\cdot\mbf)\Tr(\tqCC_I\qCC_I)\Ocal_{\mbf} \ ,
\end{equation}
which is redundant with an F-term relation~\eqref{Fterm-3} on the products $\tq q$.
In subcase~1 we are left without any relation.
In subcase~2 the relations transform in $\adj(SU(2))$ and cannot be redundant with~\eqref{Fterm-3} since those involve traces.
In subcase~3, the situation is subtler.  A syzygy between \eqref{quasi-abel-qtqO} and~\eqref{Fterm-3} boils down to finding $x\in\lie{u}(1)^{\nab}$ such that
\begin{equation}
  \sum_{I=1}^{\nCC} \Tr(\tqCC_I \qCC_I) \wab_I\cdot x
  \in \Span\{ \tqCC_J\qCC_J , \tqCC_K\qCC_K \} \ ,
\end{equation}
where $J,K$ label the hypermultiplets charged under~$\mbf$.
In particular, $\wab_I\cdot x$ must vanish for all other hypermultiplets, so that $x$ and $\mbf$ both lie into the same subspace transverse to all $\wab_I$, $I\neq J,K$.
Suitable linear combinations of the other $\lie{u}(1)^{\nab}$ generators with $\mbf$ and~$x$ ensure that they don't act on $q_J,q_K$.
Thus, the sector consisting of the two hypermultiplets $q_J,q_K$ and the $U(1)^2$ gauge group spanned by $\mbf,x$ decouples.
One easily checks that such a sector cannot be good.

Long story short, from the class of purely abelian monopole charges~$\mbf$, one gets the following number of relations,
\begin{equation}
  \crel^{\Delta(\mbf)=1,\text{abel}} = \sum_{\Delta(\mbf)=1,\mbf\in\lie{u}(1)^{\nab}} w^{\mbf}
  \begin{cases}
    0 & \text{in subcase 1,} \\
    \NCCbar_J\NCC_J - 1 & \text{in subcase 2,} \\
    1 & \text{in subcase 3.}
  \end{cases}
\end{equation}

These relations cancel nicely with~\eqref{Nst-bound-Delta1}.  Using $\nab^{\mbf}=\nab$ and the explicit differences between $\Vhyp^{\mbf}$ and~$\Vhyp$, one finds
\begin{equation}\label{Nst-D1-abel}
  \crel^{\Delta(\mbf)=1,\text{abel}}
  + \sum_{\substack{\Delta(\mbf)=1\\\mbf\in\lie{u}(1)^{\nab}}} w^{\mbf} \bigl( \singlet_{G_{\mbf}}\bigl(S^2\Vhyp^{\mbf}\bigr) - \singlet_{G}\bigl(S^2\Vhyp\bigr) - \nab^{\mbf} + \nab + 1 \bigr)
  = 0 \ .
\end{equation}
Altogether, all of the $B_1[0]^{(1,1)}$ operators in this class of monopole backgrounds factorize into an electric current times the monopole operator.
This should not come as a surprise of course: the monopole background does not break gauge symmetry, so any gauge-invariant combination of hypermultiplets in this background already exists without the monopole.

\paragraph{Case 2. Nonabelian monopole charge.}
Next, we consider the case where $\mbf$ includes a nonabelian part, hence breaks the gauge group down to a strict subgroup $G_{\mbf}\subsetneq G$.
Returning to~\eqref{Nst-bound-Delta1}, our aim is to show the positivity property
\begin{equation}
  \label{Nst-D1-nonab}
  \Bigl( \singlet_{G_{\mbf}}\bigl(S^2\Vhyp^{\mbf}\bigr) - \nab^{\mbf} + 1\Bigr)
  - \Bigl( \singlet_G\bigl(S^2\Vhyp\bigr) - \nab \Bigr) + \text{relations}
  \geq 0
\end{equation}
by describing the full set of relations between products of $\Ocal_{\mbf}$ with an electric current.
The available relations are chiral ring relations~\eqref{in the chiral ring.}, and $F$-term relations associated to $U(1)$ gauge factors in~$G_{\mbf}$.

We first discuss the latter.  Consider the gauge theory with gauge group $G_{\mbf}/U(1)_{\mbf}$ and hypermultiplets transforming in the $\Vhyp^{\mbf}$~representation.
We recall our \textbf{technical assumption}~\eqref{tech-assumption} that this theory has no free $U(1)$ vector multiplet.
In other words every $U(1)$ factor in $G_{\mbf}/U(1)_{\mbf}$ acts on some hypermultiplets of~$\Vhyp^{\mbf}$, so that the $\nab^{\mbf}-1$ F-term relations
\begin{equation}\label{F-term-with-background}
  0 = \sum_{I=1}^n \Tr(\tq_I^{\mbf} x q_I^{\mbf}) \Ocal_{\mbf} \ , \qquad x\in \lie{g}_{\mbf}/\lie{u}(1)_{\mbf} \ ,
\end{equation}
are linearly independent.  While we have written them with a parameter~$x$ acting on~$q_I^{\mbf}$, we could have split further into irreducible representations of~$G_{\mbf}/U(1)_{\mbf}$ (or even its abelian factor) and pulled out~$x$ as a scalar factor.
Thus, \eqref{F-term-with-background} are relations between dressed monopoles $\tilde{\mu}\mu\Ocal_{\mbf}$ for $\mu$~in some irreducible $G_{\mbf}$~representations of~$\Vhyp^{\mbf}$.

On the other hand, \eqref{in the chiral ring.}~states how a product $(\tq q)\Ocal_{\mbf}$ reduces to operators of the form $\tilde{\mu}\mu\Ocal_{\mbf}$.
In fact, \eqref{in the chiral ring.}~is slightly redundant, and it should rather be understood as $c_{1,0}$~relations that evaluate the product of an electric current~$J$ by~$\Ocal_{\mbf}$.
Our task is to count the number of linear combinations of \eqref{in the chiral ring.} and~\eqref{F-term-with-background} that only involve $J\Ocal_{\mbf}$ products.

Because \eqref{F-term-with-background} are linearly independent, and are linearly independent of the relations \eqref{in the chiral ring.} for independent electric currents~$J$, we have
$\nab^{\mbf} - 1 + c_{1,0}$ linearly independent relations on dressed monopoles $\tilde{\mu}\mu\Ocal_{\mbf}$ and $J$-times-$\Ocal_{\mbf}$ products.
Consider now a linear combination of these relations, and impose that the coefficient in front of all $\tilde{\mu}\mu\Ocal_{\mbf}$ vanishes.
This puts $\singlet_{G_{\mbf}}(S^2\Vhyp^{\mbf})$ constraints on the coefficients, so that there are at least
\begin{equation}
  \bigl(\nab^{\mbf} - 1 + c_{1,0}\bigr) - \singlet_{G_{\mbf}}(S^2\Vhyp^{\mbf})
\end{equation}
relations between products of electric currents with~$\Ocal_{\mbf}$.
This is precisely the statement~\eqref{Nst-D1-nonab}.

It would be interesting to furnish a more explicit description of the relations: for instance, in a quiver gauge theory (with unitary gauge groups), there is a single-trace mixed moduli whenever the set of nodes with non-zero monopole charge forms a loop.

\subsection{On genus and quivers}
\label{subsec:b11-genus}

\paragraph{Combining monopole sectors.}

We have seen that nonzero monopole sectors contribute non-negative terms to~$\Nst$, see \eqref{Nst-D1-abel} and~\eqref{Nst-D1-nonab}.
One must interpret this carefully because the calculation is done one monopole sector at a time: we do not obtain that $\Nst^{\Delta(\mbf)=1}$ would be $\geqrep 0$ in the sense of characters (or representations).
Instead, we have found that the restriction of $\Nst^{\Delta(\mbf)=1}$ to any particular power of the magnetic fugacities~$w$ is non-negative.

We focus on the most interesting power $w^0$, namely on the subspace of~$\Nst$ of weight zero under~$\Gmag$.  It is a representation of~$\Gelec$.
Monopole sectors can contribute (if all of their $U(1)_T$ charges vanish), but as shown in \eqref{Nst-D1-abel} and~\eqref{Nst-D1-nonab} their contribution is nonnegative.
The lower bound~\eqref{Nst-bound-4} on single-trace mixed moduli then yields
\begin{equation}\label{Nstw0}
  \begin{aligned}
    \Nst|_{\text{coef }w^0} & \underset{\textnormal{elec}}{\geqrep} n_T - \nsec + H_0(\Gammared) - \nna
    + \sum_{1\leq I\leq\nCC,\ \pCC_I\geq 2} (\pCC_I - 1) \NCCbar_I\NCC_I
    \\
    & \quad
    + \sum_{1\leq I\leq\nRR} \bigl(p^{\RR,S}_I S^2\NHH_I + (p^{\RR,\Lambda}_I - 1) \Lambda^2\NHH_I \bigr)
      + \sum_{1\leq I\leq\nHH} \bigl( (p^{\HH,S}_I - 1) S^2\NRR_I + p^{\HH,\Lambda}_I \Lambda^2\NRR_I \bigr) \ .
  \end{aligned}
\end{equation}
where $\underset{\textnormal{elec}}{\geqrep}$ denotes here the ordering for representations of~$\Gelec$.

\paragraph{Lower-bound on the weight-zero subspace.}

The statement of \autoref{prop:mixedmoduli} involves the coefficient of $\mu^0w^0$ in the character~$\Nst$, where $(\mu,w)$ are (electric, magnetic) flavour fugacities.
In other words, it is (the dimension of) the subspace of the representation~$\Nst$ that has zero-weight under a Cartan torus of~$\Gelec$ and under the manifest $U(1)_T^{\nab}$ symmetry\footnote{The manifest topological symmetries do not necessarily form a Cartan subalgebra.  A simple example is the case of orthosymplectic quivers, which generally have no $U(1)_T$ symmetries but can have large magnetic symmetries nevertheless.} included in~$\Gmag$.
It is bounded below by the $\mu^0$ term in~\eqref{Nstw0}.

In \autoref{app:zeroweight} we establish lower bounds on the dimension of zero-weight subspaces:
\begin{equation}\label{lower-zeroweight}
  \begin{aligned}
    \dim\bigl((\NCCbar_I\NCC_I)_{\mu^0}\bigr) & \geq \dim \NCC_I \ ,
    \\
    \dim\bigl((S^2\NHH_I)_{\mu^0}\bigr) & \geq \tfrac{1}{2} \dim\NHH_I \ , \quad
    & & \dim\bigl((\Lambda^2\NHH_I)_{\mu^0}\bigr) \geq \tfrac{1}{2} \dim\NHH_I \ ,
    \\
    \dim\bigl((S^2\NRR_I)_{\mu^0}\bigr) & \geq \tfrac{1}{2} \dim\NRR_I \ , \quad
    & & \dim\bigl(((\NRR_I)^2)_{\mu^0}\bigr) \geq \dim\NRR_I \ .
  \end{aligned}
\end{equation}
Thanks to the inequality $p^{\HH,S}_I \geq 1 + p^{\HH,\Lambda}_I$ on total complexities of quaternionic representations (an immediate consequence of the same inequality~\eqref{pHH-S-gt-Lambda} for complexities under each~$G_i$) the $\mu^0$ term in the last sum in~\eqref{Nstw0} is bounded below as
\begin{equation}
  \begin{aligned}
    \bigl((p^{\HH,S}_I - 1) S^2\NRR_I + p^{\HH,\Lambda}_I \Lambda^2\NRR_I\bigr)_{\mu^0}
    & \geq (p^{\HH,S}_I - 1 - p^{\HH,\Lambda}_I) (S^2\NRR_I)_{\mu^0} + p^{\HH,\Lambda}_I ((\NRR_I)^2)_{\mu^0} \\
    & \geq \frac{1}{2} (p^{\HH,S}_I - 1 - p^{\HH,\Lambda}_I + 2 p^{\HH,\Lambda}_I) \dim\NRR_I \ .
  \end{aligned}
\end{equation}
Inserting this in~\eqref{Nstw0} yields
\begin{equation}\label{K00-geq}
  \begin{aligned}
    & \Nst|_{\text{coef }\mu^0w^0} \\
    & \geq  n_T - \nsec + H_0(\Gammared) \\
    & \quad - \nna + \sum_{\substack{1\leq I\leq\nCC\\\pCC_I\geq 2}} (\pCC_I - 1) \dim\NCC_I
    + \frac{1}{2} \sum_{\substack{1\leq I\leq\nRR\\\pRR_I\geq 2}} (\pRR_I - 1) \dim\NHH_I
    + \frac{1}{2} \sum_{\substack{1\leq I\leq\nHH\\\pHH_I\geq 2}} (\pHH_I - 1) \dim\NRR_I \ .
  \end{aligned}
\end{equation}
We now check that this is the genus~$g$ of the graph~$\Gamma$ defined in \autoref{prop:mixedmoduli}, generalized to allow for half-hypermultiplets.

\paragraph{Genus interpretation.}

Consider first the case where there are only complex representations (for instance a quiver of unitary groups with fundamental and bifundamental matter).
Then the graph has $\nna$~vertices for the simple gauge groups and $\sum_I\dim\NCC_I$ vertices for the hypermultiplets (only those that are charged under simple gauge groups),  and $\sum_I\pCC_I\dim\NCC_I$ edges, so that the last line in~\eqref{K00-geq} is the number of edges minus the number of vertices, which is the Euler characteristic.
Combining with $H_0(\Gammared)$ yields the genus $H_1(\Gammared)$.
As a result, \eqref{K00-geq} can be written as
\begin{equation}\label{Nst00-final}
  \Nst|_{\text{coef }\mu^0w^0} \geq n_T - \nsec + H_1(\Gammared) \geq H_1(\Gammared) = g \ .
\end{equation}
Full hypermultiplets in real or quaternionic irreducible representations of~$G$ are pairs of half-hypermultiplets, which cancels the factors of $1/2$ in \eqref{K00-geq} and leads to the same genus interpretation for these representations.
To accomodate half-hypermultiplets we can simply define the genus in terms of the right-hand side of~\eqref{K00-geq}; it can then be half-integral.

\paragraph{Returning to more conventional quivers.}

The graph $\Gammared$ (or equivalently~$\Gamma$) whose genus appears in~\eqref{Nst00-final} has a large number of vertices and edges.
Only the hypermultiplets with $p_I\geq 2$ (transforming under at least two gauge factors~$G_i$, or in a sufficiently elaborate representation of a single factor) can contribute to the genus.
In addition, hypermultiplets with $p_I=2$ can be replaced by an edge joining the gauge nodes directly, without changing the genus.
This provides a more standard description of the gauge theory as a quiver in which bifundamental hypermultiplets are depicted by edges.
More precisely, to retrieve an actual gauge theory one must decorate~$\Gammared$ with the data of representations, and $U(1)$~charges, that were removed when defining~$\Gammared$.

\section{Classes of examples}
\label{sec:examples}

In \autoref{subsec:quivers} we specialize our results to a commonly studied class of theories:
\begin{equation}\label{quiver-desc}
  \text{$\prod_i U(n_i)$ quiver gauge theories without adjoint matter,}
\end{equation}
namely with only fundamental and bifundamental hypermultiplets.
For these theories we prove in \autoref{subsec:quivers-proof} that our technical assumption~\eqref{tech-assumption} holds.
We prove in \autoref{subsec:abelian} that it also holds in arbitrary $U(1)^{\nab}$ gauge theories.
To finish our journey, we work out in detail the example of circular quivers with two gauge nodes in \autoref{subsec:circular}.  This last subsection quotes some earlier results but can be read independently.

\subsection{Quiver gauge theories with unitary gauge group factors}
\label{subsec:quivers}

We denote by $Q=(V,E,n,M)$ the quiver, where $V$~is the set of gauge groups of the quiver, so that $i\in V$ in~\eqref{quiver-desc}, the edges~$E$ describe bifundamental hypermultiplets joining pairs of these gauge groups, the gauge groups are $U(n_i)$ with $n_i\geq 1$, $i\in V$, and they each have $M_i$~fundamental hypermultiplets.
We also denote by $Q'=(V',E',n,M')$ the quiver obtained by removing all gauge groups $U(n_i)$ with $n_i=1$, removing all (bi)fundamental hypermultiplets charged only under such groups, and converting bifundamental hypermultiplets of such $U(1)$ groups and of $U(n_j)$ with $n_j\geq 2$ into additional fundamental hypermultiplets.

The graph $\Gamma$ defined in \autoref{prop:mixedmoduli} is then obtained from $Q'$ by adding edges connecting each vertex $i\in V'$ to $M'_i$~new nodes representing the fundamental hypermultiplets, and subdividing each edge in~$E'$ by inserting inside it an additional vertex.
An immediate consequence is that the genus~$g$ of~$\Gamma$ is the same as the genus of the quiver~$Q'$, which is visually easier to read off.
Our general considerations above imply that the weight-zero coefficient in~$\Nst$ is at least~$g$.
For this class of theories we can be more precise and derive that this is the exact count,
\begin{equation}\label{Nst-circ-00}
  \Nst|_{\text{coef }\mu^0w^0} = g = \text{(genus of quiver without $U(1)$)} \ .
\end{equation}
The idea is to explicitly show that all products of $\phi,q,\tq$ factorize except~$g$ of them, and that monopoles cannot contribute to the weight-zero subspace.
We have not analyzed the effect of adjoint matter, but we conjecture that the equality remains valid in that case.

\paragraph{Monopoles cannot contribute.}

Monopole charges~$\mbf$ of this gauge theory have components $\mbf_{ia}$ for $1\leq i\leq\nab=\nna$ and $1\leq a\leq n_i$.
The monopole sectors that can contribute $B_1[0]^{(1,1)}$ multiplets are those with $\Delta(\mbf)=1$.
Let $\mbf^+$ denote the positive part of~$\mbf$, defined by its components $\mbf^+_{ia}=\max(0,\mbf_{ia})$.
Every term in $\Delta(\mbf)$ is an absolute value $|\mbf_{ia}|$ or $|\mbf_{ia}-\mbf_{jb}|$, and one easily checks
\begin{equation}\label{monopoles-cannot-contribute}
  \begin{aligned}
    \bigl|\mbf_{ia}\bigr| & = \bigl|\mbf^+_{ia}\bigr| + \bigl|\mbf_{ia} - \mbf^+_{ia}\bigr| \ , \\
    \bigl|\mbf_{ia}-\mbf_{jb}\bigr| & = \bigl|\mbf^+_{ia}-\mbf^+_{jb}\bigr| + \bigl|\mbf_{ia} - \mbf^+_{ia} - \mbf_{jb} + \mbf^+_{jb}\bigr| \ .
  \end{aligned}
\end{equation}
As a result,
\begin{equation}\label{Un-Delta-split-0}
  \Delta(\mbf) = \Delta(\mbf^+) + \Delta(\mbf-\mbf^+) \ .
\end{equation}
In a good theory, this can only be equal to~$1$ if one of the summands vanishes, namely if $\mbf^+=0$ or $\mbf=\mbf^+$.
Thus, either all components of~$\mbf$ are non-negative, or all are non-positive.

The $U(n_i)$ traces $\Tr_i(\mbf)$ for $1\leq i\leq\nna$ then have the same sign, all non-negative or all non-positive.
In addition, the only case where they all vanish is if $\mbf=0$, which does not have dimension~$1$.
Therefore, every $\Delta(\mbf)=1$ monopole sector has a non-trivial power of at least one magnetic flavour fugacity.\footnote{Incidentally, this proves that the Cartan subalgebra of~$\Gmag$ cannot go beyond the manifest $U(1)_T$ symmetries, so that we can equivocate between selecting the $w^0$ term and selecting the zero-weight space under~$\Gmag$.}
We deduce that monopole sectors cannot contribute to the weight zero subspace of~$\Nst$.

\paragraph{Almost all vector-hyper combinations factorize.}

The $B_1[0]^{(1,1)}$ multiplets in the zero-monopole sector arise as gauge-invariant products of the vector multiplet scalar~$\phi_i$ of a gauge group $U(n_i)$ and of two hypermultiplet scalars.
We denote by $(q_i,\tq_i)$ the scalar components of fundamental hypermultiplets, with $q_i$ and $\tq_i$ transforming in the fundamental and antifundamental representations of $U(n_i)$.
We also denote by $(q_e,\tq_e)$ the scalar components of the bifundamental hypermultiplet corresponding to the edge~$e$.
More precisely, to specify which scalar is denoted by $q_e$ and~$\tq_e$ one should fix an orientation of the edge, from a source $s(e)$ to a target $t(e)$, and fix that $q_e$ lies in the fundamental representation of $U(n_{t(e)})$ and the antifundamental of $U(n_{s(e)})$.
Besides the products of $\Tr(\phi_i)$ and of an electric current, the remaining gauge-invariant combinations are
\begin{equation}\label{Be-def}
  B_i = \tq_i \phi_i q_i \ , \qquad
  B_{ee'} = \Tr_{s(e)}(\tq_{e'} \phi_{t(e)} q_e) \ ,
\end{equation}
for any node~$i$, and for any pair of directed edges $e,e'\in E$ of the quiver~$Q$ with the same source $s(e)=s(e')$ and target $t(e)=t(e')$.
For brevity we denote $B_e=B_{ee}$: the more general $B_{ee'}$ is only relevant when the quiver has multiple edges between the same pair of nodes, and we present an example thereof in \autoref{subsec:circular}.

The product~$B_i$ transforms in the adjoint representation of the electric flavour symmetry $U(M_i)$ that rotates the $M_i$ fundamental scalars~$q_i$, which seems like it could give rise to numerous $B_1[0]^{(1,1)}$ multiplets.
However, the F-term relation for~$\tq_i$ states that $\phi_i q_i = 0$ hence $B_i=0$.
The products~$B_{ee'}$ for all edges $e,e'$ joining a given pair of nodes $i,j$ transform likewise in the adjoint representation of the $U(M_{ij})$ symmetry that rotates the $M_{ij}$ edges joining $i$ and~$j$.
For the purpose of deriving~\eqref{Nst-circ-00}, we are only interested in the zero-weight sector under $U(M_{ij})$, namely the diagonal products $B_e=B_{ee}$ for all $M_{ij}$ edges~$e$ joining $i$ and~$j$.
Altogether we can focus solely on~$B_e$ for all edges $e$ of the quiver.

A special case is that the operator $B_e$ is factorizable if $n_{t(e)}=1$, because in that case $\phi_{t(e)}$ is a scalar and can be pulled out of the trace.
We can write this as
\begin{equation}\label{Be-0}
  B_e \equiv 0 \pmod{\text{double-trace}} \ .
\end{equation}

The F-term relation for the bifundamental~$\tq_e$ states that $\phi_{t(e)} q_e=-q_e\phi_{s(e)}$, hence
\begin{equation}\label{Be-Breversed}
  B_e = \Tr_{t(e)}(\phi_{t(e)} q_e \tq_e) = - \Tr_{t(e)}(q_e \phi_{s(e)} \tq_e) = -B_{\text{reversed }e} \ .
\end{equation}
The F-term relation for a gauge node states that $\sum_{e\mid t(e)=i} q_e \tq_e = - q_i \tq_i$, and the F-term relation for~$\tq_i$ states that $\phi_i q_i = 0$ hence
\begin{equation}\label{Be-sum}
  \sum_{e\mid t(e)=i} B_e = \Tr_i(\phi_i q_i \tq_i) = 0 \ .
\end{equation}
Amusingly, these equations \eqref{Be-Breversed} and~\eqref{Be-sum} are identical to the conservation of a current flowing in the quiver seen as a circuit, which we cut at each abelian node when working modulo products of currents as in~\eqref{Be-0}.
Taking~\eqref{Be-0} into account renders the relation~\eqref{Be-sum} trivial for $n_i=1$.
The remaining redundancies (syzygies) between F-term relations are as follows: a given linear combination $\sum_i c_i \bigl(\sum_{e\mid t(e)=i} B_e\bigr)$ of these relations, for some coefficients~$c_i$, is trivial if and only if each $B_{ij}$ appears with a vanishing coefficient namely $c_i=c_j$ whenever there is an edge~$ij$.
The number of syzygies is thus the number $h'_0$ of connected components of the quiver~$Q'$.
Overall, we find that the number of (linearly independent) single-trace mixed moduli is the genus of the quiver~$Q'$, or equivalently~$\Gamma$,
\begin{equation}
  |E'| - |V'| + h'_0 = \text{genus of $Q'$} = g \ .
\end{equation}

\subsection{Proof of the technical assumption in quivers of unitary gauge groups.}
\label{subsec:quivers-proof}

\paragraph{Setup.}
We continue with the set up of \autoref{subsec:quivers}, namely $\prod_i U(n_i)$ gauge theories with fundamental and bifundamental hypermultiplets, and prove now the technical assumption~\eqref{tech-assumption}.
Fix a monopole sector~$\mbf$ with $\Delta(\mbf)=1$, and a vector $x\in\lie{h}$ such that $w\cdot x=0$ for any weight $w\preceq\Vhyp+\adj(G)$ with $w\cdot\mbf=0$.
We seek to prove that $x$ is a multiple of~$\mbf$.

\paragraph{Analysis of monopoles.}

In~\eqref{monopoles-cannot-contribute} we have decomposed monopole charges into positive and negative parts and found that the monopole dimension is additive for this decomposition.
More generally, one can split $\mbf$ into a part $\mbf_{\text{clamped}}$ whose charges are clamped to some interval $[a,b]$ with $a\leq 0\leq b$, and the rest.  The dimension is again additive for the quiver theories studied now:
\begin{equation}\label{Un-Delta-split}
  \Delta(\mbf) = \Delta(\mbf_{\text{clamped}}) + \Delta(\mbf-\mbf_{\text{clamped}}) \ .
\end{equation}
In a good theory this means that either $\Delta(\mbf)\geq 2$ or the decomposition is trivial (as happens for instance for $-a$ and~$b$ large enough, or for $a=b=0$).
For a $\Delta(\mbf)=1$ monopole, the decomposition must be trivial for arbitrary $a,b$, which requires the monopole to be ``tiny'' in the sense that all of its entries are $0$ or~$+1$, or all of its entries are $0$ or~$-1$.
Without loss of generality we can focus on the first case.

A (positive) tiny monopole charge~$\mbf$ is characterized by the traces
\begin{equation}
  0\leq \tau_i \coloneqq \Tr_i(\mbf)\leq n_i \ ,
\end{equation}
and its dimension can be recast as
\begin{equation}\label{dim-tiny}
  \Delta(\mbf) = \sum_{i\in V} (\tau_i^2 + \beta_i \tau_i) - \sum_{e\in E} \tau_{s(e)}\tau_{t(e)} \ .
\end{equation}
Here, the balance is defined as $\beta_i=-n_i + \frac{1}{2} M_i + \frac{1}{2}\sum_{(i\leftrightarrow j)\in E}n_j \in \frac{1}{2}\ZZ$ where edges starting or ending at~$i$ contribute regardless of their orientation.

\paragraph{Unlifted hypermultiplets.}

The monopole background breaks each $U(n_i)$ gauge group down to $U(n_i-\tau_i)\times U(\tau_i)$.
The unlifted vector multiplets and hypermultiplets are
\begin{equation}\label{quiver-Vhyp-mbf}
  \begin{aligned}
    \adj(G_{\mbf}) & = \sum_{i=1}^{\nna} \Bigl( \adj\bigl(U(n_i-\tau_i)\bigr) + \adj\bigl(U(\tau_i)\bigr) \Bigr) \ ,
    \\
    \Vhyp^{\mbf} & = \sum_{i=1}^{\nna} \overline{M_i} (n_i-\tau_i)
    + \sum_{e\in E} \Bigl( \overline{(n_{s(e)}-\tau_{s(e)})} (n_{t(e)}-\tau_{t(e)}) + \overline{\tau_{s(e)}}\tau_{t(e)} \Bigr) \ ,
  \end{aligned}
\end{equation}
where $M_i$ are flavour representations, and $(n_i-\tau_i)$ and $\tau_i$ denote the fundamental representations of $U(n_i-\tau_i)$ and $U(\tau_i)$.
If any of $\tau_{s(e)}$, $\tau_{t(e)}$, $n_{s(e)}-\tau_{s(e)}$, or $n_{t(e)}-\tau_{t(e)}$ vanishes, then the terms involving it in~\eqref{quiver-Vhyp-mbf} drop out.

Consider the projection $x_i$ of~$x$ in the Cartan algebra $\lie{h}_i$ of $U(n_i)$.
The vector $x\in\lie{h}$ must be orthogonal to all weights of $\adj(G_{\mbf})$, hence
\begin{equation}
  x_i = \bigl( x^0_i , \dots , x^0_i , x^1_i , \dots , x^1_i \bigr) \in \lie{h}_i \ ,
\end{equation}
where $x^0_i$ is repeated $n_i-\tau_i$ times and $x^1_i$ is repeated $\tau_i$ times.
(The superscripts $0$ and~$1$ are motivated by the corresponding components of~$\mbf$ being $0$ or~$1$.)
Notice that $x^0_i$ is undefined if $\tau_i=n_i$ and $x^1_i$ is undefined if $\tau_i=0$.
Next, we know that $x$~acts trivially on the hypermultiplets in~\eqref{quiver-Vhyp-mbf}:
for each vertex $i$, or for each edge~$e$,
\begin{itemize}
\item If $\tau_{s(e)},\tau_{t(e)}>0$ then the hypermultiplet $\overline{\tau_{s(e)}}\tau_{t(e)}$ exists, so we have $x^1_{s(e)}=x^1_{t(e)}$.
\item If $\tau_{s(e)}<n_{s(e)}$ and $\tau_{t(e)}<n_{t(e)}$ then likewise we have $x^0_{s(e)}=x^0_{t(e)}$.
\item If $\tau_i<n_i$ and $M_i>0$ then we have $x^0_i=0$.
\end{itemize}

\paragraph{Connectedness considerations.}

We know that the gauge nodes with $\tau_i>0$ are connected by edges of the quiver: if they were disconnected then clearly the dimension $\Delta(\mbf)$ would be the sum of dimensions of simpler monopoles corresponding to the connected components, which would violate goodness.
Thus, we have that all $x^1_i$ are equal to the same value $\lambda$ (for all nodes with $\tau_i>0$).
There only remains to show that all $x^0_i=0$: then we can deduce $x=\lambda\mbf$ since $\mbf$ has precisely those components equal to~$1$.

Consider next the set $V^0\subset V$ of gauge nodes such that $\tau_i<n_i$, and focus on one connected component $C\subset V^0$.
Because $C$~is connected, the orthogonality of~$x$ with weights of $\overline{(n_{s(e)}-\tau_{s(e)})} (n_{t(e)}-\tau_{t(e)})$ forces $x^0_i$~to be equal to the same value~$x^0_C$ for all $i\in C$.
We now prove that some node $i\in C$ has fundamental matter $M_i\neq 0$, which forces $x^0_C=0$.
Assume by contradiction that $M_i=0$ for all $i\in C$.
Then we exhibit a nonzero monopole charge $\widetilde{\mbf}\neq\mbf$ such that
\begin{equation}\label{widetildembf-Delta}
  \Delta(\mbf) = \Delta(\widetilde{\mbf}) + \Delta(\mbf-\widetilde{\mbf}) \ ,
\end{equation}
thus violating goodness.
Specifically, we consider the positive tiny monopole characterized by its traces
\begin{equation}
  \ttau_i = \begin{cases}
    n_i & \text{if } i\in C \ , \\
    \tau_i & \text{otherwise} \ .
  \end{cases}
\end{equation}
We establish~\eqref{widetildembf-Delta} by checking that either $w\cdot\widetilde{\mbf}$ or $w\cdot(\mbf-\widetilde{\mbf})$ vanishes for every weight $w\preceq\Vhyp+\adj(G)$.
We consider each type of weight in turn.
\begin{itemize}
\item For weights (and roots) that do not involve any $U(n_i)$ with $i\in C$, one has $w\cdot(\mbf-\widetilde{\mbf})=0$ since the two monopole charges coincide away from~$C$.
\item For roots $\alpha$ of $U(n_i)$ with $i\in C$ one has $w\cdot\widetilde{\mbf}=0$ since $\ttau_i=n_i$ means that $\widetilde{\mbf}_i=(1,1,\dots,1)$.
\item For bifundamentals of $U(n_i)$ and $U(n_j)$ with $i,j\in C$, we have $\widetilde{\mbf}_i=(1,\dots,1)$ and $\widetilde{\mbf}_j=(1,\dots,1)$ hence $w\cdot\widetilde{\mbf}=0$.
\item For bifundamentals of $U(n_i)$ and $U(n_j)$ with $i\in C$ and $j\not\in C$, we observe that $\tau_j=n_j$, as otherwise the node~$j$ would belong to the connected component~$C$.  Thus $\widetilde{\mbf}_i=(1,\dots,1)$ and $\widetilde{\mbf}_j=(1,\dots,1)$ and again $w\cdot\widetilde{\mbf}=0$.
\end{itemize}
This concludes the verification of~\eqref{widetildembf-Delta}, hence the proof by contradiction that every connected component has fundamental matter.
As a result, all $x^0_i$ vanish, so that $x=\lambda\mbf$ for some~$\lambda$, which concludes our proof of the technical assumption for quivers of unitary gauge groups without adjoint matter.

\subsection{Proof of the technical assumption in abelian theories}
\label{subsec:abelian}

\paragraph{Aim.}
We seek to prove the technical assumption~\eqref{tech-assumption} in the case where $G=U(1)^{\nab}$ is abelian.  In particular, gauge representations are complex so that we can split $\Vhyp=R+\Rbar$.
The monopole dimension is thus
\begin{equation}
  \Delta(\mbf) = \frac{1}{2} \sum_{w\preceq R} |w\cdot\mbf| \ .
\end{equation}
We wish to prove that in any sector with $\Delta(\mbf)=1$, one has
\begin{equation}\label{tech-assumption-abelian}
  \Span \bigl\{w\preceq R,w\cdot\mbf=0\bigr\} = \mbf^\perp \ .
\end{equation}

\paragraph{First case.}
One possibility is that one term is $|w_{(1)}\cdot\mbf|=2$ and all others vanish, in other words all of the weights $w\preceq R$ except $w=w_{(1)}$ are orthogonal to~$\mbf$.
Since weights of~$R$ span all of $\lie{h}^*$ (otherwise the gauge theory would have a free vector multiplet), the weights $w\neq w_{(1)}$ span at least a hyperplane, which lies in~$\mbf^\perp$ hence coincides with it.

\paragraph{Second case.}
The other possibility is that two terms are $|w_{(1)}\cdot\mbf|=|w_{(2)}\cdot\mbf|=1$ while all other weights $w\preceq R$ are orthogonal to~$\mbf$.
We assume by contradiction that they span a strict subspace $L\subsetneq\mbf^\perp\subsetneq\lie{h}^*$.
Since weights of~$R$ span all of $\lie{h}^*$ we know that $\Span(w_{(1)},w_{(2)},L)=\lie{h}^*$ namely that $L$ has codimension at most~$2$, hence exactly~$2$ because of the aforementioned strict inclusions.

The orthogonal $L^\perp\subset\lie{h}$ is a two-dimensional subspace that contains $\mbf$.  Since the theory is good and $\Delta(\mbf)=1$, the vector $\mbf$ is necessarily a primitive vector in the lattice~$\Lambdamon$ (that is, not an integer multiple of some other vector).  Thus, one can choose a second primitive vector~$\pbf$ such that $\mbf$ and $\pbf$ together form a basis of $L^\perp\cap\Lambdamon=\ZZ\mbf+\ZZ\pbf$.
The dimension of any vector $k\mbf+l\pbf$ in this two-dimensional lattice is then computed as
\begin{equation}\label{Delta-kmbf-lpbf}
  \Delta(k\mbf+l\pbf) = \frac{1}{2} \sum_{w\preceq R} \bigl|w\cdot(k\mbf+l\pbf)\bigr|
  = \frac{1}{2} \bigl|k + l w_{(1)}\cdot\pbf\bigr| + \frac{1}{2} \bigl|k + l w_{(2)}\cdot\pbf\bigr| \ .
\end{equation}
To move forward we introduce $\delta=|(w_{(1)}-w_{(2)})\cdot\pbf|$.
It cannot vanish as otherwise $w_{(1)}-w_{(2)}$ would be orthogonal to all of~$L^\perp$ hence would belong to~$L$, which would contradict $\Span(w_{(1)},w_{(2)},L)=\lie{h}^*$.
If $\delta=1$, then taking $k=-w_{(i)}\cdot\pbf$ and $l=1$ for $i=1$ or $i=2$ yields a monopole of dimension $\Delta(k\mbf+l\pbf)=1/2$, so that the abelian theory of interest has a pair of free monopoles in the infrared and is not good.
For $\delta\geq 2$ all monopoles~\eqref{Delta-kmbf-lpbf} have dimension $\Delta\geq 1$ (except for $k=l=0$ of course), yet we now prove that the theory is nevertheless not good.

\paragraph{Orbifolded free monopoles.}

The gauge group contains a subgroup $\ZZ_\delta$ generated by
\begin{equation}
  e^{2\pi i x} \in G \ , \quad x = \frac{(w_{(1)}\cdot\pbf)\mbf-\pbf}{\delta}\in\lie{h}
\end{equation}
that acts trivially on all hypermultiplets since
\begin{equation}
  w_{(1)}\cdot x = 0 \ , \qquad
  w_{(2)}\cdot x = \frac{(w_{(1)}-w_{(2)})\cdot\pbf}{\delta} = \pm 1 \in \ZZ \ ,
\end{equation}
and other weights $w\preceq R$ are orthogonal to $\mbf$ and $\pbf$ hence to~$x$.
Gauging the corresponding $\ZZ_\delta$ one-form symmetry (which acts on Wilson lines of $\ZZ_\delta\subset G$) transforms our gauge theory to a $G/\ZZ_\delta$ gauge theory with the same hypermultiplet content.
The lattice of monopole sectors becomes larger, and in particular $L^\perp\cap\Lambdamon$ grows from $\ZZ\mbf+\ZZ\pbf$ to the larger lattice
\begin{equation}
  \bigl\{ k\mbf+l\pbf \bigm| w_{(i)}\cdot(k\mbf+l\pbf)\in\ZZ \text{ for } i = 1,2 \bigr\} \ .
\end{equation}
Two sectors are particularly interesting: $(k,l)=(w_{(i)}\cdot\pbf,-1)/\delta$ for $i=1$ and for $i=2$, whose dimensions~\eqref{Delta-kmbf-lpbf} are both $\Delta=1/2$.
As a result the $G/\ZZ_\delta$ gauge theory is ugly, due to having a pair of free monopoles in the infrared.
The original $G$ gauge theory is obtained by gauging the $\ZZ_\delta$ topological (zero-form) symmetry, which orbifolds the pair of monopoles into a $\CC^2/\ZZ_\delta$ Coulomb branch.
Our notion of good theory forbids orbifolds of free fields in the infrared, hence this case $\delta\geq 2$ is ruled out, which concludes our proof of~\eqref{tech-assumption-abelian} by contradiction.

\paragraph{Concrete example theory.}

The simplest example of such a phenomenon is a $G=U(1)\times U(1)$ theory with two hypermultiplets of charges $(1,1)$ and $(1,-1)$, namely four chiral multiplets of charges $(\pm 1,\pm 1)$.  Then
\begin{equation}
  \Delta(\mbf=(m_1,m_2)) = \frac{1}{2} |m_1+m_2| + \frac{1}{2} |m_1-m_2| = \max(|m_1|,|m_2|) ,
\end{equation}
so that all (non-trivial) monopoles have dimension $\Delta\geq 1$.
The dimension~$1$ monopoles are $\mbf=(\pm 1,0)$ and $\mbf=(0,\pm 1)$ and $\mbf=(\pm 1,\pm 1)$.
By the gauging procedure mentioned above, however, the theory is equivalent to a $\ZZ_2$ orbifold of a pair of free magnetic monopoles.
From this point of view, the dimension~$1$ monopoles appear as quadratic combinations of the more ``elementary'' dimension~$1/2$ monopoles with $\mbf=(\pm 1/2,\pm 1/2)$ in the $G/\ZZ_2$ gauge theory.

\subsection{Two-node circular quivers with unitary gauge group factors}
\label{subsec:circular}

\begin{figure}[b]\centering
  \begin{tikzpicture}[scale=2,
    ->-/.style = {
      decoration = {markings, mark = at position #1 with {\arrow{stealth}}},
      postaction = {decorate}}]
    \node(A)[rounded rectangle,inner sep=2pt,draw] at (2.3,0) {\!$n_{1}$\!};
    \node(B)[rounded rectangle,inner sep=2pt,draw] at (3.6,0) {\!$n_{2}$\!};
    \node(C)[rectangle,inner sep=2pt,draw] at (1.3,-0.5) {\!$M_{1}$\!};
    \node(D)[rectangle,inner sep=2pt,draw] at (4.6,-0.5) {\!$M_{2}$\!};
    \draw[->-=.55](A) to [bend right=30] (B);
    \draw[->-=.55](B) to[bend right=30](A);
    \draw[->-=.55](B) to[bend left=45] node [midway,below]{$(q_{12},\tq_{12})$}(A);
    \draw[->-=.55](A) to[bend left=45] node [midway,above]{$(q_{21},\tq_{21})$}(B);
    \draw[->-=.55](A) to[in=60,out=120,loop] node [midway,above]{$\phi_1$} (A);
    \draw[->-=.55](B) to[in=60,out=120,loop] node [midway,above]{$\phi_2$} (B);
    \draw[->-=.55](A.215) to (C.25);
    \draw[->-=.55](C.50) to node [midway,above left=-3pt]{$(q_1,\tq_1)$} (A.190);
    \draw[->-=.55](D.145) to node [midway,above right=-3pt]{$(q_2,\tq_2)$} (B.-25);
    \draw[->-=.55](B.-50) to (D.170);
  \end{tikzpicture}
  \caption{\label{fig:two-node}Two-node circular quiver with unitary gauge nodes and flavors}
\end{figure}
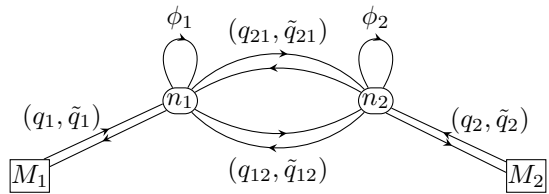

We now return to non-abelian quivers, and evaluate $\Nst$ exactly for an example family of quivers: $U(n_1)\times U(n_2)$ gauge theories (with $n_1,n_2\geq 1$) with bifundamental hypermultiplets in the $n_1\nbar_2$ and $\nbar_1n_2$ representations, and $M_i$ fundamental hypermultiplets of $U(n_i)$ for $i=1,2$.  As depicted in \autoref{fig:two-node}, we denote by $(q_{12},\tq_{12})$, $(q_{21},\tq_{21})$, $(q_1,\tq_1)$, and $(q_2,\tq_2)$ the bottom components of these hypermultiplets, with for instance $q_{12}$ and $\tq_{21}$ transforming in the $n_1\nbar_2$ representation.

\paragraph{Electric symmetries.}
The electric symmetry consists of a $U(M_i)$ symmetry rotating fundamental hypermultiplets of the $U(n_i)$ gauge group, for $i=1,2$, as well as a $U(2)$ symmetry rotating the bifundamental hypermultiplets,\footnote{For circular quivers this symmetry is specific to a two-node quiver, for which $q_{12}$ and $\tq_{21}$ transform in the same representation of the gauge groups.} modulo $U(1)\subset U(n_1)$ and $U(1)\subset U(n_2)$ gauge transformations which identify the $U(1)$ subgroups of $U(M_1)$, $U(M_2)$, and $U(2)$:
\begin{equation}
\Gelec = \bigl(U(M_1)\times U(M_2)\times U(2)\bigr)/U(1)^2 \ .
\end{equation}
The corresponding electric current multiplets are (or rather, have as bottom components) gauge-invariant products of the hypermultiplet scalars, modulo F-term relations, as follows:
\begin{center}
  \begin{tabular}{c*4{>$c<$}}
    \toprule
    & \multicolumn{3}{c}{Electric currents} & \text{F-term relations}
    \\\cmidrule(r){2-4}\cmidrule(l){5-5}
    Bottom component & \tq_1 q_1 & \tq_2 q_2 & \begin{pmatrix}\Tr(\tq_{12}q_{12}) & \Tr(\tq_{12}\tq_{21}) \\ \Tr(q_{21}q_{12}) & \Tr(q_{21}\tq_{21}) \end{pmatrix} & \begin{array}{@{}c@{}} \Tr(\tq_{12}q_{12}) + \Tr(q_{21}\tq_{21}) \\ \quad = \Tr(\tq_1 q_1) = \Tr(\tq_2 q_2) \end{array}
    \\
    \rule{0pt}{13pt}Representation & M_1 \Mbar_1 & M_2 \Mbar_2 & 2\;\overline{2} & 1 + 1
    \\\bottomrule
  \end{tabular}
\end{center}
Observe that for $M_1,M_2\geq 1$ the electric symmetry includes a $U(1)$ factor, while this factor is absent if exactly one of $M_1$ and $M_2$ vanishes.  If both $M_1$ and $M_2$ vanish, the diagonal $U(1)\subset U(n_1)\times U(n_2)$ gauge symmetry does not act on any field, hence the theory has a free vector multiplet and is bad.
We return to the well-understood issue of goodness momentarily.

\paragraph{Zero-monopole single-trace mixed moduli.}

Our aim is to evaluate $\Nst$.  We start here with the zero-monopole sector.  For general quivers we have seen near~\eqref{Be-def} that the mixed moduli to be counted take the form $B_{ee'}$ for all pairs of edges with the same source and target.  Since our present quiver has a pair of bifundamental hypermultiplets of $U(n_1)\times U(n_2)$, we get four products involving~$\phi_1$, and four involving~$\phi_2$.  They are equal up to a sign by~\eqref{Be-Breversed},
\begin{equation}
  \begin{pmatrix}\Tr(\tq_{12}\phi_1q_{12}) & \Tr(\tq_{12}\phi_1\tq_{21}) \\ \Tr(q_{21}\phi_1q_{12}) & \Tr(q_{21}\phi_1\tq_{21}) \end{pmatrix}
  = - \begin{pmatrix}\Tr(\phi_2\tq_{12}q_{12}) & \Tr(\phi_2\tq_{12}\tq_{21}) \\ \Tr(\phi_2q_{21}q_{12}) & \Tr(\phi_2q_{21}\tq_{21}) \end{pmatrix} .
\end{equation}
This quadruplet transforms in the adjoint representation of the $U(2)\subset\Gelec$ symmetry acting on the bifundamentals.
In addition, the trace vanishes thanks to the F-term condition~\eqref{Be-sum} of~$\phi_1$ or~$\phi_2$.
This leaves a triplet of mixed moduli in the adjoint of $SU(2)\subset\Gelec$.
If $n_1=1$ (resp.\ $n_2=1$), then the scalar~$\phi_1$ (resp.\ $\phi_2$) can be pulled out of the trace, so that these mixed moduli all factorize.
This means that the single-trace mixed moduli with zero weight under the magnetic symmetry transform as
\begin{equation}\label{Nst-circ-0}
  \Nst|_{\text{coef }\mu^0} = \begin{cases}
    0 & \text{if } \min(n_1,n_2) = 1 \ , \\
    \adj SU(2)_{\textnormal{elec}} & \text{if } n_1,n_2\geq 2 \ .
  \end{cases}
\end{equation}
The result can also be obtained by noting that the single-trace mixed modulus $B_{(12)} - B_{\overline{(21)}} = \Tr(\tq_{12}\phi_1q_{12}) - \Tr(q_{21}\phi_1\tq_{21})$ that was counted in the zero-weight sector in \autoref{subsec:quivers} is not invariant under $SU(2)\subset\Gelec$.

For a general quiver gauge theory with unitary gauge groups and only fundamental and bifundamental hypermultiplets, as considered in \autoref{subsec:quivers}, we expect
\begin{equation}\label{Nst-circ-0-gen}
  \Nst|_{\text{coef }\mu^0} = g' + \sum_{i<j} \adj SU(M_{ij})
\end{equation}
where $g'\leq g$ is the genus of the quiver without $U(1)$ factors and in which multiple edges are replaced by a single one, and $SU(M_{ij})$ is the flavour symmetry rotating all bifundamental hypermultiplets of $U(n_i)$ and $U(n_j)$.
The zero-weight subspace of the $\Gelec$~representation~\eqref{Nst-circ-0-gen} correctly reproduces the genus appearing in~\eqref{Nst-circ-00} since $g'+\sum_{i<j}(M_{ij}-1) = g$.

\paragraph{Goodness condition.}

Our aim is to get the complete~$\Nst$, so we consider the low-lying monopole sectors.
As a first step we rederive known conditions for the theory to be good~\cite{Gaiotto:2008ak,Assel:2012cj}.
Thanks to our additivity properties \eqref{Un-Delta-split-0} and~\eqref{Un-Delta-split} of monopole dimensions, the dimension of any monopole can be written as a sum of dimensions of tiny monopoles, namely sectors $\mbf=(\mbf_{ia})_{1\leq i\leq 2,1\leq a\leq n_i}$ such that all $\mbf_{ia}\in\{0,+1\}$ or all $\mbf_{ia}\in\{0,-1\}$.  We concentrate on the $\{0,+1\}$ sign.  Up to gauge-equivalence, tiny sectors are characterized by the traces
\begin{equation}
  \tau_i=\sum_a\mbf_{ia}\in[0,n_i] \ .
\end{equation}
For our quiver, the dimension~\eqref{dim-tiny} of tiny monopoles simplifies to
\begin{equation}\label{dim-tiny-2}
  \Delta(\mbf) = (\tau_1 - \tau_2)^2 + \beta_1 \tau_1 + \beta_2 \tau_2
\end{equation}
in terms of the balances
\begin{equation}
  \beta_1=n_2-n_1+M_1/2 \text{ and } \beta_2=n_1-n_2+M_2/2 \ .
\end{equation}

Of course this dimension vanishes for $\tau_1=\tau_2=0$.  The theory is good if and only if $\Delta\geq 1$ for any other values of $\tau_i\in[0,n_i]$.
The necessary conditions corresponding to $(\tau_1,\tau_2)$ among $(1,0)$, $(0,1)$, $(1,1)$ are
\begin{equation}\label{good-conditions-2}
  \beta_1 \geq 0 , \quad \beta_2 \geq 0 , \quad \beta_1 + \beta_2 \geq 1 \ .
\end{equation}
These conditions are also sufficient: either $\tau_1\neq\tau_2$ and the first term in~\eqref{dim-tiny-2} is at least~$1$ while the rest is non-negative, or $\tau_1=\tau_2\neq 0$ and we have $\Delta=(\beta_1+\beta_2)\tau_1\geq\tau_1\geq 1$.
Note that $\beta_1 + \beta_2 = (M_1 + M_2)/2$ so the last condition states that there are at least two fundamental flavours in total.
We henceforth assume that the theory is good, namely that \eqref{good-conditions-2}~holds.

\paragraph{Magnetic symmetries.}

For generic values of $N_1,N_2,M_1,M_2$ the magnetic symmetry group is simply $U(1)_T^2$.
There are additional magnetic symmetries whenever a monopole has $\Delta(\mbf)=1$.  Since monopole dimensions are sums of tiny monopole dimensions, which themselves are at least~$1$, the only monopoles we have to consider are the tiny ones.
We find that \eqref{dim-tiny-2} can only be equal to~$1$ if
\begin{itemize}
\item $|\tau_1-\tau_2|=1$ and $\beta_1\tau_1+\beta_2\tau_2=0$, which requires either $(\tau_1,\tau_2)=(1,0)$ and $\beta_1=0$, or $(\tau_1,\tau_2)=(0,1)$ and $\beta_2=0$,
\item or $\tau_1=\tau_2$ and $(\beta_1+\beta_2)\tau_1=1$, which requires $\beta_1+\beta_2=1$ (namely $M_1+M_2=2$) and $(\tau_1,\tau_2)=(1,1)$.
\end{itemize}
The first case corresponds to the standard $SU(2)$ flavour symmetry enhancement of the $U(1)_T$ topological symmetry of a balanced node, as in a linear quiver.
The second case leads to an $SU(2)$ flavour symmetry enhancement of the diagonal $U(1)_T$ topological symmetry of the two nodes, and is specific to circular quivers.
Both effects can coexist if $(\beta_1,\beta_2)=(1,0)$ or $(0,1)$.
To summarize, we have the following magnetic symmetry (the cases $\beta_1>\beta_2$ can be obtained by symmetry)
\begin{equation}
  \Gmag = \begin{cases}
    U(1)\times SU(2) & \text{if } \beta_1 = \beta_2 = 1/2 \ , \\
    SU(2)\times SU(2) & \text{if } (\beta_1,\beta_2) = (0,1) \ , \\
    SU(2)\times U(1) & \text{if } \beta_1 = 0 \ , \ \beta_2 \geq 3/2 \ , \\
    U(1)\times U(1) & \text{if } 1/2 \leq \beta_1 \leq \beta_2 \text{ and } (\beta_1,\beta_2) \neq (1/2,1/2) \ .
  \end{cases}
\end{equation}

\paragraph{Mixed moduli in monopole sectors, first type.}

The $B_1[0]^{(1,1)}$ multiplets can only come from the zero-monopole sector, which we have analyzed already, or sectors with $\Delta(\mbf)=1$.
The number $c_{1,1}^{\mbf}$ of mixed moduli coming from each sector is given in~\eqref{c11mbf} as the number of flavour symmetries of an auxiliary theory with gauge group $G_{\mbf}/U(1)_{\mbf}$ and hypermultiplets~$\Vhyp^{\mbf}$.
As described in~\eqref{in the chiral ring.}, some of these moduli factorize as the product of the bare monopole and an electric current of the original theory.

The first class of monopole sectors occurs for $\beta_1=0$, $\beta_2\geq 1$.
Consider the tiny monopole sector with $(\tau_1,\tau_2)=(1,0)$, namely $\mbf_{ia}=\delta_{i1}\delta_{a1}$ for $i=1,2$ and $1\leq a\leq n_i$.  This is the sector responsible for an $SU(2)$ enhancement of the $U(1)$ topological symmetry of $U(n_1)$.
One has $G_{\mbf} = U(1)\times U(n_1-1)\times U(n_2)$ and the auxiliary theory coincides with the original quiver theory with $n_1$ replaced by $n_1-1$.
If $n_1\geq 2$ the flavour symmetries are identical in the two theories, so that every mixed modulus in this sector factorizes.
If $n_1=1$, the balance condition $\beta_1=0$ imposes $n_2=1$ and $M_1=0$, so that there is no $U(M_1)$ symmetry to contend with.  In that case a change of basis in the $U(1)\times U(1)$ gauge groups shows that the theory is the tensor product of two SQED theories (with $2$ and $M_2$ flavours), for which we already know that $\Nst=0$.

\paragraph{Mixed moduli in monopole sectors, second type.}

The second class of monopole sectors occurs for $\beta_1+\beta_2=1$, that is, $M_1+M_2=2$.
Consider the sector with $(\tau_1,\tau_2)=(1,1)$, namely $\mbf_{ia}=\delta_{a1}$ for $i=1,2$ and $1\leq a\leq n_i$.
The gauge group is broken to $G_{\mbf} = U(1)\times U(n_1-1)\times U(1)\times U(n_2-1)$.
We split correspondingly the fundamental hypermultiplets as $q_1=(q_1^{\sharp},q_1^{\natural})$ with $q_1^{\sharp}$ being charged under $U(1)$ and neutral under $U(n_1-1)$ and $q_1^{\natural}$ vice-versa, and likewise for~$q_2$.
The monopole background lifts the two scalars $q_1^{\sharp}$ and~$q_2^{\sharp}$.
We introduce a similar notation $q_{12}^{\sharp}$ and $q_{12}^{\natural}$ for the unlifted parts of the bifundamental~$q_{12}$, which consists of the entry charged under both $U(1)$ factors of~$G_{\mbf}$ and the $(n_1-1)\times(n_2-1)$ representation, respectively.
With this notation, the auxiliary theory consists of two decoupled sets of fields: the original quiver with $(n_1,n_2)$ replaced by $(n_1-1,n_2-1)$ and all hypermultiplets decorated with a $\natural$ superscript, and a $T[SU(2)]$ theory consisting of the anti-diagonal $U(1)$ vector multiplet and of $q_{12}^{\sharp}$ and $q_{21}^{\sharp}$.

For $n_1,n_2\geq 2$, the auxiliary theory has an additional $SU(2)$ symmetry compared to the original theory, with the electric currents being explicitly
\begin{equation}
  \begin{aligned}
    \tq_1^{\natural} q_1^{\natural} & \in \adj(U(n_1-1)) \ ,
    & \tq_2^{\natural} q_2^{\natural} & \in \adj(U(n_2-1)) \ , \\
    \begin{pmatrix}\Tr(\tq_{12}^{\natural}q_{12}^{\natural}) & \Tr(\tq_{12}^{\natural}\tq_{21}^{\natural}) \\ \Tr(q_{21}^{\natural}q_{12}^{\natural}) & \Tr(q_{21}^{\natural}\tq_{21}^{\natural}) \end{pmatrix} & \in \adj(U(2)) \ ,
    & \begin{pmatrix}\tq_{12}^{\sharp}q_{12}^{\sharp} & \tq_{12}^{\sharp}\tq_{21}^{\sharp} \\ q_{21}^{\sharp}q_{12}^{\sharp} & q_{21}^{\sharp}\tq_{21}^{\sharp} \end{pmatrix} & \in \adj(U(2)) \ ,
  \end{aligned}
\end{equation}
subject to F-term relations~\eqref{F-term-with-background}, which read
\begin{equation}
  \Tr(\tq_1^{\natural} q_1^{\natural}) = \Tr(\tq_2^{\natural} q_2^{\natural})
  = \Tr(\tq_{12}^{\natural}q_{12}^{\natural}) + \Tr(q_{21}^{\natural}\tq_{21}^{\natural}) \ ,
  \qquad
  \Tr(\tq_{12}^{\sharp}q_{12}^{\sharp}) + \Tr(q_{21}^{\sharp}\tq_{21}^{\sharp}) = 0 \ .
\end{equation}
As described in~\eqref{in the chiral ring.}, the product of the bare monopole with a current is a the monopole dressed by the image of the current under the projection from $\Vhyp$ to $\Vhyp^{\mbf}$.
This projection has $q_i^{\mbf} = q_i^{\natural}$ and $q_e^{\mbf} = q_e^{\natural} + q_e^{\sharp}$.
Thus, the dressed monopoles $\Ocal_{\mbf}\tq_i^{\natural}q_i^{\natural}$ are factorized, while only the sum of the two $\adj(SU(2))$ dressed monopoles factorizes.
There remains a non-factorizable $B_1[0]^{(1,1)}$ multiplet in the $\adj(SU(2))$ representation of~$\Gelec$.

For $n_1=1$ and $n_2\geq 2$, the hypermultiplets $q_1^{\natural}$, $q_{12}^{\natural}$ and $q_{21}^{\natural}$ disappear, so that the auxiliary theory only has $\adj(SU(n_2-1))+\adj(SU(2))$ flavour symmetry (here we took into account F-term relations).
The dressed monopoles all factorize in that case.
Likewise, for $n_2=1$ the auxiliary theory has a reduced flavour symmetry, so that all $B_1[0]^{(1,1)}$ multiplets (coming from this monopole sector) factorize.

\paragraph{Single-trace mixed moduli in the two-node circular quiver.}

We are ready to summarize our conclusion.  For our quiver, $\Nst$ receives a monopole contribution precisely when $n_1,n_2\geq 2$.
Combining with \eqref{Nst-circ-0} we find
\begin{equation}\label{Nst-circ}
  \Nst = \begin{cases}
    0 & \text{if } \min(n_1,n_2) = 1 \ , \\
    \adj SU(2)_{\textnormal{elec}} & \text{if } n_1,n_2\geq 2 \text{ and } M_1+M_2\geq 3 \ , \\
    \adj SU(2)_{\textnormal{elec}} \times \adj SU(2)_{\textnormal{mag}} & \text{if } n_1,n_2\geq 2 \text{ and } M_1+M_2 = 2 \ ,
  \end{cases}
\end{equation}
where $SU(2)_{\textnormal{elec}}$ is the electric symmetry acting on the bifundamental hypermultiplets, and $SU(2)_{\textnormal{mag}}$ is the magnetic symmetry that contains the $U(1)_T$ topological symmetry of the diagonal subgroup $U(1)\subset U(n_1)\times U(n_2)$.

For a circular quiver with $k\geq 3$ nodes, the quiver no longer has edges with multiplicity, hence the $\adj SU(2)_{\textnormal{elec}}$ symmetry disappears and $\Nst$ does not receive an enhancement from the zero-monopole sector.
As for the two-node case, there are two classes of monopole sectors with $\Delta(\mbf)=1$.
\begin{itemize}
\item The first arises from one or more consecutive balanced nodes, and leads to an auxiliary theory $(G_{\mbf}/U(1)_{\mbf},\Vhyp^{\mbf})$ with the same (or lower) flavour symmetry as the original quiver, so that the resulting $B_1[0]^{(1,1)}$ multiplets are factorized.
\item The second arises from the tiny monopole with $\tau_1=\tau_2=\dots=1$, which has $\Delta(\mbf)=(M_1+M_2+\dots)/2=1$ provided the quiver has exactly two fundamental hypermultiplets in total.  The auxiliary theory is given by the original quiver with all $n_i$ replaced by $n_i-1$, together with the $U(1)^{k-1}$ theory that is mirror to SQED with $k$ flavours.  This leads to a single non-factorizable $B_1[0]^{(1,1)}$ multiplet in this sector, unless some $n_i=1$.  This last condition arises exactly as in the two-node case.
\end{itemize}
Taking into account the $\mbf$ and $-\mbf$ sectors and the zero-monopole sector, we obtain an $\adj SU(2)_{\textnormal{mag}}$ worth of single-trace mixed moduli.
This establishes~\eqref{Nst-circular}.

The formula~\eqref{Nst-circular} for $\Nst$ must be invariant under mirror symmetry, but the condition $\min(n_i)\geq 2$ seems worrisome at first sight.  Thankfully, circular quivers related by mirror symmetry have the same value of $\min(n_i)$, which in the Hanany--Witten brane construction is the number of D3 branes that fully wrap the circle direction, denoted by~$L$ in~\cite[Section 2.2]{Assel:2012cj}.

This completes our discussion of circular quivers.  It would be interesting to determine the precise expression of $\Nst$ for a general quiver, including all possible magnetic enhancements due to $\Delta=1$ monopole sectors.

% ===============================================

\paragraph{Acknowledgements.}
We thank Costas Bachas for a closely related collaboration and discussions, and Antoine Bourget for useful comments and examples.
B.L.F.~was employed by the Institut Philippe Meyer (\'Ecole Normale Sup\'erieure, Paris, France) for a large proportion of the work. I.L.~is supported by the Deutsche Forschungsgemeinschaft under Germany's Excellence Strategy --- EXC-2094\,--\,390783311 (the ORIGINS Excellence Cluster).  I.L.~thanks the LPENS and LPTHE for the hospitality during the final stages of this work.

\appendix

% ===============================================
% 					SECTION APPENDIX A
% ===============================================

\section{Representation theory statements}
\label{app:representations}

\subsection{Tensor products with the adjoint representation}
\label{app:tensor-adjoint}

\paragraph{Set-up and main lesson.}

In the main text we seek lower bounds on the term $\singlet_G(\adj(G)\otimes S^2\Vhyp)$ appearing in the index.
Upon decomposing $G$ into factors and $\Vhyp$ into irreducible representation, the question boils down to analyzing, for a (non-abelian) simple compact Lie group~$K$,
\begin{itemize}
\item $\singlet_K(\adj(K)\otimes V_\mu\otimes V_\nu)$ for irreducible representations of highest weights $\mu$ and~$\nu$;
\item $\singlet_K(\adj(K)\otimes S^2V_\mu)$ and $\singlet_K(\adj(K)\otimes\Lambda^2V_\mu)$, especially for real or quaternionic~$V_\mu$.
\end{itemize}
The notion of highest weight requires a splitting of roots into positive and negative roots, and we denote by $\alpha_j$ the corresponding simple roots of~$K$.
Here and in \autoref{app:tensor-adjoint-explicit}, respectively, we outline how these questions are addressed in mathematical work to appear by the second author and I.~Smilga~\cite{LeFloch-Smilga}, and deduce some concrete results used in the main text.\footnote{We acknowledge use of the Mathematica package \texttt{LieART}~\cite{Feger:2019tvk} to improve our intuition and conjecture the results.}

We recall that the Weyl group is generated by reflections
\begin{equation}
  w_\alpha \colon \lambda \mapsto
  \lambda - \langle\lambda,\alpha^\vee\rangle \alpha \ , \qquad
  \alpha^\vee \coloneqq \frac{2}{\langle\alpha,\alpha\rangle} \alpha \ ,
\end{equation}
associated to roots~$\alpha$ of~$K$, where $\alpha^\vee$ is the corresponding coroot.
The (closed) fundamental Weyl chamber, defined as the set of weights~$\lambda$ such that $\langle\lambda,\alpha_j\rangle\geq 0$ for all~$j$, is a fundamental domain of the Weyl group, in the sense that each orbit under the Weyl group has exactly one point in this chamber.

We begin with the singlets in $\adj(K)\otimes V_\mu\otimes V_\nu$.
As we justify shortly, there are exactly two cases where such gauge singlets occur~\cite{LeFloch-Smilga}:
\begin{itemize}
\item if $\mu=\nubar$, there is at least~$1$ gauge singlet (except for $\mu=\nubar=0$), and up to~$\rank(K)$ (the typical case): the precise number is the number of simple roots~$\alpha_j$ such that $\langle\mu,\alpha_j\rangle>0$;
\item if $\mu-\nubar$ is a root of~$K$, there is $0$ or $1$ gauge singlet, for instance there is one if both $\mu$ and $\nu$ lie strictly inside the fundamental Weyl chamber.
\end{itemize}
To exactly determine the number of singlets in $\singlet_G(\adj(G)\otimes S^2\Vhyp)$, needed to count single-trace $B_1[0]^{(1,1)}$ multiplets exactly, one must fully track both types of gauge singlets.
Instead, our lower bound in the main text is obtained by only considering the gauge singlets with $\mu=\nubar$, and ignoring those with $\mu\neq\nubar$.
In the case $\mu=\nu$, a natural question is whether these gauge singlets occur in $\adj(K)\otimes S^2V_\mu$ or $\adj(K)\otimes\Lambda^2V_\mu$.
The question is richest for $\mu=\nu=\nubar$, namely for real and quaternionic representations, and we return to it in \autoref{app:tensor-adjoint-explicit}, ignoring the singlets with $\mu-\mubar\in\roots(K)$, which is why we will only write lower bounds.

\paragraph{Weyl character formula.}

By Schur's lemma, a product $W\otimes V$ of irreducible representations of~$K$ has exactly one gauge singlet if $W=\bar{V}$, and otherwise has none.
More generally, for any representation~$W$, the singlets in $W\otimes V_\nu$ are in one-to-one correspondence with copies of~$V_{\nubar}$ inside~$W$.
Thus, the problem of interest boils down to decomposing $\adj(K)\otimes V_\mu$ into irreducible representations.

We recall the character of the adjoint representation,
\begin{equation}\label{char-adj}
  \chi_{\adj(K)}(x) = \rank(K) + \sum_{\alpha\in\roots(K)} e^{\alpha\cdot x} \ ,
\end{equation}
and the Weyl character formula\footnote{A word of warning: the Vandermonde determinant here can be recast as a product over half of the roots only (for instance, positive roots), whereas $\ZVand$ is a product over all roots.}
\begin{equation}\label{Weyl-char}
  \chi_{V_\mu}(x) = \chi_\mu \coloneqq \frac{\sum_{\sigma\in W} (-1)^\sigma e^{(\sigma(\rho+\mu))\cdot x}}{\text{Vandermonde}} \ , \qquad \text{Vandermonde} = \sum_{\sigma\in W} (-1)^\sigma e^{(\sigma(\rho))\cdot x} \ ,
\end{equation}
in which the Weyl group elements~$\sigma$ act on weights~$\mu$ by Weyl reflections parallel to roots, and $\rho$ is the Weyl vector, defined equivalently as the half-sum of all positive roots or as the sum of fundamental weights, so that
\begin{equation}
  \langle\rho,\alpha_j^\vee\rangle=1 , \qquad
  \langle\rho,\alpha_j\rangle=\langle\alpha_j,\alpha_j\rangle/2
  \quad\text{for}\quad 1\leq j\leq\rank(K) \ .
\end{equation}
The Weyl character formula naturally extends beyond dominant weights~$\mu$ (namely highest weights of some representation), and we use the same notation~$\chi_\mu$ for it.

\paragraph{Characters with non-dominant~$\mu$.}

To understand $\chi_\mu$ for a weight~$\mu$ that is not dominant, one should map $\rho+\mu$ to the Weyl chamber using one or more Weyl reflections, and use the following relation for any Weyl group element $\sigma\in W$:
\begin{equation}\label{Weyl-reflection-characters}
  \chi_\mu
  = \frac{\sum_{\sigma'\in W} (-1)^{\sigma'} e^{(\sigma'(\rho+\mu))\cdot x}}{\text{Vandermonde}}
  = (-1)^\sigma \frac{\sum_{\sigma''\in W} (-1)^{\sigma''} e^{(\sigma''\circ\sigma(\rho+\mu))\cdot x}}{\text{Vandermonde}}
  = (-1)^\sigma \chi_{\sigma(\rho+\mu)-\rho} \ .
\end{equation}
Let us distinguish three cases when analysing~$\chi_\mu$.
We will use at this point that $\langle\rho,\alpha_j^\vee\rangle=1$, and that $\langle\lambda,\alpha_j^\vee\rangle$ are integers for any weight~$\lambda$ of a finite-dimensional representation.

If $\rho+\mu$ is strictly inside the fundamental Weyl chamber, then all $\langle\rho+\mu,\alpha_j^\vee\rangle$ are strictly positive integers, namely are at least~$1$.
In that case, $\langle\mu,\alpha_j^\vee\rangle\geq 0$, namely $\mu$ is the highest weight of some representation.  The character $\chi_\mu$ defined by the Weyl character formula is the character of some genuine highest-weight representation~$V_\mu$.

If $\rho+\mu$ is on the boundary of the fundamental Weyl chamber, then $\langle\rho+\mu,\alpha_j^\vee\rangle=0$ for some~$j$.  The Weyl reflection $\sigma=w_{\alpha_j}$ (which has $(-1)^\sigma=-1$) maps it to itself since
\begin{equation}
  \sigma(\rho+\mu) = \rho+\mu - \langle\rho+\mu,\alpha_j^\vee\rangle\alpha_j = \rho+\mu \ .
\end{equation}
Thus, $\chi_\mu=-\chi_\mu$, namely this character vanishes.
In fact, we have $\chi_\mu=0$ more generally whenever $\rho+\mu$ is on any fixed plane of any Weyl reflection, namely if $\rho+\mu$ is on the boundary of any Weyl chamber.

Finally, if $\rho+\mu$ is strictly outside the fundamental Weyl chamber, then it must be mapped to the fundamental Weyl chamber (or its boundary) after some Weyl reflections using~\eqref{Weyl-reflection-characters}.

\paragraph{Generic situation versus Racah--Speiser algorithm.}

In view of~\eqref{char-adj} and~\eqref{Weyl-char},
the character of $\adj(K)\otimes V_\mu$ is
\begin{equation}\label{char-adj-Vmu}
  \begin{aligned}
    \chi_{\adj(K)\otimes V_\mu}(x) & = \chi_{\adj(K)}(x) \chi_\mu(x) \\
    & = \biggl( \rank(K) + \sum_{\alpha\in\roots(K)} e^{\alpha\cdot x} \biggr)
    \frac{\sum_{\sigma\in W} (-1)^\sigma e^{(\sigma(\rho+\mu))\cdot x}}{\text{Vandermonde}}
    \\
    & = \rank(K) \chi_\mu(x)
    + \sum_{\alpha\in\roots(K)} \frac{\sum_{\sigma\in W} (-1)^\sigma e^{(\sigma(\rho+\mu+\alpha))\cdot x}}{\text{Vandermonde}} \ ,
  \end{aligned}
\end{equation}
where we have used that the set of roots is invariant under the Weyl group to change $e^{\alpha\cdot x}$ to~$e^{\sigma(\alpha)\cdot x}$ in the sum.
This reduces to a sum of $\dim K$ characters:
\begin{equation}\label{char-adj-Vmu-2}
  \chi_{\adj(K)\otimes V_\mu}(x)
  = \rank(K) \chi_\mu(x) + \sum_{\alpha\in\roots(K)} \chi_{\mu+\alpha}(x) \ .
\end{equation}
For $\mu$ deep enough in the fundamental Weyl chamber, this yields the decomposition into irreducible representations,
\begin{equation}\label{adjGammaVmu}
  \adj(K)\otimes V_\mu = (V_\mu)^{\rank(K)} \oplus \bigoplus_{\alpha\in\roots(K)} V_{\mu+\alpha} \ , \qquad
  \text{for generic $\mu$} \ .
\end{equation}

For $\mu$ near the boundary of the Weyl chamber, some of the $\mu+\alpha$ may be outside the Weyl chamber.
In that case, they fail to be dominant weights.
The Racah--Speiser algorithm consists of listing all summands in~\eqref{adjGammaVmu}, mapping their (incorrect) ``highest weights'' to dominant weights using Weyl reflections, and cancelling characters that appear with opposite signs, as well as characters whose $\rho+\nu$ lies on a Weyl chamber boundary.
A case-by-case analysis of ABCDEFG simple Lie algebras in~\cite{LeFloch-Smilga} proves that in all cases, the number of copies of~$V_\mu$ that remain after applying the algorithm to~\eqref{adjGammaVmu} is precisely as we stated early in this section, namely equal to the number of simple roots~$\alpha_j$ such that $\langle\mu,\alpha_j\rangle>0$.

\subsection{Describing singlets and their symmetries}
\label{app:tensor-adjoint-explicit}

\paragraph{Expliciting copies of the adjoint representation.}

The second type of singlets to analyse are those in $\adj(K)\otimes S^2V_\mu$ and $\adj(K)\otimes\Lambda^2V_\mu$.
As these sit in $\adj(K)\otimes V_\mu\otimes V_\mu$, they are particular examples of the singlets we have found so far, and they can only exist if $\mu=\mubar$ or $\mu-\mubar\in\roots(K)$.
Our lower bounds in the main text only take into account the singlets with $\mu=\mubar$, namely for real or quaternionic~$V_\mu$.
For this case, the number of singlets in $\adj(K)\otimes V_\mu\otimes V_\mu$ is the number of simple roots~$\alpha_j$ such that $\langle\alpha_j,\mu\rangle>0$, and we must determine which ones are symmetric or antisymmetric.

Decompose $\mu=\sum_j\mu_j\varpi_j$ on the basis of fundamental weights~$\varpi_j$ (which is dual to the coroot basis~$\alpha^\vee_j$).
In the tensor product $V_{\mu_1\varpi_1}\otimes\cdots\otimes V_{\mu_k\varpi_k}$, where $k=\rank(K)$, the irreducible summand of highest weight is~$V_\mu$.
For any $j$ with $\mu_j\neq 0$, the non-trivial action of the Lie algebra $\lie{k}$ on $V_{\mu_j\varpi_j}$ defines a singlet of $\Hom(\adj(K)\otimes V_{\mu_j\varpi_j},V_{\mu_j\varpi_j})$, and tensoring with the identity defines a singlet
\begin{equation}
  T_j \colon \left\{
    \begin{array}{@{}r@{}l@{}}
      \adj(K) \otimes V_{\mu_1\varpi_1}\otimes\cdots\otimes V_{\mu_k\varpi_k} & {} \to V_{\mu_1\varpi_1}\otimes\quad\ \cdots\ \quad\otimes V_{\mu_k\varpi_k} \\
      x \ \otimes \ \ \, v_1 \, \ \ \otimes\cdots\otimes \ v_k \quad\ \quad & {} \mapsto v_1\otimes\dots\otimes(xv_j)\otimes\dots\otimes v_k
    \end{array}
  \right.
\end{equation}
We also denote by $T_j$ the projection to a singlet in $\Hom(\adj(K)\otimes V_\mu,V_\mu)$.
It can be shown~\cite{LeFloch-Smilga} that the $T_j$ (for $\mu_j\neq 0$) are linearly independent, so that they provide precisely the expected number of singlets in $\Hom(\adj(K)\otimes V_\mu,V_\mu)\simeq \adj(K)\otimes V_\mu\otimes\Vbar_\mu\simeq \adj(K)\otimes V_\mu\otimes V_\mu$.
There only remains to understand the symmetry properties of these singlets.

\paragraph{Inequalities between (anti)symmetric complexities.}

To determine how swapping the two $V_\mu$ factors acts on~$T_j$ one must take into account the isomorphisms $\adj(K)\otimes V_\mu\otimes\Vbar_\mu\simeq \adj(K)\otimes V_\mu\otimes V_\mu$.
One finds that the swap acts as $T_j\mapsto \pm T_{f(j)}$ where the involution $f\colon\{1,\dots,k\}\to\{1,\dots,k\}$ is defined by how conjugation acts on fundamental weights, namely $\Vbar_{\varpi_j}=V_{\varpi_{f(j)}}$, and the sign is more subtle to determine, see~\cite{LeFloch-Smilga}.
\begin{itemize}
\item Whenever $f(j)\neq j$, the linear combinations $T_j\pm T_{f(j)}$ have opposite eigenvalues under the swap, hence provide one singlet in each of $\adj(K)\otimes S^2V_\mu$ and $\adj(K)\otimes\Lambda^2V_\mu$.
\item Whenever $f(j)=j$, the singlet $T_j$ is symmetric or antisymmetric according to whether $V_\mu$ is quaternionic or real, respectively.
\end{itemize}
Thus, there are more singlets in $\adj(K)\otimes S^2V_\mu$ or $\adj(K)\otimes\Lambda^2V_\mu$ for quaternionic or real~$V_\mu$.
The equality case can only happen if the only non-zero $\mu_j$ have $f(j)\neq j$, but in this case the tensor product $V_{\mu_1\varpi_1}\otimes\cdots\otimes V_{\mu_k\varpi_k}$ reduces to a tensor product of pairs $V_{\mu_j\varpi_j}\otimes V_{\mu_{f(j)}\varpi_{f(j)}}$ (with necessarily $\mu_{f(j)}=\mu_j$) which are all real representations, so that the tensor product is a real representation and its subrepresentation $V_\mu$ is real as well.
Thus, the equality case is excluded for quaternionic representations:
\begin{equation}
  \begin{aligned}
    \singlet_K(\adj(K)\otimes\Lambda^2V_\mu) & \geq \singlet_K(\adj(K)\otimes S^2V_\mu)
    && \text{for real } V_\mu \ ,
    \\
    \singlet_K(\adj(K)\otimes S^2V_\mu) & > \singlet_K(\adj(K)\otimes\Lambda^2V_\mu)
    && \text{for quaternionic } V_\mu \ .
  \end{aligned}
\end{equation}
In addition, for real~$V_\mu$, either $V_\mu$ is trivial and there are no singlets whatsoever, or it is not, and there is at least one singlet in $\singlet_K(\adj(K)\otimes V_\mu\otimes V_\mu)$, hence at least one in the left-hand side $\singlet_K(\adj(K)\otimes\Lambda^2V_\mu)$ of the inequality.
We have collected these conclusions as~\eqref{pHH-S-gt-Lambda} in the main text.

\subsection{A bound on zero-weight spaces}
\label{app:zeroweight}

Here we switch gears completely and establish lower bounds~\eqref{lower-zeroweight} on the dimension of zero-weight subspaces of $\NCCbar_I\otimes\NCC_I$ and of (anti)symmetric squares of $\NHH_I$ and~$\NRR_I$.  Specifically, we prove
\begin{equation}
  \begin{aligned}
    \dim\bigl((\Nbar\otimes N)_{\mu^0}\bigr) & \geq \dim N \quad
    & & \text{for $N$ among } \NCC_I,\NHH_I,\NRR_I \ ,
    \\
    \dim\bigl((S^2N)_{\mu^0}\bigr) & \geq \tfrac{1}{2} \dim N
    & & \text{for $N$ among } \NHH_I,\NRR_I \ ,
    \\
    \dim\bigl((\Lambda^2\NHH_I)_{\mu^0}\bigr) & \geq \tfrac{1}{2} \dim\NHH_I \ .
  \end{aligned}
\end{equation}
By choosing a basis $(e_\alpha)_{1\leq\alpha\leq\dim N}$ of $N$ with definite weights, and the dual basis $e^*$ of $\Nbar$, it is clear that $e^{*\alpha}\otimes e_\alpha$, $1\leq\alpha\leq\dim N$, are linearly independent vectors in the zero-weight space of $\Nbar\otimes N$, yielding the first lower bound.

Next, consider $N$ to be real or quaternionic.
Given a basis $e_\alpha$ of~$N$ with definite weights~$\lambda_\alpha$, the elements $e_\alpha\otimes e_\beta+e_\beta\otimes e_\alpha$ for $\alpha\leq\beta$ form a basis of $S^2N$, and those with $\lambda_\beta=-\lambda_\alpha$ are in the zero-weight space.
Distinguishing the zero-weight subspace of~$N$ from the rest, we find
\begin{equation}
  \begin{aligned}
    \dim\bigl((S^2N)_{\text{weight }0}\bigr)
    & = \sum_{\alpha\leq\beta,\lambda_\alpha=\lambda_\beta=0} 1 + \sum_{\alpha<\beta,\lambda_\alpha=-\lambda_\beta\neq 0} 1
    \\
    & = \binom{\dim N_{\text{weight }0}+1}{2}
    + \frac{1}{2} \sum_{\lambda\neq 0} \bigl(\dim N_{\text{weight }\lambda}\bigr)^2
    \\
    & \geq \frac{1}{2} \sum_\lambda \dim N_{\text{weight }\lambda}
    = \frac{1}{2} \dim N \ ,
  \end{aligned}
\end{equation}
To get the inequality, we used that $\frac{1}{2}(n+1)n\geq\frac{1}{2}n$ for all integers~$n\geq 0$.

Finally, we turn to a quaternionic representation~$\NHH_I$ (corresponding to real~$\RRR_I$).
The dimension of the zero-weight space of $\NHH_I$ cannot be~$1$: because the invariant tensor $\epsilon\in\Lambda^2\NHH$ reduces to a non-degenerate antisymmetric pairing on the zero-weight space, that space must have an even dimension.
With the same logic as above one finds
\begin{equation}
  \begin{aligned}
    \dim\bigl((\Lambda^2\NHH_I)_{\text{weight }0}\bigr)
    & = \binom{\dim (\NHH_I)_{\text{weight }0}-1}{2}
    + \frac{1}{2} \sum_{\lambda\neq 0} \bigl(\dim (\NHH_I)_{\text{weight }\lambda}\bigr)^2
    \\
    & \geq \frac{1}{2} \sum_\lambda \dim (\NHH_I)_{\text{weight }\lambda}
    = \frac{1}{2} \dim\NHH_I \ ,
  \end{aligned}
\end{equation}
where crucially we used that $\frac{1}{2}(n-1)n\geq\frac{1}{2}n$ for any integer $n\geq 0$ except $n=1$.

% ===============================================

\end{document}